\documentclass[sigconf,screen,nonacm]{acmart}

\usepackage{graphicx}
\usepackage{amsmath}
\usepackage{tabularray}
\UseTblrLibrary{booktabs}
\usepackage{xcolor} 
\usepackage{colortbl}
\usepackage{makecell}
\usepackage{algpseudocode}
\usepackage{transparent}
\usepackage{pifont}  
\usepackage{subcaption}
\usepackage{cleveref}
\usepackage[linesnumbered,ruled,vlined]{algorithm2e}
\usepackage{tabularx}
\usepackage[english]{babel}
\usepackage{multirow}
\usepackage{pifont}
\usepackage{circledsteps}
\usepackage{adjustbox}
\usepackage[frozencache,cachedir=.]{minted}
\usepackage{tcolorbox}
\tcbuselibrary{listingsutf8} 
\usepackage{enumitem}
\usepackage{tikz} 
\definecolor{applegreen}{rgb}{0.55, 0.71, 0.0}

\newcommand{\bcheckmark}{\ding{51}} 
\newcommand{\btimes}{\ding{55}}     
\newcommand{\bslash}{$\boldsymbol{/}$}
\crefformat{section}{\S#2#1#3} 
\crefformat{subsection}{\S#2#1#3}
\crefformat{subsubsection}{\S#2#1#3}

\definecolor{table_blue}{HTML}{1071e5}
\definecolor{boxheadbgcol}{HTML}{D5CCC3}
\definecolor{boxheadcol}{HTML}{7E3F3F}

\definecolor{category1}{HTML}{EDF5FF}  
\definecolor{category2}{HTML}{FCFCCA}  

\definecolor{green1}{HTML}{018A67}
\definecolor{blue1}{HTML}{1868B2}
\definecolor{red1}{HTML}{DE582B}
\definecolor{orange1}{HTML}{F3A332}
\definecolor{blue2}{HTML}{F1F9FE}
\definecolor{orange2}{HTML}{F8F3E4}

\newcommand{\circled}[1]{%
    \tikz[baseline=(char.base)]{%
        \node[draw=white, 
              circle,
              fill=black, 
              inner sep=1pt, 
              text=white, 
              font=\bfseries 
              ] (char) {#1};%
    }%
}

\AtBeginDocument{%
  }

\newcommand{\eg}{\textit{e.g.},\ }
\newcommand{\ie}{\textit{i.e.},\ }

\definecolor{customblue}{HTML}{91BFFA}
\colorlet{customblue_transparent}{customblue!45}

\newcommand{\framework}{How2Compress\xspace}
\newcommand{\abbvemphasis}{\textit{golden emphasis}\xspace}
\newcommand{\proxyemtarget}{\textit{proxy emphasis target}\xspace}

\begin{document}

\title{\textsc{\framework}: Scalable and Efficient Edge Video Analytics via Adaptive Granular Video Compression}

\author{Yuheng Wu}
\email{yuhengwu@kaist.ac.kr}
\affiliation{%
  \institution{KAIST}
  \city{Daejeon}
  \country{Republic of Korea}
}
\author{Thanh-Tung Nguyen}
\email{tungnt@kaist.ac.kr}
\affiliation{%
  \institution{KAIST}
  \city{Daejeon}
  \country{Republic of Korea}
}
\author{Lucas Liebe}
\email{lucasliebe@kaist.ac.kr}
\affiliation{%
  \institution{KAIST}
  \city{Daejeon}
  \country{Republic of Korea}
}
\author{Quang Tau}
\email{quangntau1223@kaist.ac.kr}
\affiliation{%
  \institution{KAIST}
  \city{Daejeon}
  \country{Republic of Korea}
}
\author{Pablo Espinosa Campos}
\email{pabloe@kaist.ac.kr}
\affiliation{%
  \institution{KAIST}
  \city{Daejeon}
  \country{Republic of Korea}
}
\author{Jinghan Cheng}
\email{chengjh@kaist.ac.kr}
\affiliation{%
  \institution{KAIST}
  \city{Daejeon}
  \country{Republic of Korea}
}
\author{Dongman Lee}
\authornote{Corresponding Author}
\email{dlee@kaist.ac.kr}
\affiliation{%
  \institution{KAIST}
  \city{Daejeon}
  \country{Republic of Korea}
}

\renewcommand{\shortauthors}{Yuheng Wu et al.}

\begin{abstract}

With the rapid proliferation of the Internet of Things, video analytics has become a cornerstone application in wireless multimedia sensor networks. To support such applications under bandwidth constraints, learning-based adaptive quantization for video compression have demonstrated strong potential in reducing bitrate while maintaining analytical accuracy. However, existing frameworks often fail to fully exploit the fine-grained quality control enabled by modern blockbased video codecs, leaving significant compression efficiency untapped.

In this paper, we present \framework, a simple yet effective framework designed to enhance video compression efficiency through \textit{precise, fine-grained} quality control at the macroblock level. \framework is a plug-and-play module and can be seamlessly integrated into any existing edge video analytics pipelines. We implement \framework on the H.264 codec and evaluate its performance across diverse real-world scenarios. Experimental results show that \framework achieves up to $50.4\%$ bitrate savings and outperforms baselines by up to $3.01\times$ without compromising accuracy, demonstrating its practical effectiveness and efficiency.
\textit{Code is available at \href{https://github.com/wyhallenwu/how2compress}{link} and a reproducible docker image at \href{https://hub.docker.com/r/wuyuheng/how2compress}{link}.}
\end{abstract}




\maketitle

\section{Introduction}
\label{sec:intro}

\begin{figure}[t!] 
    \centering
    \includegraphics[width=0.98\linewidth]{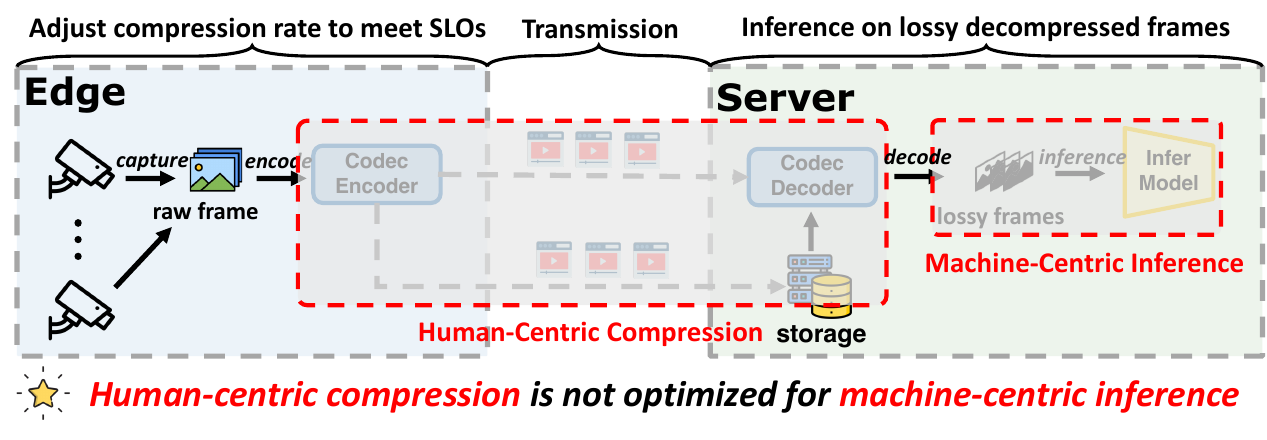}
    \vspace{-1.5em}
    \caption{Machine-centric video compression is a key component to unleash wireless sensor network efficiency.} 
    \label{fig:application}
\end{figure}

Video analytics has emerged as a critical application in edge computing, enabling intelligent services such as traffic management~\cite{nguyen2025octopinf, control,intro2,management1} and urban surveillance~\cite{ILCAS,intro1,monitoring1,monitoring2, nguyen2023preacto}. As illustrated in Fig.~\ref{fig:application}, typical edge video analytics systems compress video data at edge cameras and offload it to nearby edge clusters for inference~\cite{accmpeg,casva,elf,reducto,ILCAS,cross-cam,entro}. These systems must operate under stringent bandwidth constraints and fluctuating scene complexities~\cite{casva,ILCAS,intro2, nguyen2025octopinf}, while simultaneously adhering to strict Service Level Objectives (SLOs) that demand both high inference accuracy and low end-to-end latency~\cite{reducto,casva,ILCAS,context-img-offload1,accmpeg,entro, nguyen2025octopinf}. To meet these requirements,  video compression becomes a critical enabler. 
To meet these requirements,  video compression becomes a critical enabler. However, there is a key misalignment in current edge video analytics system. Video codec is primarily designed for human visual perception and often overlook semantically important features that are essential for machine-centric analytics tasks.

\definecolor{green1}{HTML}{018A67}
\definecolor{blue1}{HTML}{1868B2}
\definecolor{red1}{HTML}{DE582B}
\definecolor{orange1}{HTML}{F3A332}
\definecolor{blue2}{HTML}{F5FFFF}
\definecolor{orange2}{HTML}{FEEBD5}

\begin{figure*}[thb]
    \centering

    \begin{minipage}{0.23\textwidth}
        \raggedright
        \includegraphics[width=0.95\textwidth]{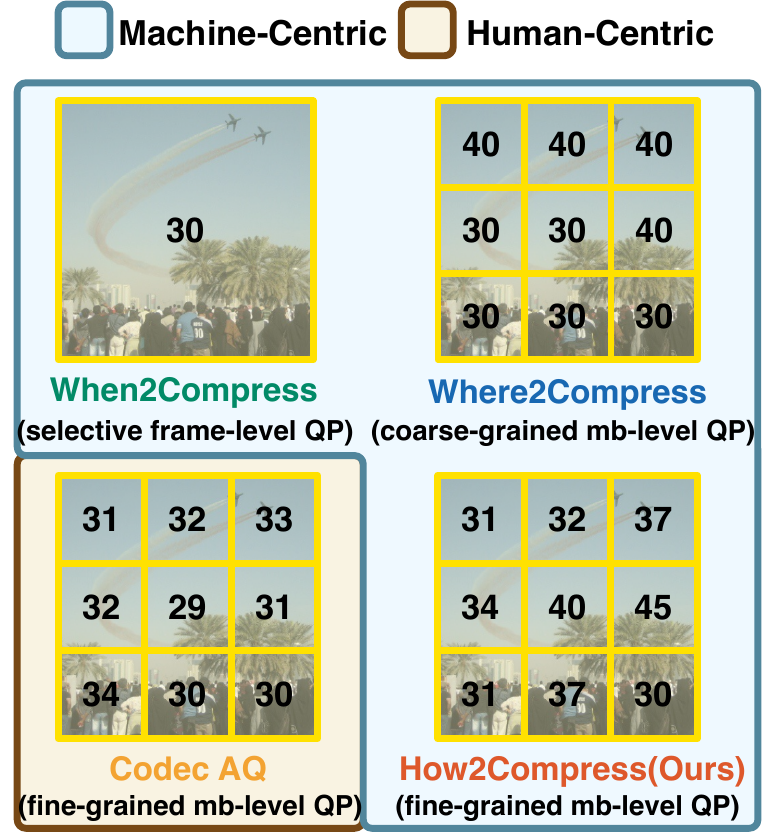}
    \end{minipage}%
    \begin{minipage}{0.74\textwidth}
        \raggedright
        \captionof{table}{Comparison between our proposed framework and prior works.}
        \vspace{-1.0em}
        \label{tab:comparison}
    
        \resizebox{0.99\textwidth}{!}{
        \renewcommand{\arraystretch}{1.6} 

        \begin{tabular}{c|c|c c c c c}
        \toprule
       \textbf{Category} & \textbf{Framework} & \makecell{\textbf{Macroblock-level}  \\ \textbf{Assignment}} &  \makecell{\textbf{Fine-grained} \\ \textbf{Assigment}} & \makecell{\textbf{Low Training} \\ \textbf{Overhead}$^*$} & \makecell{\textbf{Low 
 Inference} \\ \textbf{Overhead}$^*$$^\dagger$} & \makecell{\textbf{Machine-} \\ \textbf{Centric}}\\
        \midrule
        \multirow{2}{*}{\textbf{\makecell{Uniform QP \\ \textcolor{green1}{(When2Compress)}}}} 
        & \textbf{CASVA}~\cite{casva} & \cellcolor{orange2}\btimes & \cellcolor{blue2}\bcheckmark  & \cellcolor{orange2}\btimes  & \cellcolor{blue2}\bcheckmark & \cellcolor{blue2}\bcheckmark   \\

        & \textbf{ILCAS}~\cite{ILCAS} & \cellcolor{orange2}\btimes & \cellcolor{blue2}\bcheckmark  & \cellcolor{blue2}\bcheckmark   & \cellcolor{orange2}\btimes & \cellcolor{blue2}\bcheckmark   \\ \midrule
        
        \textbf{\makecell{Coarse-grained \\ \textcolor{blue1}{(Where2Compress)}}} & \textbf{AccMPEG}~\cite{accmpeg} & \cellcolor{blue2}\bcheckmark  & \cellcolor{orange2}\btimes & \cellcolor{orange2}\btimes & \cellcolor{orange2}\btimes & \cellcolor{blue2}\bcheckmark  \\ \midrule
        
        \textbf{\makecell{Adaptive Quant \\ \textcolor{orange1}{(Codec AQ)}}} & 
        \makecell{\textbf{Blockbased Codec}~\cite{h264,nvaq,h265,h266,vp9,aom_av1} \\ (H.264, H.265, H.266, VP9, AV1)} & \cellcolor{blue2}\bcheckmark & \cellcolor{blue2}\bcheckmark & \bslash & \bslash & \cellcolor{orange2}\btimes \\ \midrule
        
        \textbf{\makecell{Fine-grained \\ \textcolor{red1}{(How2Compress)}}} & \textbf{\framework (Ours)} & \cellcolor{blue2}\bcheckmark  & \cellcolor{blue2}\bcheckmark  & \cellcolor{blue2}\bcheckmark  & \cellcolor{blue2}\bcheckmark & \cellcolor{blue2}\bcheckmark   \\
        \bottomrule
    \end{tabular}
    }
    \begin{flushleft}
        \small ${^*}$ evaluated on 1080p resolution videos\\
        \small $^\dagger$ e2e overhead latency $\leq$ 33.3ms (30fps) on the edge-level device (\eg Nvidia Jetson Orin Nano)
    \end{flushleft}

    \end{minipage}
    \caption{Comparison between our proposed framework and prior works. \framework is a superset of both \textit{When2Compress} and \textit{Where2Compress}. Each macroblock is $16 \times 16$ pixels.} 
    \label{fig:granularity-comparison}
    \vspace{-1.2em}
\end{figure*}

To address this gap, prior work has explored machine-centric adaptive quality assignment~\cite{casva,ILCAS,accmpeg,accelir}, differing in spatial granularity. As shown in Fig.~\ref{fig:granularity-comparison}, Frame-level methods~\cite{casva,ILCAS} apply a uniform Quantization Parameter (QP) per frame, deciding \textbf{when to compress} based on global scene importance. In contrast, macroblock-level methods~\cite{accmpeg,accelir} distinguish salient from non-salient regions, assigning binary (high/low) QPs per macroblock to decide \textbf{where to compress}.
Modern codecs~\cite{h264,h265,vp9,vp8,vvc,aom_av1}, however, support multi-level QP control at the macroblock level, enabling finer-grained and more expressive quality modulation. Yet, the exponentially large decision space at this scale poses a significant challenge, limiting existing methods from fully leveraging the codec's configurability.

\noindent\textbf{Goal and Insight.} In this work, we aim to improve video compression through fine-grained macroblock-level quality assignment while preserving downstream analytical accuracy. Our key insight is that macroblocks contribute \textit{unequally} to analytical performance and exhibit \textit{content-dependent resilience} to compression. By exploring this variability, we can assign \textit{just enough quality} to each macroblock  (\textbf{how to compress}), minimizing bitrate without sacrificing accuracy. Specifically, in this paper, we first assign a low QP as a base and then selectively apply fine-grained \textit{emphasis} (QP offset) to adjust macroblock-level quality. We refer to the optimal \textit{emphasis} decision for each macroblock as \abbvemphasis.

\noindent\textbf{Challenges.} We identify two key and non-trivial challenges:

\textbf{1) How can we determine a content adaptive but task agnostic signal?} Each macroblock contributes differently to downstream task performance, depending on its content and context. This creates a local-to-global optimization problem, where per-macroblock quality decisions collectively influence frame-level detection accuracy. However, the relationship between local quality and global task utility is non-linear and content-dependent, making it infeasible to predetermine macroblock importance through deterministic heuristics~\cite{accelir,scencode}.

\textbf{2) How can we efficiently navigate quality assignment in the exponential decision space?} Encoding a 1080p frame involves $\sim8160$ macroblocks ($M$), each with $N$ (\eg 5) possible emphasis levels. This leads to an exponential decision space of size $N^M$. A natural formulation is to treat this as a Markov Decision Process (MDP) and use planning or RL to discover optimal emphasis assignments. However, this approach faces three major difficulties: 1) \textit{Long-horizon decisions}~\cite{longrange-1,longrange-2,longrange-3}, where a full sequence of 8160 decisions must be made before evaluating the outcome. 2) \textit{Sparse reward signal}~\cite{sparse-1,sparse-2,sparse-3,sparse-4}, as performance feedback is only available after the entire assignment is completed. 3) \textit{Expensive exploration}, exploring such a large space requires extensive trial-and-error, yet the encoding process is computationally expensive. The inefficiency and scalability bottlenecks of MDP-based approaches in our setting necessitate a different solution strategy.

\noindent\textbf{Proposed Framework.} In this paper, we propose \framework, a self-supervised video compression framework that performs fine-grained quality assignment at the macroblock level. Rather than formulating emphasis assignment as a sequential decision-making problem, \framework reframes it as a segmentation task, which enables parallel and scalable inference. The core component of \framework is the Region-aware Emphasis Routing (RER) module, which predicts the optimal quality emphasis for each macroblock without requiring manual labels or exhaustive preprofiling~\cite{accelir}. RER is co-trained with a \textit{proxy emphasis target}, which serves as a dynamic pseudo-label that evolves based on downstream task feedback and spatiotemporal priors. This co-evolutionary training strategy allows the emphasis assignment model to efficiently explore the quality assignment space and converge toward high-utility compression strategies.

We integrate \framework as a plug-and-play module into the standard H.264 pipeline, which is the most widely adopted codec in practical edge deployments, and validate its performance across diverse real-world settings.

\noindent\textbf{Contributions.} Our main contributions are as follows:

\noindent1) We introduce the \textbf{Proxy Emphasis Target}, a dynamic supervisory signal that evolves during training and enables fine-grained quality assignment without relying on pre-profiling.
 
\noindent2) We design \textbf{Region-aware Emphasis Routing}, a self-supervised module that exploits spatial and temporal priors to efficiently assign compression quality across macroblocks in an exponential decision space.

\noindent3) We integrate \framework into the H.264 codec and demonstrate up to 50.4\% bitrate reduction and a $3.01\times$ improvement over state-of-the-art baselines, with negligible computational overhead on edge hardware.

\section{Background}

\framework is designed to be codec-agnostic, provided the codec supports encoding pixel blocks at varying quality levels (\eg H.264~\cite{h264}, H.265 (HEVC)~\cite{h265}, VP8~\cite{vp8}, VP9~\cite{vp9}, H.266 (VVC)~\cite{vvc} and AV1~\cite{aom_av1}). In this paper, we integrate \framework with the H.264 codec and evaluate it on object detection, tracking and keypoint detection (pose estimation). Both are most common practice in edge computing. To motivate our design, we first analyze the compression mechanism of H.264, and then examine the limitations of prior approaches in enabling fine-grained quality control at the macroblock level.

\begin{figure}[t]
    \centering
    \includegraphics[width=0.45\textwidth]{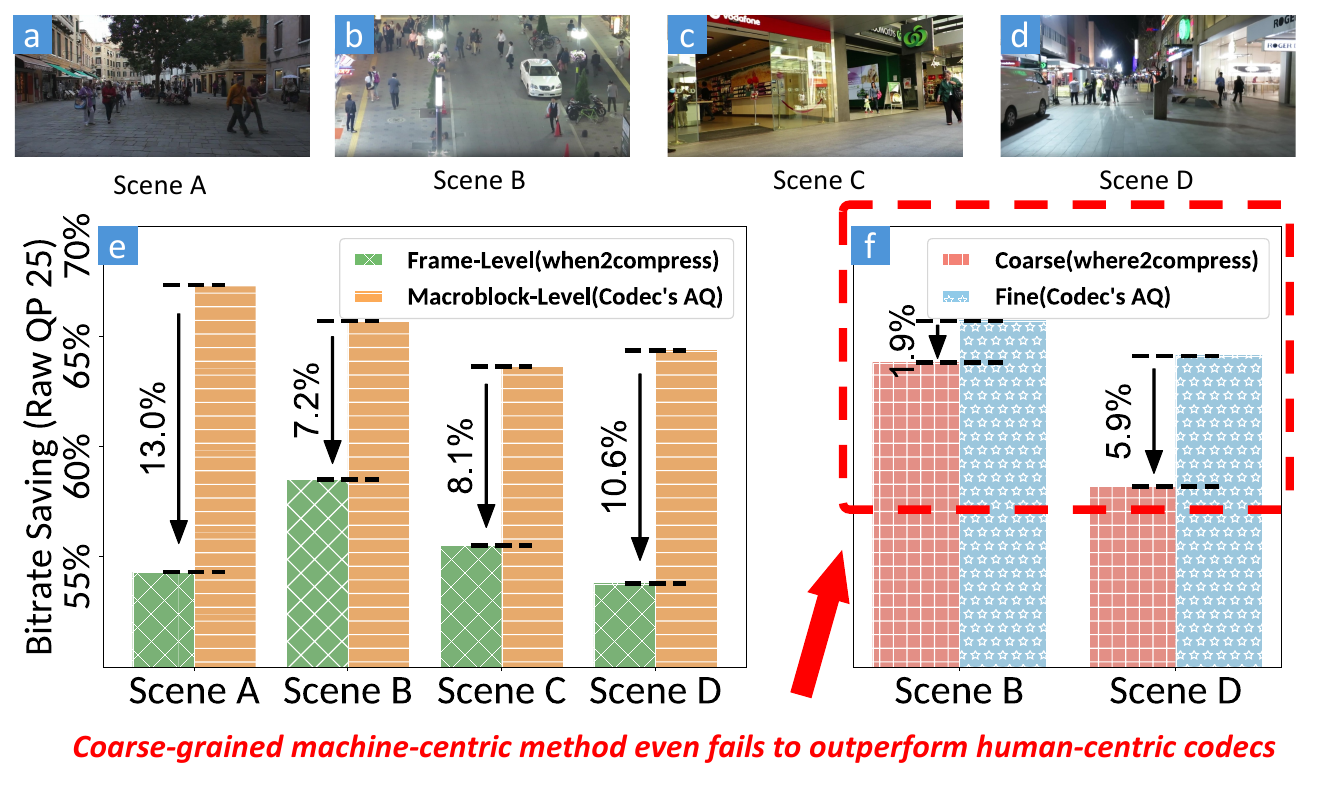}
    \vspace{-1em}
    \caption{Samples from MOT dataset (a-d). Bitrate savings over raw video using human-centric fine-grained (e) and machine-centric coarse-grained (f) compression.}
    \label{fig:motivation}
\end{figure}

\subsection{Preliminaries on Video Compression}
\label{sec:nvcodec}
Video compression aims to reduce spatial and temporal redundancies while preserving visual quality. The encoding process of H.264 consists of three main stages. First, each frame is \textit{typically} divided into $16\times16$ macroblocks. Each macroblock is predicted using either intra (within the current frame) or inter (motion-compensated from reference frames) modes. The difference between the predicted and actual macroblock forms the residual. Next, the residual undergoes a Discrete Cosine Transform (DCT), converting the data from the spatial domain to the frequency domain, concentrating the energy into fewer significant coefficients. Quantization is then applied to these coefficients, where they are divided by a quantization step, reducing precision. The degree of quantization is controlled by the QP, allowing a balance between compression efficiency and quality at the macroblock level. Finally, the quantized coefficients are further compressed through lossless entropy coding.

\subsection{Related Work}
\label{sec:related-work}

Prior efforts to reduce bandwidth in video analytics span three main strategies: (1) \textit{spatial pruning} (\eg cropping RoIs)\cite{vabus,elf,cross-cam,tvm,flexpatch,tileclipper}, (2) \textit{temporal pruning} (\eg frame filtering or semantic skipping)\cite{reducto,mm23-kfe,cova}, and (3) \textit{tuning global quality knobs} (\eg framerate, resolution, CRF)~\cite{casva,ILCAS}. While effective, these approaches are often tightly coupled to specific downstream models or system pipelines, limiting their generalizability.

In contrast, our work focuses solely on fine-grained quality assignment which adjusts macroblock-level QP to reduce bitrate while remaining agnostic to downstream tasks and models. This strategy can be easily integrated with other system-level optimizations for additional gains.

\noindent\textbf{\#When2Compress: Frame-level quality assignment.}
Methods like CASVA~\cite{casva} and ILCAS~\cite{ILCAS} dynamically adjust frame-level video settings (\eg resolution, framerate, QP) to balance latency and accuracy. However, they lack spatial granularity, often preserving high-quality macroblocks that are irrelevant to task performance, resulting in inefficient compression.

\noindent\textbf{\#Where2Compress: Region- or Macroblock-level quality assignment.} The most closely related work to ours is research that has explored quality assignment at the macroblock level~\cite{accmpeg,accelir}. Compared to frame-level approaches, this method offers significant advantages. Ideally, it allows for reducing the quality of macroblocks that are not directly associated with performance, while assigning necessary higher quality to those that influence the performance. Despite this potential, existing approaches fall short, failing to fully exploit the granular quality levels available at the macroblock level. For instance, AccMPEG~\cite{accmpeg} estimates the impact of each macroblock on accuracy and assigns quality based on its importance. But its reliance on an \textit{accuracy gradient} limits the quality assignment to a coarse binary decision (\ie high (QP 30) or low (QP 40)), thereby underutilizing the codec’s full range of quantization levels. AccelIR~\cite{accelir} targets on image restoration and conducts exhaustive pre-profiling of the benefit of each macroblock. While effective in restoration tasks, this approach is impractical for analytics tasks, where complex local-to-global dependencies cannot be captured by such pre-profiling.

\section{Measurement Study}
\label{sec:motivation}

\noindent\textbf{Experimental Setup}. All measurement experiments are conducted on NVIDIA RTX 3090 GPU. For video encoding, we use FFMPEG with libx264~\cite{ffmpeg}. To explore macroblock-level fine-grained quality adjustments, we activate the AQ mode in the \textit{variance} setting, which enables dynamic refinement of the QP at macroblock level based on content variance. Each video chunk has a framerate of 30 FPS and a GOP size of 30, consisting of 1 I-frame and up to 3 consecutive B-frames per group. Four representative video sequences from the MOT dataset~\cite{MOT17} are selected as shown in Fig.~\ref{fig:motivation}(a-d) to represent a diverse range of conditions, including varying view angles (horizontal \textit{vs.} vertical), object density (number of objects in a frame), luminance levels (daytime \textit{vs.} nighttime), and camera dynamics (stationary \textit{vs.} moving). To ensure a robust comparison, we first conduct a thorough search of the frame-level QP for each video sequence, identifying the highest QP value. Raw videos are encoded at QP 25 and all parameters are optimized to ensure that accuracy fluctuations remain within a margin of $2\%$.

\begin{figure*}[thb] \centering
    \includegraphics[width=0.98\textwidth]{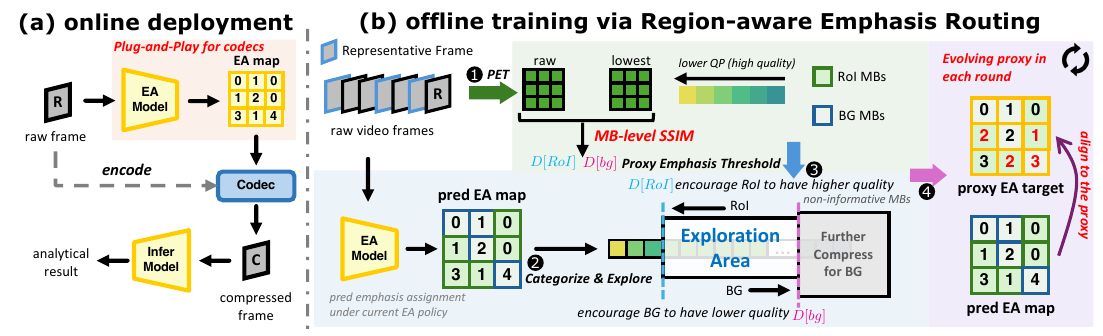}
    \vspace{-1em}
    \caption{Overview of \framework.} \label{fig:overview}
    \vspace{-1em}
\end{figure*}

\noindent\textbf{Observation \#1: Human-centric fine-grained AQ improves compression but struggles with content dynamics}. 
As shown in Fig.~\ref{fig:motivation}e, our results indicate that H.264’s built-in AQ enables more aggressive compression while maintaining comparable accuracy. However, its effectiveness diminishes in videos containing large, low-complexity flat regions (\eg Fig.~\ref{fig:motivation}b), where visual quality is more sensitive to degradation and thus requires additional bits to preserve perceptual fidelity. This limitation stems from AQ’s original design~\cite{h264-aq}, which prioritizes human visual perception rather than task-relevant machine perception. As a result, it may allocate bits inefficiently from an analytics perspective.

\noindent\textbf{Observation \#2: Potential of machine-centric, finer-grained macroblock-level quality assignment is promising but remains underexplored}. As shown in Fig.~\ref{fig:motivation}f, machine-centric coarse-grained methods (\ie AccMPEG) even fail to outperform built-in codec's AQ. This inefficiency underscores an untapped opportunity for macroblock-level, task-aware quality assignment to enhance both compression efficiency and task accuracy.

\noindent\textbf{Motivation.} These observations reveal a critical gap: existing built-in AQ techniques are misaligned with the needs of video analytics, and coarse-grained approaches fail to leverage the codec’s expressive capabilities. This motivates our design of a machine-centric, fine-grained macroblock-level quality assignment framework.

\section{Problem Formulation}
\label{sec:problem-formulation}

Given an input image of resolution $(H, W)$, we divide it into a grid of $n_w \times n_h$ non-overlapping macroblocks, where $n_w = \left\lceil \frac{H}{16} \right\rceil$ and $n_h = \left\lceil \frac{W}{16} \right\rceil$. Each macroblock covers a region of $16 \times 16$ pixels. Instead of predicting a unique QP for each macroblock, we employ a fixed base QP (\ie 45 in our setup) and assign an emphasis level $q_{i,j} \in \{0, 1, 2, 3, 4\}$ to each macroblock $M_{i,j}$\footnote{This formulation is designed to be compatible with the \textit{Emphasis Map} feature of NVIDIA’s Video Codec SDK~\cite{nvcodec}. (More design rationale in Appendix~\ref{apx:codec-impl})}. This emphasis level serves as a proxy for quality refinement: a higher $q_{i,j}$ implies a more negative offset to the base QP, thus improving the quality of that macroblock. 

We define the resulting emphasis configuration over the frame as an emphasis map $EM \in \mathbb{R}^{n_w \times n_h}$, where each entry $EM(i, j)$ denotes the emphasis level assigned to macroblock $M_{i,j}$. The optimization objective is to reduce the total emphasis cost (a surrogate for bitrate) while maintaining the application-level accuracy within a margin $\tau$ of the ground-truth performance under maximum quality. Formally, the objective is:

\begin{equation}
    \begin{aligned}
        \max_{\mathbf{q}} \, \mathbb{E}[A(\mathbf{q})] \quad 
        \text{s.t.} \quad \sum_{i=1}^{n_w} \sum_{j=1}^{n_h} EM(i,j) \text{ is minimized}, \\
        \text{and} \quad |\mathbb{E}(A(\mathbf{q})) - \mathbb{E}(G)| \leq \tau,
    \end{aligned}
    \label{eq:goal}
\end{equation}

Here, $A(\mathbf{q})$ denotes the application-level accuracy for frames decoded using the predicted emphasis map $\mathbf{q}$, while $\mathbf{E}(G)$ is the reference accuracy computed from frames encoded at the highest possible quality. The constraint ensures that the degradation in accuracy remains within an acceptable bound $\tau$. Intuitively, the goal is to learn an emphasis assignment strategy that minimally allocates quality to accuracy-insensitive macroblocks, while preserving or boosting quality in regions critical to the application’s performance.

\section{Methodology}
\textbf{Overview}. As shown in Fig.~\ref{fig:overview}, \framework consists of two stages: (a) online deployment and (b) offline training via Region-aware Emphasis Routing. In the deployment stage (Fig.~\ref{fig:overview}a), the \textit{Emphasis Assignment (EA) Model} (\cref{ssec:model}) predicts macroblock-level emphasis in parallel, producing an \textit{emphasis map}. This map guides QP refinement to control bitrate. The encoded video is then offloaded to a nearby edge server for inference. During training (Fig.~\ref{fig:overview}b), we first sample representative frames and compress them at the lowest quality to estimate per-macroblock SSIM. Based on this, two \textit{proxy emphasis thresholds} are derived to guide the exploration~\circled{1}. The EA map is then partitioned into RoI and Background regions~\circled{2}, enabling the \textit{Region-aware Emphasis Routing (RER)} module (\cref{subsec:emphasis-routing}) to apply \textit{region-aware exploration}~\circled{3}, generating a \textit{proxy emphasis target} that evolves over iterations. This proxy target serves as adaptive supervision~\circled{4}, encouraging the model to assign lower quality to non-informative regions while preserving task-relevant details. Through iterative updates, the EA Model learns to make fine-grained, task-aware quality decisions.

\subsection{Emphasis Assignment Model (EA Model)}
\label{ssec:model}

As discussed in \cref{sec:intro}, similar to the approach in~\cite{accmpeg}, the emphasis assignment problem can be cast as a segmentation task, where each $16 \times 16$ macroblock serves as an independent prediction unit. Solving this task requires two core capabilities: 1) the ability to capture both short-range and long-range semantic context~\cite{deeplab,segformer,mobileseg}, and 2) the computational efficiency necessary for real-time inference on edge devices (\eg sustaining about 30fps even for high-resolution inputs such as 1080p)~\cite{casva,accelir,accmpeg}. Therefore, the backbone network must strike a balance between computational efficiency and performance.

To this end, \framework is designed to be backbone-agnostic as long as the model meets above requirements. Leveraging recent advancements in lightweight architectures~\cite{deeplab,deeplabv3,convnext,transformer,vit,detr,segformer,topformer,mobilevitv2}, we adopt MobileViT v2 as our backbone due to two key advantages: 1) $16\times16$ pixel macroblocks aligns naturally with the patch-based design of vision transformers backbones, facilitating the modeling of relationships between these macroblocks and 2) its optimization tailored to mobile platforms. The Emphasis Assignment Model is trained offline on a centralized server using pre-encoded videos and distributed to edge devices for deployment.

\begin{figure}[t] \centering
    \includegraphics[width=0.48\textwidth]{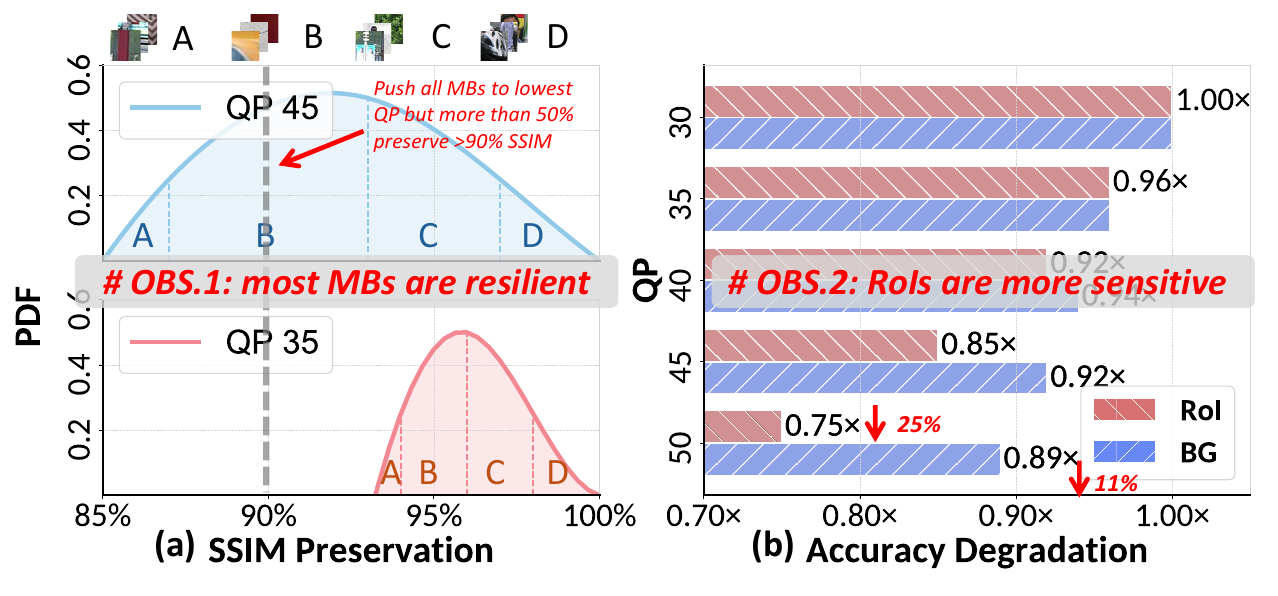}
    \vspace{-2.5em}
    \caption{Design motivation: (a) Different macroblocks ex-
hibit varying resilience to compression based on content
characteristics. (b) Macroblocks in RoI are more sensitive.} \label{fig:design-motivation}
\end{figure}

\vspace{-1em}

\begin{table}[t]
\centering
\caption{Notation used in Region-aware Emphasis Routing}
\label{tab:notation}
\begin{tabular}{ll}
\toprule
\textbf{Symbol} & \textbf{Description} \\
\midrule
$I_r$, $I_c$ & Raw and compressed input frames \\
$raw\_mb$, $low\_mb$ & Macroblocks from $I_r$ and $I_c$ \\
$EM$ & Emphasis Assignment model \\
$em[i,j]$ & Predicted emphasis level at macroblock $(i,j)$ \\
$proxy\_em[i,j]$ & Proxy target emphasis at macroblock $(i,j)$ \\
$p$ & Exploration probability \\
$\tau_{roi}$, $\tau_{bg}$ & Thresholds for RoI and BG macroblocks \\
$\mathcal{E}(a,b)$ & Exponential sampling between $a$ and $b$ \\
$Acc_r$, $Acc_c$ & Accuracy on raw and compressed input \\
$\lambda_1$, $\lambda_2$ & Weights for loss terms \\
$penalty$ & Scaling factor for alignment loss \\
\bottomrule
\end{tabular}
\end{table}

\subsection{Region-aware Emphasis Routing (RER)}
\label{subsec:emphasis-routing}

\textit{Region-aware Emphasis Routing} (RER) is a simple yet effective mechanism for fine-grained, macroblock-level emphasis assignment in video compression. It is designed to tackle two key challenges outlined in \cref{sec:intro}:
\textbf{C1}: how to determine a content-aware but task-agnostic \proxyemtarget\ for each macroblock, and
\textbf{C2}: how to efficiently navigate the large decision space of possible emphasis configurations.
To address these challenges, RER builds on the exploration-exploitation paradigm, with two core designs that accelerate and structure this process:
(1) \textit{Proxy Emphasis Threshold}, which narrows the exploration area by filtering out trivially saturated macroblocks. (2) \textit{Region-aware Exploration}, which focuses exploration differently across foreground (RoI) and background regions to efficiently search the QP assignment space. To further improve training efficiency, we utilize hardware-optimized codec implementations during offline training. We ensure that the QP assignment behavior is aligned with standard codec semantics, enabling seamless deployment across heterogeneous edge devices.

\subsubsection{Proxy Emphasis Threshold (PET)}
\label{ssec:pet}
\phantom{x}

\noindent\textbf{Challenge \#1: How to identify a content-aware yet task-agnostic supervisory signal?} In real-world scenarios, switching between tasks or analytical models is common. To accommodate this, we seek a supervision strategy that remains broadly applicable across diverse tasks and models, without requiring task-specific annotations or retraining.

\noindent\textbf{Observation \#1: Structural and textural information is generally useful, and most macroblocks are resilient to compression.}
Prior studies~\cite{texture1,texture2,texture3,texture4,texture5,texture6,texture7,texture8} emphasize that preserving fine-grained structural and textural details is essential for sustaining the performance of downstream tasks such as object detection. Motivated by this, our design aims to retain such cues during compression. To approximate their preservation, we adopt SSIM~\cite{ssim} as a proxy, recognizing that while it does not perfectly capture task relevance, it provides a reasonable content-aware signal. Importantly, the overall compression extent is ultimately guided by downstream task performance (see Appendix~\ref{apx:why-ssim}). Furthermore, as shown in Fig.~\ref{fig:design-motivation}a, we observe that even when macroblocks are compressed at the lowest quality, a substantial portion of structural cues remains intact. This suggests that many macroblocks are inherently resilient to aggressive compression. This insight enables us to shrink the exploration space by focusing efforts on the smaller subset of MBs that are more sensitive to quality loss.

\begin{algorithm}[t]
    \caption{Proxy Emphasis Threshold}
    \label{algo:proxy-em-threshold}
    \SetKwInOut{KwDef}{Input}
    \SetKwInOut{KwHyper}{Config}
    \SetKwProg{Fn}{Function}{:}{}
    \SetKwFunction{FProxyEmphasisThreshold}{ProxyEmphasisThreshold}
    
    \KwDef{$I_r$, $I_c$, $raw\_mb$, $low\_mb$, $EM$, $em$}
    \KwHyper{$\tau_{roi}$, $\tau_{bg}$}
    
    \BlankLine
    
    \Fn{\FProxyEmphasisThreshold{$I_r, I_c$}}{
        \textcolor{blue}{\tcp{SSIM threshold for RoI and BG}}
        $\text{D} \gets \left[ ~ \right]$\;
        \For{$i = 0, 1, \dots$}{
            \For{$j =0, 1, \dots$} {
                 $D \gets D \cup \{ \text{SSIM}(raw\_mb[i, j]$, $low\_mb[i, j]) \}$\;
            }
        }
        $ D \gets \text{sort}(D)$\;
        \Return{$D[\tau_{roi}]$, $D[\tau_{bg}]$}\;
    }
\end{algorithm}

\noindent\textbf{Design \#1: Proxy Emphasis Threshold.}
To operationalize this insight, we introduce the \textit{Proxy Emphasis Threshold} mechanism. As shown in Algorithm~\ref{algo:proxy-em-threshold} and Fig.~\ref{fig:overview}, we begin by compressing a representative raw frame to its lowest quality and compute SSIM scores for each macroblock to estimate its compression resilience~\circled{1}. These scores are sorted to define two percentile-based thresholds: one for regions of interest (RoI) and another for background (BG) regions. The thresholds determine which macroblocks are structurally important and guide the quality assignment during training, encouraging higher quality (lower QP) for RoI and more aggressive compression for resilient background blocks~\circled{3} later.

This threshold-driven process provides a content-aware and task-agnostic supervisory signal, enabling adaptive bitrate allocation without exhaustive search. Moreover, to ensure computational efficiency, this profiling step is performed only once on a small set of representative frames. The resulting thresholds are then reused across the video stream, taking advantage of the spatial and temporal redundancy common in edge video streams.

\subsubsection{Region-aware Dual Exploration} 
\label{ssec:rer}
\phantom{x}

\noindent\textbf{Challenge \#2: How to efficiently explore and exploit the macroblock decision space?}
Not all macroblocks contribute equally to accuracy. Efficient exploration requires a strategy that adapts emphasis decisions based on the task relevance of each macroblock, ensuring computational effort is directed where it matters most.

\noindent\textbf{Observation \#2: RoI MBs are generally more critical than those in Background.} Empirical analysis (Fig.~\ref{fig:design-motivation}b) shows that compression artifacts in RoI regions significantly degrade accuracy, while aggressive compression in BG regions has less effect. Prioritizing RoI macroblocks during exploration can potentially improve efficiency and accuracy retention.

\begin{algorithm}[t]
\caption{Region-aware Emphasis Routing}
\label{algo:em-routing}
\SetKwInOut{KwDef}{Input}
\SetKwInOut{KwHyper}{Config}
\SetKwProg{Fn}{Function}{:}{}
\SetKwFunction{FProxyEmphasisThreshold}{ProxyEmphasisThreshold}
\SetKwFunction{FEmRouting}{RegionAwareEmphasisRouting}

\KwDef{$I_r$, $I_c$, $raw\_mb$, $low\_mb$, $EM$, $em$}
\KwHyper{$p$, $\tau_{roi}$, $\tau_{bg}$, $\lambda_1$, $\lambda_2$, $penalty$}

\BlankLine

\Fn{\FEmRouting{$I_r, I_c$}}{
    \textcolor{blue}{\tcp{\small efficiently explore decision space}}
    $\tau_{roi}, \tau_{bg} \gets \FProxyEmphasisThreshold(I_r, I_c)$\;
    \Repeat{accuracy requirement not met}{
        $em[i, j] \gets EM(I_r)$\;
        \textcolor{blue}{\tcp{\small Dual-Exploration for RoI mb}}
        \For{$mb \in RoI$}{
            \If{$p' \sim \mathcal{U} < p \land em[i, j] \leq \tau_{roi}$} {
                $proxy\_em[i, j] \gets \mathcal{E}\left( (em[i,j], high \right)$\;
            }
        }
        \textcolor{blue}{\tcp{\small Dual-Exploration for BG mb}}
        \For{$mb \in BG$}{
            \If{$p' \sim \mathcal{U} < p \land em[i, j] \geq \tau_{bg}$}{
                $proxy\_em[i, j] \gets \mathcal{E}\left( low, em[i, j] \right)$\;
            }
            \Else{
                $proxy\_em[i, j] \gets proxy\_em[i, j] - 1$\;
            }
        }
        $loss1 \gets \mid Acc_{c} - Acc_{r}\mid$\;
        $loss2 \gets CE\left( em, proxy\_em \right) \cdot penalty$\;
        $loss \gets \lambda_1\cdot loss1 + \lambda_2\cdot loss2$\;
        Update $EM$\;
        decay $p$\;
    }
}
\end{algorithm}

\noindent\textbf{Design \#2: Region-aware Dual Exploration Strategy.}
To accelerate macroblock-level search, we adopt a region-aware dual exploration strategy that treats RoI and BG macroblocks differently. RoI regions, typically more critical for analysis, are guided toward higher quality via upward sampling, while BG regions are steered toward more aggressive compression~\circled{2}.

This region-aware design serves only as a heuristic to improve exploration efficiency. It remains broadly applicable across diverse tasks, as RoI-like regions (\eg objects, humans) tend to be structurally important regardless of the specific objective (\eg detection, tracking, pose estimation). Even in the absence of ground-truth boxes, pseudo-RoIs can be extracted using any pretrained object detectors, making the approach practical and generalizable. The final emphasis decisions are not hardcoded by region type but are supervised by the downstream task performance. As shown in Algorithm~\ref{algo:em-routing}, the EA model iteratively refines its emphasis map based on proxy thresholds and region-specific sampling, while a combined loss ensures the learned policy preserves task accuracy~\circled{4}. We refer to Appendix~\ref{apx:proof} for the theoretical justification of RER.



\section{Evaluation}

We evaluate \framework with a comprehensive real-world evaluation across 1) diverse edge devices, 2) varying content dynamics, 3) all baseline categories and 4) thorough ablations, revealing the following key findings:

1) \framework is compression-efficient. It achieves up to \textbf{50.4\%} bitrate reduction and up to \textbf{3.01$\times$} improvement over baselines without compromising accuracy by preserving task-relevant structure and discarding redundant detail (\cref{result:contents}).

2) \framework is lightweight and scalable. Its pipelined design interleaves QP assignment and encoding, keeping latency under 6 ms per 1080p frame and enabling real-time performance on resource-constrained devices (\cref{result:devices}).

3) \framework is robust and generalizable. Region-aware Emphasis Routing adapts across backbones, downstream models, and hyperparameters, enabling stable convergence and effective navigation of the fine-grained emphasis decision space (\cref{sec:ablation}).

\begin{table*}[t]
\centering
\caption{Bitrate (Mbps) comparison over MOT, Nvidia AI City, and VisDrone datasets. The accuracy drop remains within 2\% for detection and tracking tasks, and within 5\% for keypoint detection. \textbf{Bold}/\textcolor{red}{Red} highlights the best compression performance. \underline{Values} indicate the second-best compression.}
\resizebox{\textwidth}{!}{
\renewcommand{\arraystretch}{2}
\begin{tabular}{c|c|c|c|c|c|c|c|c|c|c|c|c|c|c|c|c}
\hline
\multirow{2}{*}{\Large\textbf{Method}} & \multicolumn{6}{c|}{\Large\textbf{Multiple Object Tracking~\cite{MOT17}}} & \multicolumn{3}{c|}{\Large\textbf{Nvidia AI City~\cite{aicity}}} & \multicolumn{7}{c}{\Large\textbf{VisDrone~\cite{visdrone}}} \\
\cline{2-17}
& \textbf{1702} & \textbf{1704} & \textbf{1709} & \textbf{1710} & \textbf{1711} & \textbf{1713} & \textbf{S01} & \textbf{S02} & \textbf{S03} & \textbf{086} & \textbf{117} & \textbf{137} & \textbf{182} & \textbf{268} & \textbf{305} & \textbf{339} \\
\hline
\multicolumn{17}{c}{\cellcolor{gray!10}\large \textit{\textbf{Frame-Level (baseline)}}} \\
\hline
\makecell{\large\textbf{Uniform QP}~\cite{casva,ILCAS} \\ \textcolor{green1}{(When2Compress)}}  & 4.076 & 1.646 & 3.986 & 5.027 & 5.117 & 3.711 & 3.323 & 4.313 & 7.439 & 4.27 & 3.59 & 5.53 & 8.52 & 9.08 & 3.81 & 2.52 \\
\hline
\multicolumn{17}{c}{\cellcolor{gray!10}\large\textit{\textbf{Macroblock-Level}}} \\
\hline
\makecell{\large\textbf{AccMPEG$^*$}~\cite{accmpeg} \\ \textcolor{blue1}{(Where2Compress)}} & \makecell{3.354 \\ (17.7\%$\downarrow$)} & \makecell{1.569 \\ (4.7\%$\downarrow$)} & \makecell{3.622 \\ (9.1\%$\downarrow$)} & \makecell{4.518 \\ (10.1\%$\downarrow$)} & \makecell{4.469 \\ (12.7\%$\downarrow$)} & \makecell{3.300 \\ (11.1\%$\downarrow$)} & \makecell{\underline{2.707} \\ (18.5\%$\downarrow$)} & \makecell{\underline{3.446} \\ (20.1\%$\downarrow$)} & \makecell{\underline{6.457} \\ (13.2\%$\downarrow$)} & \makecell{3.61 \\ (15.5\%$\downarrow$)} & \makecell{3.02 \\ (15.9\%$\downarrow$)} & \makecell{4.66 \\ (15.7\%$\downarrow$)} & \makecell{7.13 \\ (16.3\%$\downarrow$)} & \makecell{7.21 \\ (20.6\%$\downarrow$)} & \makecell{3.31 \\ (13.1\%$\downarrow$)} & \makecell{2.14 \\ (15.1\%$\downarrow$)} \\
\makecell{\large\textbf{Adaptive Quantization$^*$}~\cite{h264-aq,nvaq} \\ \textcolor{orange1}{(Codec AQ)}} & \makecell{\underline{2.914} \\ (28.5\%$\downarrow$)} & \makecell{\textbf{1.362} \\ (\textcolor{red}{17.3\%}$\downarrow$)} & \makecell{\underline{3.258} \\ (18.3\%$\downarrow$)} & \makecell{\underline{3.878} \\ (22.9\%$\downarrow$)} & \makecell{\underline{4.101} \\ (19.9\%$\downarrow$)} & \makecell{\underline{3.144} \\ (15.3\%$\downarrow$)} & \makecell{2.971 \\ (10.6\%$\downarrow$)} & \makecell{3.912 \\ (9.3\%$\downarrow$)} & \makecell{6.662 \\ (10.4\%$\downarrow$)} & \makecell{\underline{2.91} \\ (31.9\%$\downarrow$)} & \makecell{\underline{2.92} \\ (18.7\%$\downarrow$)} & \makecell{\underline{3.72} \\ (32.7\%$\downarrow$)} & \makecell{\underline{5.88} \\ (31.0\%$\downarrow$)} & \makecell{\textbf{4.64} \\ (\textcolor{red}{48.9\%}$\downarrow$)} & \makecell{\underline{3.09} \\ (18.9\%$\downarrow$)} & \makecell{\underline{2.06} \\ (18.3\%$\downarrow$)} \\
\makecell{\large\textbf{\framework} \\ \textcolor{red1}{(Machine-centric fine-grained)}} & \makecell{\textbf{2.023} \\ (\textcolor{red}{50.4\%}$\downarrow$)} & \makecell{\underline{1.388} \\ (15.7\%$\downarrow$)} & \makecell{\textbf{2.943} \\ (\textcolor{red}{26.2\%}$\downarrow$)} & \makecell{\textbf{3.469} \\ (\textcolor{red}{31.0\%}$\downarrow$)} & \makecell{\textbf{3.488} \\ (\textcolor{red}{31.8\%}$\downarrow$)} & \makecell{\textbf{2.759} \\ (\textcolor{red}{25.7\%}$\downarrow$)} & \makecell{\textbf{2.202} \\ (\textcolor{red}{33.7\%}$\downarrow$)} & \makecell{\textbf{3.182} \\ (\textcolor{red}{26.2\%}$\downarrow$)} & \makecell{\textbf{4.475} \\ (\textcolor{red}{39.8\%}$\downarrow$)} & \makecell{\textbf{2.53} \\ (\textcolor{red}{40.7\%}$\downarrow$)} & \makecell{\textbf{2.30} \\ (\textcolor{red}{35.9\%}$\downarrow$)} & \makecell{\textbf{3.39} \\ (\textcolor{red}{38.7\%}$\downarrow$)} & \makecell{\textbf{4.38} \\ (\textcolor{red}{48.6\%}$\downarrow$)} & \makecell{\underline{5.47} \\ (39.8\%$\downarrow$)} & \makecell{\textbf{2.25} \\ (\textcolor{red}{40.9\%}$\downarrow$)} & \makecell{\textbf{1.60} \\ (\textcolor{red}{36.5\%}$\downarrow$)} \\
\hline
\makecell{$\delta$ bitrate saving \\ over \underline{second-best}} & $1.77\times$ & $0.91\times$ & $1.43\times$ & $1.35\times$ & $1.60\times$ & $1.68\times$ & $1.82\times$ & $1.30\times$ & $3.01\times$ & $1.28\times$ & $1.91\times$ & $1.18\times$ & $1.57\times$ & $0.81\times$ & $2.16\times$ & $1.99\times$
\\
\hline
\end{tabular}
}
\label{tab:result}
\end{table*}

\subsection{Experimental Setup}

\noindent\textbf{Task and Datasets.} We evaluate analytical accuracy across three representative tasks: object detection, multi-object tracking, and keypoint detection. We mainly report YOLOv8-X~\cite{yolov8} for detection and tracking, and YOLOv8-Pose for keypoint estimation. Experiments are conducted on the MOT17~\cite{MOT17}, NVIDIA AI City~\cite{aicity}, and VisDrone~\cite{visdrone} datasets.

\noindent\textbf{Implementation.} We adopt two H.264 codec implementations: Nvidia Video SDK~\cite{nvcodec} and FFMPEG with libx264~\cite{h264}. Nvidia Video SDK support Advanced Emphasis Map feature~\cite{nvem-map}, which provides five emphasis levels to adjust quality at the macroblock level. From our empirical observations, setting a base QP of 45 and a minimum QP of 30 aligns Nvidia’s five emphasis levels approximately to libx264 QP values of $\left[ 45, 43, 37, 34, 30 \right]$. We adopt this empirical alignment to unify the behavior of both implementations. Note that \framework is a plug-and-play module that is agnostic to specific codec implementations. We utilize Nvidia Video SDK mainly for accelerated video encoding during training, while deploying libx264 for evaluation to ensure robust and practical performance across diverse edge devices. Additional details are in Appendix~\ref{apx:codec-impl}.

\noindent\textbf{Evaluation Metrics.} We evaluate performance across four dimensions: 1) computational cost, measured in MACs per pixel~\cite{compute-constrain}, 2) latency overhead, 3) bitrate savings and 4) training efficiency. Accuracy is measured by standard task-specific metrics: mAP for object detection, MOTA/MOTP/IDF1 for tracking, and OKS, PCK@0.2, and PCK@0.5 for keypoint detection. In scenarios where ground-truth annotations are unavailable, we treat results obtained on the raw (uncompressed) video as the reference baseline.

\noindent\textbf{Training Details.} Refer to Appendix~\ref{apx:training-details}.

\noindent\textbf{Baselines.} We benchmark \framework with three categories of methods. 1) \textit{When2Compress}: We adapt all methods to a frame-level QP assignment approach, while other video configurations (\eg resolution, framerate) remain orthogonal and can serve as complementary optimizations. 2) \textit{Where2Compress}: AccMPEG is the state-of-the-art method which directly assign binary QP configurations to distinguished (non)-informative regions. 3) \textit{H.264 Adaptive Quantization mode}: A human-centric fine-grained QP assignment mechanism. In our experiments, for When2Compress, we exhaustively search for the minimum frame-level QP that satisfies the accuracy requirement. For Where2Compress, we use the SOTA AccMPEG~\cite{accmpeg}. For H.264 AQ mode, we adopt the \textit{variance} mode, which dynamically adjusts the QP based on the variance of macroblocks to improve SSIM~\cite{h264-aq}. For all other advanced codecs (\ie H.265, VP9, H.266 and AV1), we use their most practical configurations as recommended for real-world deployment (Appendix~\ref{apx:other-codecs}).

\begin{table}[t]
    \centering
    \caption{Post-compression SSIM comparisons of MOT dataset.}
    \vspace{-1em}
    \resizebox{0.48\textwidth}{!}{
    \large
        \begin{tabular}{*{1}{l}||*{6}{c}}
                    \toprule
                    Method & 1702 & 1704 & 1709 & 1710 & 1711 & 1713 \\
                    \midrule
                    H.264 AQ~\cite{ffmpeg} & 0.981 & 0.983 & 0.971 & 0.984 & 0.982 & 0.985\\
                    AccMPEG~\cite{accmpeg} & 0.945 & 0.949 & 0.900 & 0.948 & 0.930 & 0.930 \\
                    \framework & \textbf{0.907 } & \textbf{0.948} & \textbf{0.871} & \textbf{0.939} & \textbf{0.919} & \textbf{0.928} \\
                    \bottomrule
        \end{tabular}
    }
    \label{tab:ssim}
\end{table}

\vspace{-1em}

\subsection{Evaluation Results}

\subsubsection{Better Bitrate Savings across Diverse Scenes}
\label{result:contents}

We evaluate \framework across a variety of video scenes to validate its robustness and show its consistent compression efficiency.

\begin{figure}[t] \centering
    \includegraphics[width=0.99\linewidth]{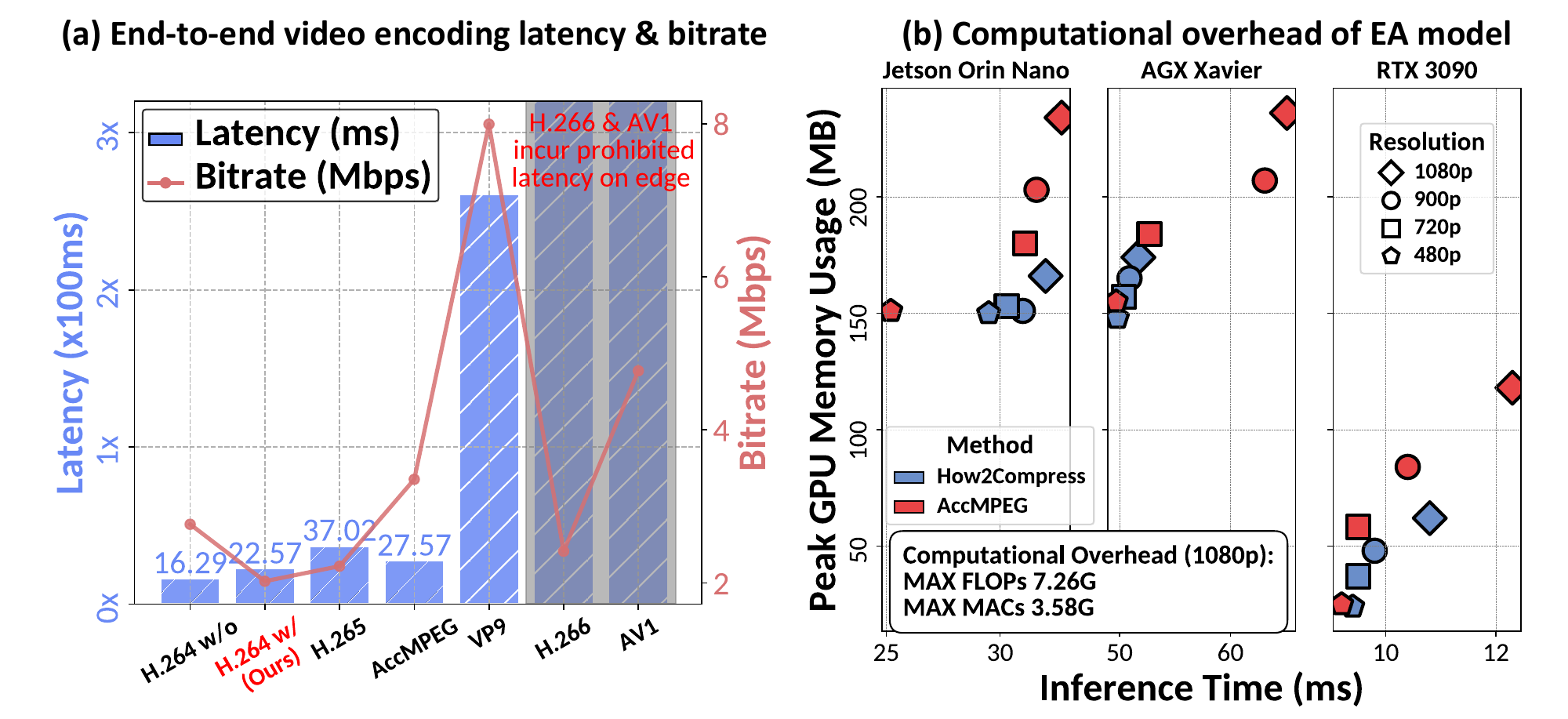}
    \vspace{-1em}
    \caption{(a) E2E video encoding latency and bitrate comparison of advanced codecs. (b) Computational resource and latency overhead of EA model.}
    \label{fig:overhead}
\end{figure}

\noindent\textbf{Bitrate Savings.} As shown in Table~\ref{tab:result}, \framework achieves a bitrate reduction of up to $50.4\%$, outperforming the baseline methods by as much as $3.01\times$. Figure~\ref{fig:overhead}a further shows that, when paired with vanilla H.264, \framework matches H.265 (HEVC)~\cite{h265} and surpasses advanced codecs like VP9~\cite{vp9}, H.266 (VVC)~\cite{vvc}, and AV1~\cite{aom_av1}. We integrate \framework with H.264 due to its widespread adoption in edge video analytics scenarios. Since it only refines codec's QP assignment, integrating it with more advanced codecs is expected to yield even greater compression gains.

\noindent\textbf{Structural and Textural Information Preservation.} The primary objective of \framework is to eliminate redundant structural and textural information that does not affect the performance of the detection task. As presented in Table~\ref{tab:ssim}, \framework yields a lower SSIM compared to AccMPEG~\cite{accmpeg} and the codec’s AQ while achieving comparable accuracy. This result indicates that \framework more effectively retains the essential structural features required by the downstream model while discarding non-essential details, thereby optimizing bit allocation for accuracy-critical macroblocks.

\subsubsection{Overhead}
\label{result:devices}
We evaluate it across heterogeneous devices to assess its scalability and suitability for real-world deployment.

\noindent\textbf{Computational Latency Overhead.} In standard video analytics pipelines, cameras typically operate at 30 frames per second (fps), resulting in an inter-frame interval of approximately 33.3 milliseconds.  \framework is explicitly implemented to interleave emphasis assignment within this interval. This interleaved architecture enables emphasis assignment decision-making and encoding to be performed in parallel with frame acquisition, thereby amortizing additional latency overhead. As shown in Fig.~\ref{fig:overhead}, integrating with H.264, our framework incurs no more than 6 milliseconds of latency per frame in average. This is lower than that of advanced codecs while achieving superior bitrate savings. 
\begin{figure}[t]
    \centering
    \includegraphics[width=0.45\textwidth,height=0.13\textwidth]{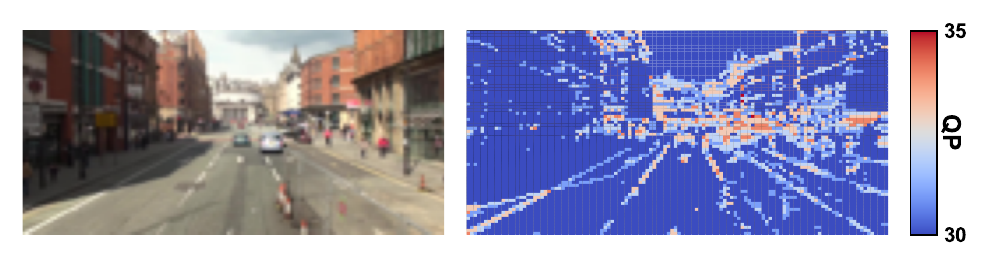} \\
    \vspace{-1em}
    \makebox[0.18\textwidth]{\small \qquad(a) Sample Image from Scene F}
    \makebox[0.26\textwidth]{\small (b) AQ QP allocation}
    \includegraphics[width=0.45\textwidth,height=0.13\textwidth]{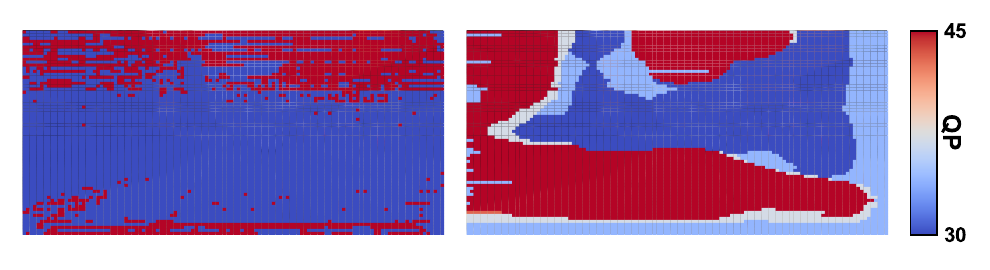} \\
    \vspace{-1em}
    \makebox[0.18\textwidth]{\small \qquad(c) Where2Compress allocation}
    \makebox[0.26\textwidth]{\small (d) \framework allocation}
    \vspace{-1em}
    \caption{QP allocation heatmaps for different methods across macroblocks. (zoom in for better visualization.)}
    \label{fig:case-study}
    \vspace{-1em}
\end{figure}

\begin{figure}[t]
    \centering
    \includegraphics[width=0.97\linewidth]{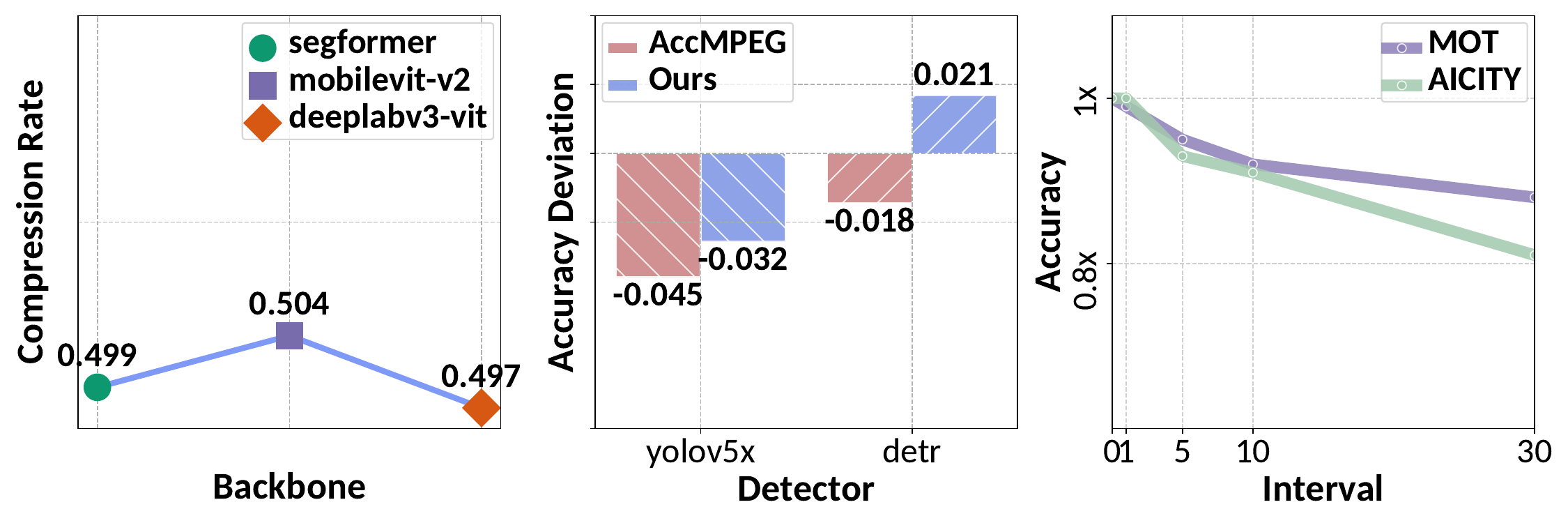}
    \caption{Ablation study on (a) Different Emphasis Assignment Model backbones. (b) Different detection backends. (c) Different execution intervals.}
    \label{fig:ablation}
\end{figure}

\noindent\textbf{Computational Resource Overhead.} As shown in Fig.~\ref{fig:overhead}b, our \framework requires only up to 7.26 GFLOPS (Floating Point Operations per Second) and 3.58 GMACs (Multiply-Add Operations per Second) when processing 1080p video input. The per-pixel computational cost is limited to 1726 MACs, which is significantly below the hard constraint of 2000 MACs per pixel for mobile platforms~\cite{compute-constrain}. 

\subsubsection{Emphasis Assignment In-depth Analysis}

In this case study, we compare the quality assignment strategies of different methods and present qualitative results that highlight the advantages of \framework. As shown in Fig.~\ref{fig:case-study},we visualize the QP allocation on a sampled I-frame. Figure~\ref{fig:case-study}b shows the QP allocation of the standard H.264 codec using the AQ variance mode~\cite{h264-aq}, which employs a conservative adjustment strategy designed to preserve human visual perception. Figure~\ref{fig:case-study}c illustrates the allocation pattern of \textit{Where2Compress}, which primarily lowers the quality in non-object regions. In contrast, Fig.~\ref{fig:case-study}d demonstrates that \framework not only identifies \textit{where} to compress but also determines \textit{how much} compression each macroblock can tolerate. This fine-grained and adaptive strategy leads to higher compression efficiency without sacrificing detection accuracy. Additional analysis of Table~\ref{tab:result} and Table~\ref{tab:ssim}, along with qualitative results, is provided in Appendix~\ref{apx:in-depth-analysis}~\&~\ref{apx:qualitative}.

\subsection{Ablation Study}
\label{sec:ablation}

We validate the effectiveness of our key contribution, RER, by ablating \framework under four settings: 1) different Emphasis Assignment Model backbones, 2) detection backends, 3) execution intervals, and 4) with \textit{vs.} without RER.

\noindent\textbf{Backbone-agnostic.} As shown in Fig.~\ref{fig:ablation}a, \framework consistently captures macroblock importance across different backbones, validating that RER generalizes well and is agnostic to the choice of Emphasis Assignment Model architecture.

\noindent\textbf{Backend-agnostic.} We pretrain the Emphasis Assignment Model and evaluate the accuracy on different object detectors (\ie DETR~\cite{detr} and YOLOv5~\cite{yolov5}) without retraining EA model. As illustrated in Fig.~\ref{fig:ablation}b, \framework maintains high detection accuracy across backends, suggesting the generalizability of using the \proxyemtarget as a proxy backend-agnostic supervision signal.

\begin{table}[t]
    \centering
    \caption{Effect of exploration probability decay and RER mechanism. \colorbox{gray!40}{gray} indicates accuracy deviation $\geq10\%$.}
    \vspace{-1em}
    \resizebox{0.48\textwidth}{!}{
    \large
        \begin{tabular}{*{1}{c}|*{1}{l}|*{5}{c}}
            \toprule
            & & \multicolumn{5}{c}{Probability Decay}       \\
            Method & Perf$^*$ & 0.1 & 0.15 & 0.2 & 0.25 & 0.3  \\
            \midrule
            \multirow{2}{*}{\makecell{\framework \\ (\small w/ RER)}} 
                & Comp$^\dagger$ & $1\times$  & $0.98\times$ & $0.99\times$ & $0.96\times$ & $0.97\times$  \\
                & Time$^\ddagger$ & $1\times$  & $1\times$ & $1.3\times$ & $1.6\times$ & $1.6\times$ \\ 
            \midrule
            \multirow{2}{*}{\makecell{\framework \\ (\small w/o RER)}} 
                & Comp$^\dagger$ & \cellcolor{gray!40} $0.83\times$ & \cellcolor{gray!40}$0.87\times$ & \cellcolor{gray!40}$0.81\times$ & \cellcolor{gray!40}$0.83\times$ & \cellcolor{gray!40}$0.88\times$  \\
                & Time$^\ddagger$ & $2\times$  & $2\times$ & $2\times$ & $2\times$ & $2\times$ \\ 
            \bottomrule
        \end{tabular}
    }
        \begin{flushleft}
                \small ${^*}$ Perf: performance metric category. \\
                \small $^\dagger$ Comp: bitrate after compression (normalized). \\
                \small $^\ddagger$ Time: \# epochs to convergence (normalized). \\
        \end{flushleft}
    \label{tab:ablation-p-rer}
\end{table}

\noindent\textbf{Adaptive Temporal Execution.} Video streams often exhibit strong temporal coherence, especially in slow-motion scenarios. We leverage this property by selectively recomputing the emphasis assignment at adaptive intervals. This strategy aligns with the principle behind \textit{When2Compress} and can be viewed as a complementary mechanism.  As shown in Fig.~\ref{fig:ablation}c, in slow-motion scenarios such as those in MOT datasets, increasing the execution interval ($\leq5$) does not significantly degrade accuracy.

\noindent\textbf{Effect of RER.} Table~\ref{tab:ablation-p-rer} shows that removing RER and only use \proxyemtarget leads to a significant drop in accuracy ($\ge10\%$), despite achieving higher compression via fine-grained QP assignment. This is because without RER, the model fails to emphasize critical regions thereby treating all macroblocks equally. This results in the loss of informative regional features. Moreover, RER proves robust to changes in the exploration probability decay schedule, indicating that it is stable and hyperparameter-insensitive.

\section{Discussion}
\label{sec:discussion}

\noindent\textbf{Generalization to Other Tasks.} Although the EA model is pretrained on detection, it generalizes well to tracking (within 2\% accuracy drop) and keypoint detection (within 5\%), due to the shared focus on foreground regions. Quantitative results are provided in Appendix~\ref{apx:performance-other-tasks}.

\noindent\textbf{Generalization to Other Blockbased Codecs.}  A wide range of widely adopted video codecs, including H.264~\cite{h264}, H.265~\cite{h265}, VP8~\cite{vp8}, VP9~\cite{vp9}, H.266~\cite{h266} and AV1~\cite{aom_av1}, rely on blockbased compression algorithms. \framework is inherently compatible with these codecs, as it does not alter the fundamental encoding process but instead refines the quality assignment strategy. We provide the implementation examples in Appendix~\ref{apx:generalization}.

\section{Conclusion}
We present \framework, a novel plug-and-play module that enhances video compression for analytics through fine-grained, macroblock-level quality assignment. It dynamically allocates the minimum necessary quality to each macroblock via a progressive, self-guided \textit{Region-aware Emphasis Routing} mechanism. Integrated with the H.264 codec, \framework achieves up to $50.4\%$ bitrate savings and outperforms baselines by up to $3.01\times$, all while preserving task performance. Extensive evaluations prove that \framework is lightweight, backbone-agnostic, backend-agnostic, and training efficient, making it both practical and scalable for real world deployment in edge video analytics systems.

\begin{acks}
This work is supported by the Institute of Information \& Communications Technology Planning \& Evaluation (IITP), funded by the Ministry of Science and ICT (MSIT) of the Republic of Korea, under grants No. RS-2019-II191126 and No. RS-2024-00459749.
\end{acks}

\bibliographystyle{ACM-Reference-Format}
\bibliography{ref}

\clearpage
\appendix

\setcounter{page}{1}
\twocolumn[
{
\centering
\Huge

\textbf{\textsc{\framework}: Efficient Edge Video Analytics via Adaptive Granular Video Compression}\\
\vspace{0.5em}
Supplementary Material \\
\vspace{1.0em}
}
]

\section{Artifacts}
\label{artifacts}

To ensure full reproducibility of our experiments, we provide both the source code and the complete software environment. The source code is available at \url{https://github.com/wyhallenwu/how2compress}. Additionally, we provide a Docker image, available at dockerhub \url{https://hub.docker.com/r/wuyuheng/how2compress}, which contains all necessary dependencies, and scripts for reproducing our results.

\subsection{Hardware}

\framework is trained on a compute cluster and evaluated on real-world deployable edge devices.

The compute cluster used for training is configured as follows: 

\begin{itemize} 
    \item \textbf{CPU:} Dual Intel(R) Xeon(R) Silver 4210R @ 2.40GHz 
    \item \textbf{GPU:} 4× NVIDIA RTX 3090 (only 3 GPUs are used due to limitations of the Nvidia Video Codec) \item \textbf{Memory:} 192 GB RAM
    \item \textbf{Cuda:} CUDA 12.6 and Driver 560.28.03
\end{itemize}

For edge deployment evaluation, we use the NVIDIA Jetson Orin Nano 4GB. Detailed specifications for this device can be found at: \url{https://www.nvidia.com/en-us/autonomous-machines/embedded-systems/jetson-orin/}.

\subsection{Software}
\label{apx:codec-impl}

The implementation of H.264 used in our work is a customized version of the libx264 encoder that supports macroblock-level Quantization Parameter (QP) assignment. This customized encoder is built upon the repository available at \url{https://github.com/Alex-q-z/myh264}. We use FFMPEG~\cite{ffmpeg} version 3.4.8 and libx264 version 0.163.x (paired with libswscale version 4.8.100).

To accelerate training, we utilize the NVIDIA Video Codec SDK v11.0.10, which supports GPU-accelerated video encoding and includes an \textit{Emphasis Map} feature. Detailed documentation is available at: \url{https://docs.nvidia.com/video-technologies/video-codec-sdk/11.1/nvenc-video-encoder-api-prog-guide/index.html#emphasis-map}. Although the SDK provides five emphasis levels, their behavior differs slightly from that of libx264. Through extensive empirical testing, we determine that, when using a base QP of 45 and a minimum QP of 30, these emphasis levels approximately correspond to effective QP values of $[45, 43, 37, 34, 30]$.

The compiled binary tools used in our experiments are included in the repository. For completeness, we also evaluate the encoder without the use of the \textit{Emphasis Map} feature.

\subsection{Datasets}
\label{apx:datasets}

\framework is evaluated on a large-scale collection of diverse, real-world edge video stream datasets.

\noindent\textbf{Multiple Object Tracking Benchmark.} The Multiple Object Tracking 2017 (MOT17) dataset is a widely used benchmark in the computer vision community, particularly for evaluating multi-object tracking algorithms. It consists of seven distinct video sequences recorded in both indoor and outdoor public environments, with a primary focus on pedestrian tracking. Each sequence is divided into training and testing subsets, enabling comprehensive algorithm development and evaluation under diverse real-world conditions. The dataset poses various challenges such as varying crowd densities, illumination changes, and camera motion. In total, MOT17 comprises 15,948 frames, corresponding to approximately 645 seconds (or 10.75 minutes) of video data. Details are available at \url{https://motchallenge.net/data/MOT17Det/}.

\noindent\textbf{NVIDIA AI City Challenge.} We primarily utilize the dataset from the 2021 edition of the NVIDIA AI City Challenge, a large-scale urban traffic video benchmark. This dataset comprises approximately 9 hours of video footage captured from 20 distinct urban locations, including single-direction intersection approaches, full intersections, highway segments, and city streets. The dataset reflects a broad range of environmental conditions such as different times of day (\eg dawn). Details are available at  \url{https://www.aicitychallenge.org/2021-data-and-evaluation/}.

\noindent\textbf{VisDrone.} While the previous two datasets utilize static cameras, the VisDrone dataset~\cite{visdrone} features cameras mounted on drones, thereby capturing more dynamic scenes and diverse visual content. This enables evaluation under more realistic and challenging conditions involving camera motion. For further details, please refer to the official repository at \url{https://github.com/VisDrone/VisDrone-Dataset}.

\subsection{Downstream Models}
\label{apx:detectors}

We employ three representative object detectors in our evaluation: YOLOv5~\cite{yolov5}, YOLOv8~\cite{yolov8}, and DETR~\cite{detr}. These models span both one-stage convolutional architectures and transformer-based paradigms, offering a comprehensive basis for assessing the generalization of our framework across diverse detector designs.

Specifically, we fine-tune YOLOv5 and YOLOv8 on the two datasets introduced above using high-quality video frames (encoded with a Quantization Parameter of 25) to establish strong baseline detection performance. During the training and evaluation of our compression framework, these fine-tuned detectors are kept fixed (frozen), while input frames are re-encoded at lower quality levels using various baseline and proposed methods. Our objective is to maximize compression efficiency while minimizing any degradation in detection accuracy with respect to these fixed detectors.

For pose estimation tasks, we employ various versions of YOLO-Pose provided by Ultralytics, and for tracking, we utilize their corresponding tracking models. We primarily focus on detection and tracking tasks, as they represent the most widely adopted applications in real-world scenarios.

\begin{tcolorbox}[title={Encoding Configurations}, colbacktitle=boxheadbgcol, coltitle=boxheadcol, fonttitle=\bfseries, colframe=black, colback=white, boxrule=1pt, sharp corners]
\begin{minted}[fontsize=\small,breaklines=true]{bash}
# uniform QP (When2Compress)
ffmpeg -i <input> -c:v libx264 -qp <qp> -x264-params aq-mode=0 <output>
# AccMPEG (Where2Compress)
# qp is placeholder
# and it will load QP assignment from <qp_matrix_file>
/myh264/ffmpeg -y -i <input> -qp <useless placeholder> -pix_fmt yuv420p <output> 
# Codec AQ (Codec's How2Compress)
ffmpeg -i <input> -c:v libx264 -qp <qp> -x264-params aq-mode=<aq-mode> <output> 
# How2Compress (Ours)
# load QP assignment from <qp_matrix_file>
/myh264/ffmpeg -y -i <input> -qp <useless placeholder> -pix_fmt yuv420p <output> 
# x265
ffmpeg -benchmark -framerate 30 -i <input> -c:v libx265 -tune zerolatency -preset veryfast -x265-params "pools=1:qp=<qp>:aq-mode=<mode>" -threads 1 <output>
# VP9
ffmpeg -benchmark -framerate 30 -i <input> -c:v libvpx-vp9 -crf <config> -deadline realtime  -threads 1 <output>
# VVC encoding
ffmpeg -benchmark -framerate 30 -i <input> -c:v libvvenc -preset 0 -qp <qp>  -pix_fmt yuv420p -threads 1 <output>
# AV1 encoding
ffmpeg -benchmark -framerate 30 -i <input> -c:v libaom-av1 -crf <config> -b:v 0 -cpu-used 4 -usage realtime -row-mt 1 -threads 1 -g 30 -pix_fmt yuv420p <output>
\end{minted}
\label{box:encoding-cmd}
\end{tcolorbox}

\subsection{Emphasis Assignment Model Implementation}
\label{apx:ea-model-impl}

It is important to note that \framework is not tied to a specific backbone architecture. Our core contribution lies in the proposed \textbf{Region-aware Emphasis Routing} mechanism, which efficiently guides fine-grained macroblock-level QP assignment. The main results reported in the paper are based on the MobileViTv2 backbone, and we conduct ablation studies with alternative backbones to demonstrate the robustness and generality of our approach.

For the Emphasis Assignment model, we implement a prediction head on top of MobileViTv2, SegFormer, or DeepLabV3 backbones. The head consists of a single Conv2D layer that produces a probability distribution of shape $(B, H, W, 5)$, where $(H, W)$ corresponds to the number of macroblocks, and the last dimension represents the five discrete emphasis levels.

To accommodate the constraints of various edge devices, the input resolution can be downsampled by a factor of $D$ ($\left[1, 4\right]$) along both spatial dimensions. If $D>1$, the resulting low-resolution emphasis map is then upscaled to the original resolution using bilinear interpolation to match the macroblock grid.

\subsection{Training Details}
\label{apx:training-details}
The training procedure, detailed in Algorithm~\ref{algo:em-routing}, employs the following default hyperparameters: the exploration probability \( p \) starts at 0.8 and decays by 0.1 per epoch; SSIM threshold percentiles are set at 90\% for \( \tau_{\text{roi}} \) and 50\% for \( \tau_{\text{bg}} \). The model is optimized using AdamW with a CosineAnnealing learning rate schedule, decaying from \(1 \times 10^{-3}\) to \(1 \times 10^{-6}\). The loss function is balanced with weights \( \lambda_1 = 10 \) and \( \lambda_2 = 5 \), and emphasis levels are penalized progressively with factors \([1,\, 1.3,\, 1.6,\, 1.9,\, 2.2]\) to discourage excessive quality allocation.

For training and evaluation, the dataset is split into training, validation, and test sets using a 70:20:10 ratio. To avoid temporal dependencies, each frame is encoded as an I-frame with its corresponding predicted emphasis map during training, ensuring the model focuses on spatial frame-level decisions. In deployment, the video stream follows a GOP structure of 30 frames, comprising 1 I-frame and up to 3 B-frames between P-frames.

\section{Proof}
\label{apx:proof}
We provide a \textit{soft theoretical bound for convergence and compression efficiency.} 
Let $\mathcal{E}^{(t)} \in \mathbb{Z}_+^{n_w \times n_h}$ denote the predicted emphasis map at iteration $t$, where each entry $\mathcal{E}^{(t)}_{i,j} \in \{0,1,\dots,K-1\}$ corresponds to the QP offset (\ie emphasis level) assigned to macroblock $M_{i,j}$. Let $\mathcal{P}^{(t)}$ denote the proxy emphasis target at iteration $t$, obtained through percentile-based PET thresholds and Region-aware Emphasis Routing (RER). The training loss is defined as:
\[
\mathcal{L}^{(t)} = \lambda_1 \cdot \text{AccLoss}^{(t)} + \lambda_2 \cdot \text{AlignLoss}^{(t)},
\]
where $\text{AccLoss}^{(t)} := \left| \text{Acc}(\mathcal{E}^{(t)}) - \text{Acc}(\mathcal{E}^{\text{max}}) \right|$ quantifies the task performance deviation from maximum quality, and $\text{AlignLoss}^{(t)} := \text{CE}(\mathcal{E}^{(t)}, \mathcal{P}^{(t)})$ measures the divergence between the model output and the proxy.

\smallskip
\noindent
\textbf{Assumptions.}
\begin{enumerate}
    \item \textit{Proxy Improvement:} The proxy $\mathcal{P}^{(t)}$ improves over time, \ie $\text{CE}(\mathcal{P}^{(t)}, \mathcal{E}^{*}) \to 0$ as $t \to T$, where $\mathcal{E}^{*}$ is the (unknown) optimal emphasis map.
    \item \textit{Temporal Smoothness:} For adjacent frames, macroblock assignments evolve slowly: $\| \mathcal{E}^{(t)} - \mathcal{E}^{(t-1)} \|_1 \leq \delta$.
    \item \textit{Loss Contraction:} The model learns to align with the proxy at each round:
    \[
    \mathcal{L}^{(t+1)} \leq \mathcal{L}^{(t)} - \eta \cdot \| \mathcal{E}^{(t+1)} - \mathcal{P}^{(t)} \|_1,
    \]
    for some $\eta > 0$.
\end{enumerate}

\smallskip
\noindent
\textbf{Proposition.}
Under the assumptions above, for any accuracy tolerance $\tau > 0$, the number of rounds $T$ needed to obtain an emphasis map $\mathcal{E}^{(T)}$ satisfying
\[
\left| \text{Acc}(\mathcal{E}^{(T)}) - \text{Acc}(\mathcal{E}^{\text{max}}) \right| \leq \tau
\quad \text{and} \quad
\mathbb{E}[R(\mathcal{E}^{(T)})] \geq R_{\text{min}}(\tau),
\]
is bounded by
\[
T \leq \frac{\mathcal{L}^{(0)}}{\eta \cdot \epsilon},
\]
where $\epsilon$ denotes the minimal improvement in alignment per round, and $R(\cdot)$ is the achieved bitrate reduction.

\smallskip
\noindent
\textbf{Interpretation.}
This bound shows that convergence is linear in the initial loss and inversely proportional to the progress made in each round. In practice, due to temporal consistency and the design of RER, we empirically observe convergence in only 1--3 rounds.

\section{Comparison with advanced Codecs}
\label{apx:other-codecs}

\framework is designed to be codec-agnostic, assuming the underlying codec supports block-level quality control \ie the ability to encode pixel blocks at varying quality levels. This includes widely adopted codecs such as H.264~\cite{h264}, H.265~\cite{h265}, VP9~\cite{vp9}, H.266 (VVC)~\cite{h266} and AV1~\cite{aom_av1}. In this work, we primarily use H.264 as a case study, as it remains the most prevalent video codec in edge computing scenarios due to its widespread hardware support and deployment maturity. To ensure experimental reproducibility, the encoding configurations used for each codec are provided in Box~\ref{box:encoding-cmd}. All codecs were evaluated using the FFMPEG toolchain~\cite{ffmpeg}, with default parameters unless otherwise specified. It is important to note that we deliberately disable multi-threaded encoding during our comparisons. This decision reflects a realistic edge deployment scenario, where devices typically feature limited CPU cores (2–6 threads) and are concurrently burdened with computationally intensive tasks, such as real-time video analytics or other deep learning inference workloads.

\section{Constant Rate Factor (CRF)}
\label{apx:crf}

Some research~\cite{crf1,crf2} leverage CRF in modern video encoders to enable implicit macroblock-level QP adjustment. It operates by targeting a perceptually constant visual quality across frames and spatial regions. Internally, it modulates the QP at the macroblock level based on content characteristics such as texture complexity, luminance masking, and motion. Regions deemed perceptually less sensitive (\eg flat or low-motion areas) are assigned higher QPs, while visually salient areas are preserved with lower QPs. This implicit adjustment is guided by heuristics rooted in human visual system modeling.

Despite its adaptive nature, CRF is not considered suitable for edge video analytics scenarios. First, CRF is inherently optimized for human visual perception rather than machine-centric tasks such as object detection or activity recognition. Consequently, it may prioritize quality in visually salient but analytically irrelevant regions, while under-allocating bitrate to semantically important areas (\eg small moving objects). Second, the mechanism is opaque and nondeterministic from external control. It does not expose a clear interface for frame- or region-level quality manipulation, making it ill-suited for latency-sensitive edge systems that require direct and responsive QP control in reaction to scene dynamics or detection confidence.

In contrast, offering explicit macroblock-level QP assignment facilitates fine-grained and machine-centric quality adaptation. It allows encoders to prioritize bitrate allocation to analytically critical regions—identified using objectness maps, motion vectors, or model-informed saliency while compressing less relevant background areas more aggressively. This capability is particularly beneficial in edge-cloud collaborative systems, where uplink bandwidth is limited and quality-budget precision is essential to maintain detection performance across a broad range of video scenes. Moreover, such direct control improves system responsiveness, enabling real-time adaptation to content changes and ensuring robust operation under stringent resource constraints.

\begin{figure}[t]
    \centering
     \includegraphics[width=0.48\textwidth,height=0.13\textwidth]{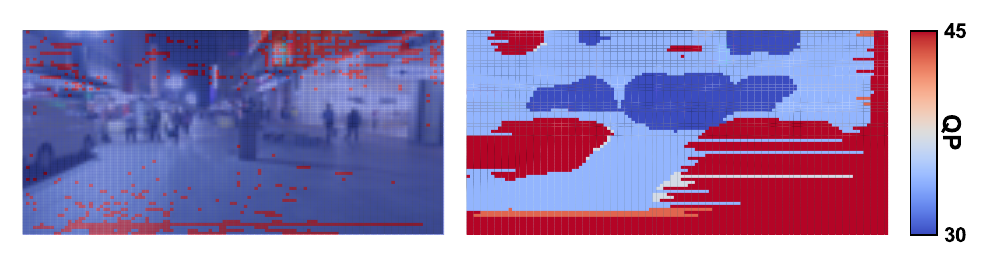} 
     \\
    \makebox[0.18\textwidth]{\makecell{\small (a) AccMPEG QP allocation \\ (Sence D)}}
    \makebox[0.24\textwidth]{\makecell{\small (b) Ours QP allocation \\ (Scene D)}}
\caption{Comparison of QP allocation between AccMPEG and our method. AccMPEG requires careful online threshold tuning to prevent over-allocation of high-quality regions in object-sparse scenes.}
\label{fig:apx-decision-compare}
\end{figure}

\section{PSNR \textit{\small vs.} SSIM}
\label{apx:why-ssim}

Peak Signal-to-Noise Ratio (PSNR) and Structural Similarity Index Measure (SSIM) are two widely adopted metrics for evaluating image and video quality. While both aim to quantify the degradation introduced by lossy compression, they differ fundamentally in their design principles and perceptual alignment.

PSNR is a pixel-wise fidelity metric that computes the logarithmic ratio between the maximum possible pixel value and the mean squared error (MSE) between the original and compressed images~\cite{psnr,psnr-ssim}. It assumes an independent and identically distributed error model across pixels, making it analytically convenient and computationally efficient. However, PSNR is agnostic to spatial correlations and human perceptual sensitivity. As a result, it often fails to capture structural distortions or texture degradation that are critical to downstream tasks.

In contrast, SSIM evaluates image similarity by modeling luminance, contrast, and structural information in localized regions~\cite{ssim}. It correlates more strongly with human visual perception by emphasizing structural consistency rather than absolute pixel accuracy. Despite being originally designed for perceptual quality assessment, SSIM better captures the degradation patterns relevant to DNN-based video analytics, particularly in the presence of structural artifacts and compression-induced distortions.

\textbf{It is important to note that neither PSNR nor SSIM is designed to measure the specific feature retention of DNNs used in video analytics.} DNNs rely not on perceptual quality \textit{per se}, but on the preservation of task-relevant features. In this regard, both PSNR and SSIM may misrepresent the true impact of compression on detection accuracy. PSNR can over-penalize visually insignificant pixel differences, while SSIM may overemphasize structures that are perceptually relevant but semantically irrelevant to the analytical models.

Nevertheless, among the two, SSIM remains more adoptable in analytics-aware compression systems. Its emphasis on structural preservation provides a better proxy for semantic fidelity, especially in scenarios where object shapes and boundaries are critical. Empirically, in our experiments, SSIM tends to correlate more strongly than PSNR with downstream performance metrics such as mean Average Precision (mAP) in object detection tasks. Furthermore, SSIM's localized and differentiable structure~\cite{psnr-ssim} makes it useful for guiding encoder decisions or supervising quality estimation models when model-in-the-loop feedback is computationally expensive or unavailable.

\section{Further In-depth Analysis}
\label{apx:in-depth-analysis}

\subsection{Disentangling optimization of streaming and inference}
\label{apx:disentangling-optimization}

As discussed in \cref{sec:related-work}, beyond controlling video quality, there exist several alternative strategies to reduce video bitrate, including: \textcolor{blue}{1)} filtering redundant frames (temporal redundancy), \textcolor{blue}{2)} reducing video resolution (spatial downsampling), and \textcolor{blue}{3)} cropping regions of interest (RoI) for targeted inference.

Differentiate from them, \framework focuses exclusively on adjusting quality within each frame, \ie exploring the quality dimension at a fine-grained level. This design choice offers three primary advantages:

\begin{itemize}[label=\bcheckmark]
    \item \textbf{Modularity and decoupled optimization:} \framework is a plug-and-play module that decouples the optimization of the frontend (edge device) from the backend (edge cluster). In comparison, strategies such as \textcolor{blue}{2)} resolution reduction and \textcolor{blue}{3)} RoI cropping often generate video frames of varying resolutions. To maintain high throughput in downstream inference (\eg with batch processing) such variability necessitates joint optimization between edge devices and analytical backends. In contrast, \framework preserves consistent frame structure, simplifying deployment and is compatible with backend inference optimization.
    \item \textbf{Complementarity:} \framework is orthogonal to the above methods. It can be seamlessly integrated with \textcolor{blue}{1)}, \textcolor{blue}{2)}, and \textcolor{blue}{3)} to further enhance bitrate reduction, offering additive benefits when combined with existing system optimization.
    \item \textbf{Effectiveness in densely populated scenes:} In scenarios where frames contain numerous objects, spatial pruning methods (\eg resolution reduction or RoI cropping) often become less effective, as most regions are deemed important. In such cases, \framework remains effective by adaptively reducing information along the quality dimension, selectively lowering the fidelity of less critical regions within crowded frames.
\end{itemize}

\begin{figure}[t] \centering
    \includegraphics[width=0.23\textwidth]{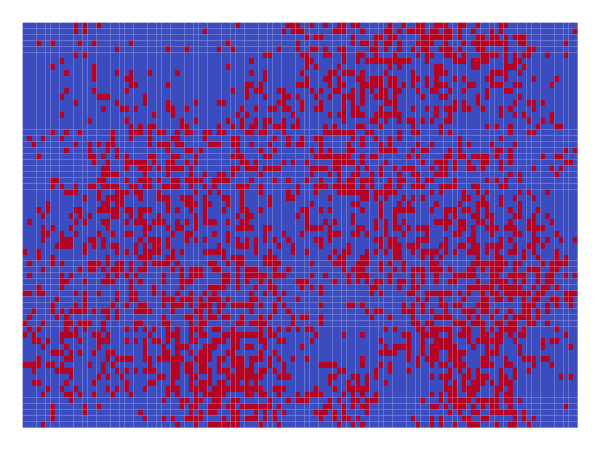}
    \includegraphics[width=0.23\textwidth]{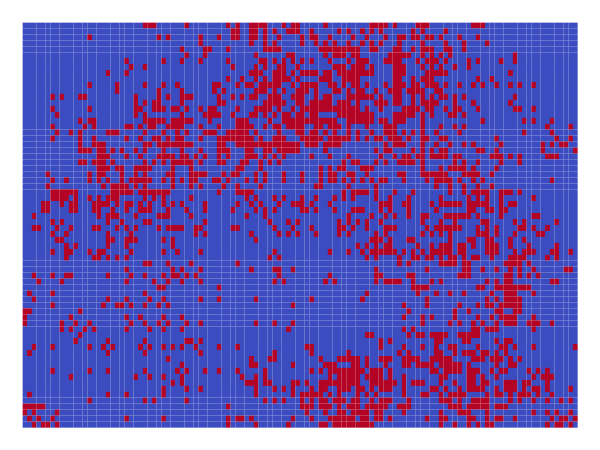}
    \\
    \makebox[0.23\textwidth]{YOLO}
    \makebox[0.23\textwidth]{DETR}
    \caption{Macroblocks identified as important by AccMPEG using accuracy gradients from different detection models. The variation indicates model-specific dependence, necessitating re-profiling for each detector.} \label{fig:apx-accuracy-gradient}
\end{figure}

\subsection{Comparison with AccMPEG and Its Potential Extensions}
\label{apx:advantages-where2compress}

Our proposed framework, \textit{How2Compress}, is fundamentally distinct from AccMPEG in both methodological philosophy and implementation strategy. Below, we outline two core differences that enable \textit{How2Compress} to achieve better scalability, generalization, and runtime efficiency compared to AccMPEG.

\begin{enumerate}
    \item \textbf{Model/Task Independence.} AccMPEG requires extensive offline gradient profiling for each specific model and task to estimate the relative importance of individual macroblocks (MBs) for downstream performance. This process involves backpropagating gradients from the downstream task (\eg object detection or tracking) through the video processing pipeline, which is computationally expensive and must be repeated for every new model, dataset, or task configuration. Consequently, AccMPEG’s applicability is limited to fixed, pre-profiled scenarios.

    In contrast, our framework eliminates the need for any downstream task gradients by leveraging \textit{structural cues} (\eg edges, textures, and spatial layouts) that are inherently present in video content. These cues serve as a general and task-agnostic signal to guide quality allocation decisions. As a result, \textit{How2Compress} is naturally \textbf{model- and task-independent}, allowing it to generalize seamlessly across various analytical models and tasks without re-profiling. This advantage is especially valuable in real-world edge deployment scenarios where the underlying models may change over time or differ across devices.    
    \item \textbf{Global Correlation Modeling.} Another key distinction lies in how inter-macroblock relationships are handled. AccMPEG uses \textit{AccGrad}, a heuristic metric that evaluates each macroblock's contribution to performance in isolation, based solely on the magnitude of its associated gradient. This per-MB analysis neglects the complex correlations and contextual dependencies between macroblocks, such as object continuity, motion coherence, or scene layout consistency.

    By contrast, our framework adopts a learning-based strategy that models \textbf{global correlations across macroblocks}. Rather than assigning quality in a greedy or local manner, \textit{How2Compress} employs a self-supervised, attention-guided mechanism that captures long-range dependencies. This enables the system to allocate bitrate in a globally optimized fashion, taking into account the overall impact on scene understanding and task accuracy.
\end{enumerate}

\begin{table*}[t] \centering
    \caption{Benchmark Results Across Different Hardware Platforms (Mean Time in ms)}
    \label{tab:benchmark_results}
    \resizebox{\textwidth}{!}{
    \Huge
        \begin{tabular}{c|*{3}{c}|*{3}{c}|*{3}{c}|*{3}{c}}
        \toprule
        & \multicolumn{3}{c|}{RTX3090} 
                               & \multicolumn{3}{c|}{Orin Nano1} 
                               & \multicolumn{3}{c|}{Xavier NX1} 
                               & \multicolumn{3}{c}{AGX Xavier1}  \\
                               & NumPy & Numba & PyTorch 
                               & NumPy & Numba & PyTorch 
                               & NumPy & Numba & PyTorch 
                               & NumPy & Numba & PyTorch \\
        \midrule
        $480\times 640$ (480p)    & 18.06 & 24.72 & 3.47 & 24.96 & 16.41 & 5.61 & 37.13 & 25.98 & 7.96 & 26.10 & 15.31 & 15.79 \\
        $720\times 1280$ (720p)   & 54.76 & 55.61 & 8.81 & 70.77 & 41.89 & 13.60 & 110.04 & 65.76 & 14.21 & 72.81 & 45.39 & 10.64 \\
        $900\times 1600$ (900p)   & 89.05 & 59.77 & 6.83 & 115.76 & 70.86 & 19.74 & 161.78 & 68.37 & 19.68 & 102.76 & 49.20 & 15.14 \\
        $1080\times 1920$ (1080p) & 129.45 & 73.55 & 7.13 & 167.92 & 103.00 & 27.43 & 230.56 & 99.13 & 19.22 & 148.38 & 63.38 & 22.00 \\
        \bottomrule
    \end{tabular}
    }
\end{table*}

\begin{table*}[t] \centering
    \caption{Benchmark Results Across Different Hardware Platforms (Standard Deviation in ms)}
    \label{tab:benchmark_results_std}
    \resizebox{\textwidth}{!}{
    \Huge
        \begin{tabular}{c|*{3}{c}|*{3}{c}|*{3}{c}|*{3}{c}}
        \toprule
        & \multicolumn{3}{c|}{RTX3090} 
                               & \multicolumn{3}{c|}{Orin Nano1} 
                               & \multicolumn{3}{c|}{Xavier NX1} 
                               & \multicolumn{3}{c}{AGX Xavier1}  \\
                               & NumPy & Numba & PyTorch 
                               & NumPy & Numba & PyTorch 
                               & NumPy & Numba & PyTorch 
                               & NumPy & Numba & PyTorch \\
        \midrule
        $480\times 640$ (480p)    & 4.42 & 8.98 & 1.23 & 0.10 & 0.21 & 0.12 & 1.06 & 1.90 & 1.14 & 3.14 & 1.71 & 0.78 \\
        $720\times 1280$ (720p)   & 1.92 & 5.59 & 1.38 & 0.09 & 0.16 & 0.18 & 2.46 & 14.72 & 0.23 & 2.05 & 12.83 & 0.49 \\
        $900\times 1600$ (900p)   & 1.20 & 6.18 & 2.14 & 0.08 & 0.42 & 0.08 & 1.72 & 1.30 & 5.33 & 2.36 & 4.16 & 1.13 \\
        $1080\times 1920$ (1080p) & 2.94 & 3.91 & 0.54 & 0.14 & 0.21 & 0.07 & 1.16 & 1.09 & 0.39 & 2.36 & 0.84 & 1.76 \\
        \bottomrule
    \end{tabular}
    }
\end{table*}

\begin{table*}[t] \centering
    \caption{Normalized Performance Comparison (Lower is Better)}
    \label{tab:normalized_performance}
    \resizebox{\textwidth}{!}{
    \Huge
        \begin{tabular}{c|*{3}{c}|*{3}{c}|*{3}{c}|*{3}{c}}
        \toprule
        & \multicolumn{3}{c|}{RTX3090} 
                               & \multicolumn{3}{c|}{Orin Nano1} 
                               & \multicolumn{3}{c|}{Xavier NX1} 
                               & \multicolumn{3}{c}{AGX Xavier1}  \\
                               & NumPy & Numba & PyTorch 
                               & NumPy & Numba & PyTorch 
                               & NumPy & Numba & PyTorch 
                               & NumPy & Numba & PyTorch \\
        \midrule
        $480\times 640$ (480p)    & 5.2× & 7.1× & \textbf{1.0×} & 4.4× & 2.9× & \textbf{1.0×} & 4.7× & 3.3× & \textbf{1.0×} & 1.7× & \textbf{1.0×} & 1.0× \\
        $720\times 1280$ (720p)   & 6.2× & 6.3× & \textbf{1.0×} & 5.2× & 3.1× & \textbf{1.0×} & 7.7× & 4.6× & \textbf{1.0×} & 6.8× & 4.3× & \textbf{1.0×} \\
        $900\times 1600$ (900p)   & 13.0× & 8.8× & \textbf{1.0×} & 5.9× & 3.6× & \textbf{1.0×} & 8.2× & 3.5× & \textbf{1.0×} & 6.8× & 3.2× & \textbf{1.0×} \\
        $1080\times 1920$ (1080p) & 18.2× & 10.3× & \textbf{1.0×} & 6.1× & 3.8× & \textbf{1.0×} & 12.0× & 5.2× & \textbf{1.0×} & 6.7× & 2.9× & \textbf{1.0×} \\
        \bottomrule
    \end{tabular}
    }
\end{table*}

\noindent\textbf{On Potential Extensions to AccMPEG.}  
In an attempt to reduce AccMPEG's complexity, one may consider the following two extensions that discretize gradients into QP bins:

\begin{itemize}
    \item \textbf{(Ext@1): Fixed Scalar Thresholds.} One straightforward extension involves defining a set of scalar thresholds to partition the gradient values into discrete QP bins. For instance, using 5 equally spaced thresholds across the observed gradient range \([ \text{min}, \text{max} ]\), one could assign coarser or finer compression quality per macroblock accordingly. However, such an approach suffers from severe practical limitations: scalar thresholds must be hand-tuned for each scene, model, and task. Without careful tuning, the bin boundaries may poorly reflect the actual semantic importance of content regions.

    To empirically validate this limitation, we implemented a variant of AccMPEG using 5-bin discretization over the per-scene gradient range. The resulting bitrate performance (measured in Megabits) is reported below for the MOT17 and AI CITY datasets:

    \begin{center}
    \begin{tabular}{l|l}
    \toprule
    \textbf{Dataset} & \textbf{Bitrate (Mbits)} \\
    \midrule
    MOT17     & [2.732, 1.441, 3.312, 4.132, 3.818, 2.918] \\
    AI CITY   & [2.544, 3.289, 5.452] \\
    \bottomrule
    \end{tabular}
    \end{center}

    These results demonstrate clear performance degradation when compared to our method, which uses a learned, soft-assignment strategy without requiring manual threshold tuning.

    \item \textbf{(Ext@2): Quantile-Based Thresholds.} An alternative approach replaces fixed scalar thresholds with quantiles derived from the empirical distribution of gradients. This method computes percentiles (\eg 20th, 40th, 60th, and 80th) to dynamically form QP bins, adapting to the actual gradient distribution shape rather than relying on arbitrary fixed boundaries. While this strategy offers improved theoretical robustness by naturally accounting for gradient distribution characteristics, it introduces prohibitive computational overhead that severely limits practical deployment. Our empirical evaluation on the Jetson Orin Nano platform reveals substantial runtime penalties for quantile computation per frame. As shown in our benchmark results (Table~\ref{tab:benchmark_results}), quantile computation requires 27.43 ms when implemented using PyTorch on GPU and 103.00 ms using CPU with Numba acceleration. These latencies represent significant bottlenecks that violate real-time processing constraints essential for edge video analytics pipelines. For context, at 30 FPS video processing, the total frame budget is only 33.33 ms, making the quantile computation overhead alone consume 82\% of the available processing time on GPU or exceed the frame budget by $3\times$ on CPU.
    Furthermore, the computational complexity scales poorly with gradient tensor size, creating additional challenges for higher resolution inputs. The overhead becomes even more pronounced when considering that quantile computation must be performed for every frame in the video stream, leading to cumulative performance degradation that renders this approach impractical for real-time edge deployment scenarios.
\end{itemize}

In summary, both proposed extensions to AccMPEG fail to meet practical requirements: (Ext@1) lacks generality due to the need for manual tuning, while (Ext@2) is too computationally expensive for real-time applications. By contrast, \textit{How2Compress} \textbf{circumvents both limitations} by learning macroblock-level emphasis directly from structural content, without relying on downstream gradients or expensive runtime operations. This enables fast, generalizable, and efficient video compression that is well-suited for modern edge intelligence systems.

\subsection{Advantages over Codec's Adaptive Quantization}
\label{apx:adv-over-codec}

An interesting observation, which may initially appear paradoxical, arises from Table~\ref{tab:result} and Fig.~\ref{fig:case-study}: AccMPEG consistently outperforms the codec’s AQ across all scenarios of the NVIDIA AI City dataset, yet underperforms compared to AQ on the MOT17 dataset. In the QP assignment heatmap, the QP range produced by the codec's AQ ($\left[30, 35\right]$) is noticeably narrower than that of AccMPEG ($\left[30,45\right]$). Nonetheless, codec AQ achieves superior compression efficiency in MOT17. This might cause by two reasons: (1) differences in object density across the datasets, and (2) the codec’s AQ's more effective exploitation of skip-mode macroblocks.

The NVIDIA AI City dataset generally exhibits low object density, with most frames containing fewer than five detectable objects and many frames containing none. In such cases, AccMPEG (\textit{Where2Compress}) can efficiently assign lower quality to the majority of macroblocks, focusing quality on a sparse set of informative regions, which leads to greater compression efficiency than the codec's AQ. In contrast, the MOT17 dataset features a significantly higher object density, with frequent occurrences of densely packed targets within a frame. This necessitates a larger number of high-quality macroblocks to maintain detection accuracy. While AccMPEG does allocate high quality to critical regions, the codec's AQ often designates even more macroblocks with higher quality. However, when AQ is enabled in the H.264 codec, it adjusts the QP by increasing it in flat or low-activity regions. This raises the likelihood that these regions will closely match their motion predictions, resulting in negligible residuals. Consequently, the encoder can exploit skip mode, which omits residual coding altogether and significantly reduces the number of bits required. This leads to increased sparsity, which in turn enhances the efficiency of entropy coding, thereby lowering the bitrate even if the QP is relatively low.

Our proposed method, How2Compress, consistently outperforms both AccMPEG and the codec's AQ across datasets in most scenarios, due to its capacity for more aggressive, fine-grained, and region-aware QP modulation. Unlike traditional heuristic-based approaches or coarse quantization control, How2Compress dynamically allocates quality at the macroblock level based on the localized visual importance of regions. This allows it to preserve object-relevant areas with higher fidelity while aggressively compressing uninformative background regions. As a result, it can also leverage the benefits of skip-mode macroblocks and achieves superior accuracy–bitrate trade-offs across diverse video scenarios.

\section{Further Results}
\label{apx:more-results}

\subsection{Generalize to Other Tasks}
\label{apx:performance-other-tasks}

See Table~\ref{tab:pose-estimation} and Table~\ref{tab:multi-object-tracking}.
\begin{table}[thb]
\centering
\caption{Pose estimation performance comparison. The table presents evaluation results using standard pose estimation metrics: Object Keypoint Similarity (OKS), Percentage of Correct Keypoints at 0.2 threshold (PCK@0.2), and Percentage of Correct Keypoints at 0.5 threshold (PCK@0.5).}
\label{tab:pose-estimation}
\resizebox{\columnwidth}{!}{
\renewcommand{\arraystretch}{1.4}
\begin{tabular}{l|c|c|c}
\hline
\multicolumn{4}{c}{\cellcolor{gray!20}\textbf{Pose Estimation (OKS / PCK@0.2 / PCK@0.5)}} \\
\hline
\textbf{Sequence} & \textbf{AQ} & \textbf{ACCMPEG} & \textbf{Ours} \\
\hline
MOT17-02 & 0.956 / 0.862 / 0.940 & 0.954 / 0.854 / 0.936 & 0.947 / 0.839 / 0.924 \\
MOT17-04 & 0.968 / 0.921 / 0.966 & 0.948 / 0.894 / 0.951 & 0.939 / 0.880 / 0.943 \\
MOT17-09 & 0.970 / 0.900 / 0.948 & 0.964 / 0.892 / 0.944 & 0.954 / 0.886 / 0.937 \\
MOT17-10 & 0.927 / 0.898 / 0.929 & 0.926 / 0.895 / 0.932 & 0.919 / 0.886 / 0.934 \\
MOT17-11 & 0.975 / 0.948 / 0.965 & 0.975 / 0.948 / 0.966 & 0.976 / 0.952 / 0.973 \\
MOT17-13 & 0.950 / 0.893 / 0.946 & 0.933 / 0.870 / 0.934 & 0.935 / 0.875 / 0.941 \\
\hline
\end{tabular}
}
\end{table}

\begin{table}[thb]
\centering
\caption{Multi-object tracking performance comparison. The table presents evaluation results using standard tracking metrics: Multiple Object Tracking Accuracy (MOTA), Multiple Object Tracking Precision (MOTP), and ID F1 Score (IDF1).}
\label{tab:multi-object-tracking}
\resizebox{\columnwidth}{!}{
\renewcommand{\arraystretch}{1.4}
\begin{tabular}{l|c|c|c}
\hline
\multicolumn{4}{c}{\cellcolor{gray!20}\textbf{Multi-Object Tracking (MOTA / MOTP / IDF1)}} \\
\hline
\textbf{Sequence} & \textbf{AQ} & \textbf{ACCMPEG} & \textbf{Ours} \\
\hline
MOT17-02 & 0.910 / 0.949 / 0.938 & 0.882 / 0.949 / 0.934 & 0.891 / 0.941 / 0.927 \\
MOT17-04 & 0.982 / 0.971 / 0.989 & 0.983 / 0.962 / 0.985 & 0.982 / 0.961 / 0.982 \\
MOT17-09 & 0.978 / 0.977 / 0.982 & 0.974 / 0.971 / 0.976 & 0.972 / 0.967 / 0.964 \\
MOT17-10 & 0.930 / 0.955 / 0.950 & 0.909 / 0.951 / 0.942 & 0.909 / 0.944 / 0.935 \\
MOT17-11 & 0.970 / 0.973 / 0.972 & 0.960 / 0.965 / 0.961 & 0.965 / 0.963 / 0.945 \\
MOT17-13 & 0.928 / 0.950 / 0.941 & 0.923 / 0.943 / 0.932 & 0.919 / 0.949 / 0.911 \\
\hline
\end{tabular}
}
\end{table}

\subsection{Overhead of Different Resolution Input}
\label{apx:different-rs-input-overhead}

\begin{figure}
    \centering
    \includegraphics[width=0.5\textwidth]{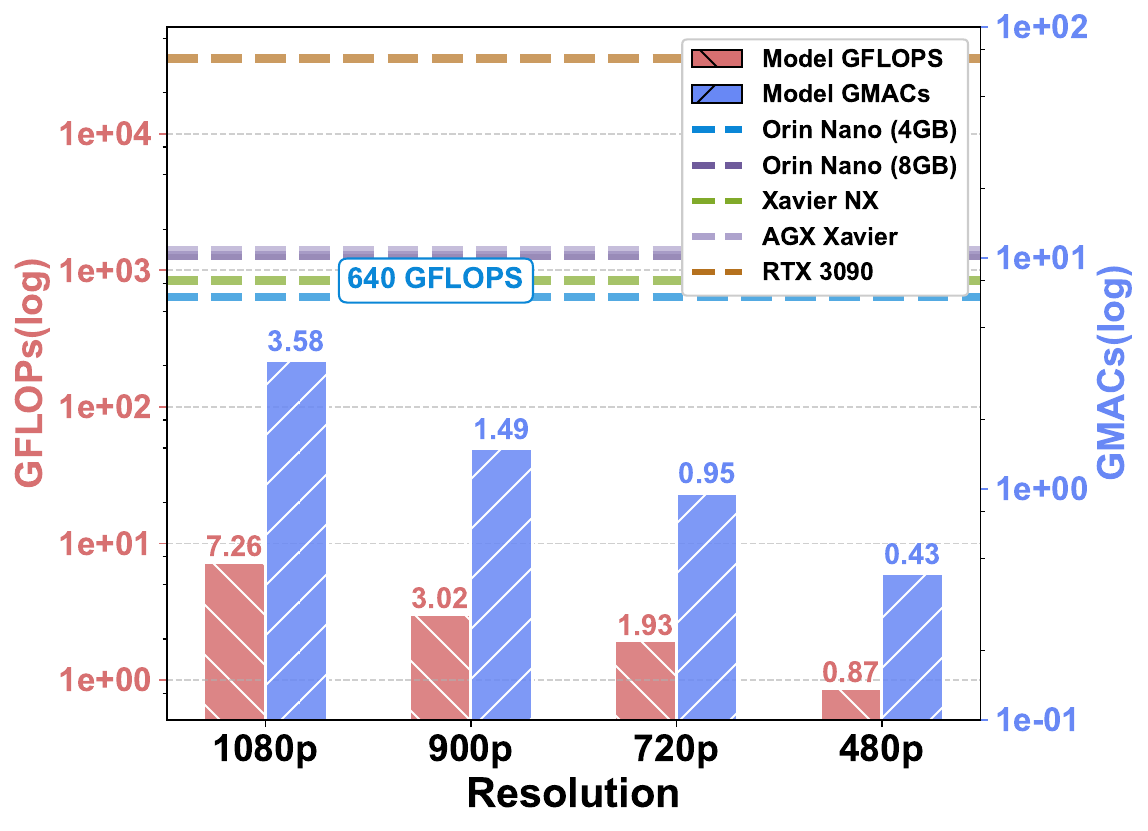}
    \caption{Computational overhead across varying input resolutions and device power profiles}
    \label{fig:apx-compute-cost-eval}
\end{figure}

We evaluate \framework across different input resolutions (1080p, 900p, 720p, and 480p). As shown in Fig.~\ref{fig:apx-compute-cost-eval}, the model requires only 7.21 GFLOPS even for 1080p videos, rendering it suitable for resource-constrained edge devices. The model exhibits low computational complexity, with only 3.56 GMACs. In comparison to the lightweight detection model YOLOV5-s~\cite{yolov5}, the Emphasis Assignment model uses just $0.4\times$ the resources, highlighting its efficient across different platforms.

\subsection{Qualitative Result}
\label{apx:qualitative}
\begin{figure*}[t] \centering
    \makebox[0.012\textwidth]{}
    \makebox[0.15\textwidth]{\small MOT1702}
    \makebox[0.15\textwidth]{\small MOT1704}
    \makebox[0.15\textwidth]{\small MOT1709}
    \makebox[0.15\textwidth]{\small MOT1710}
    \makebox[0.15\textwidth]{\small MOT1711}
    \makebox[0.15\textwidth]{\small MOT1713}
    \\
    \raisebox{0.1\height}{\makebox[0.018\textwidth]{\rotatebox{90}{\makecell{\tiny \makecell{Uniform QP \\ (when2compress)}}}}}
    \includegraphics[width=0.15\textwidth]{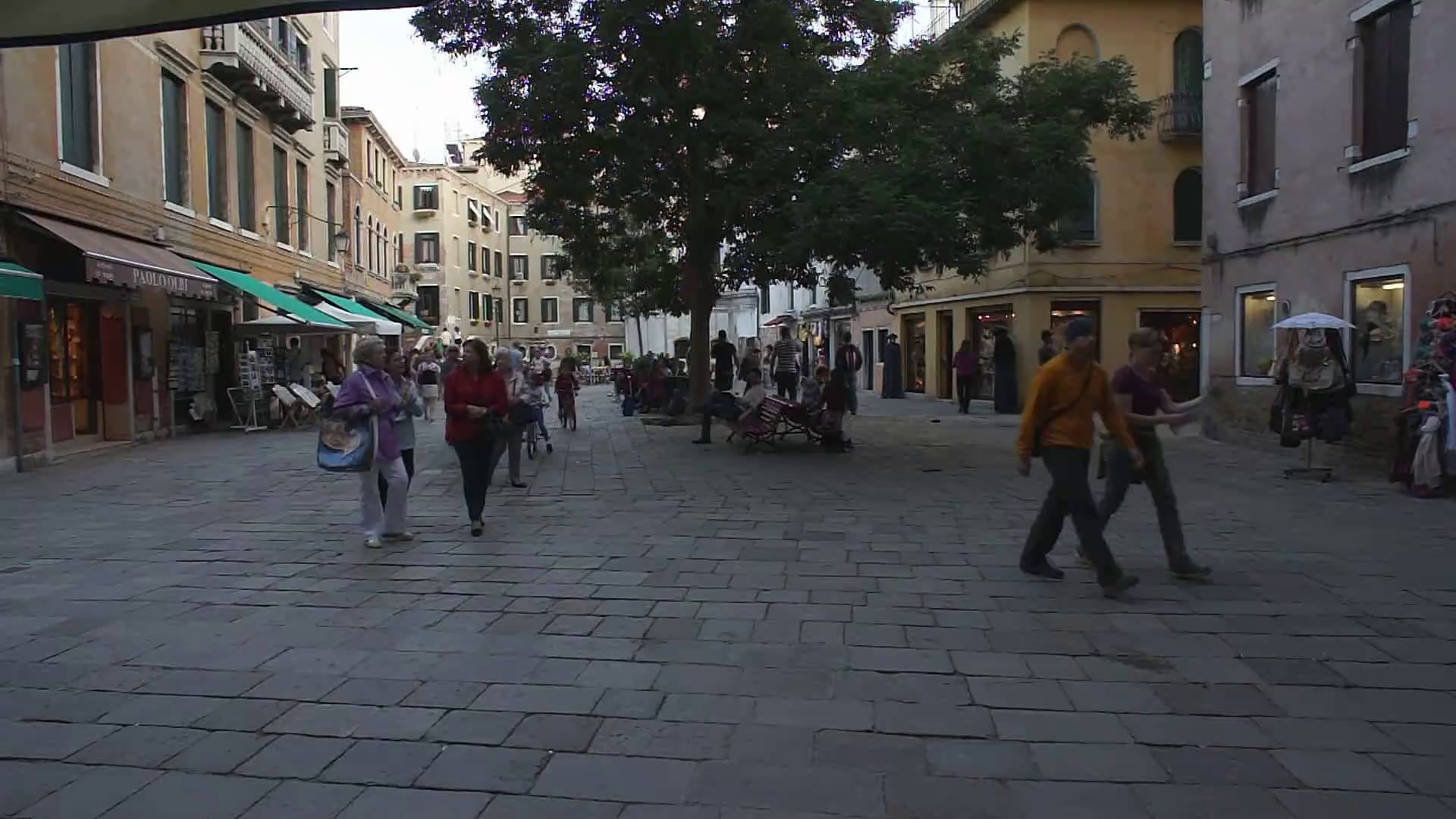}
    \includegraphics[width=0.15\textwidth]{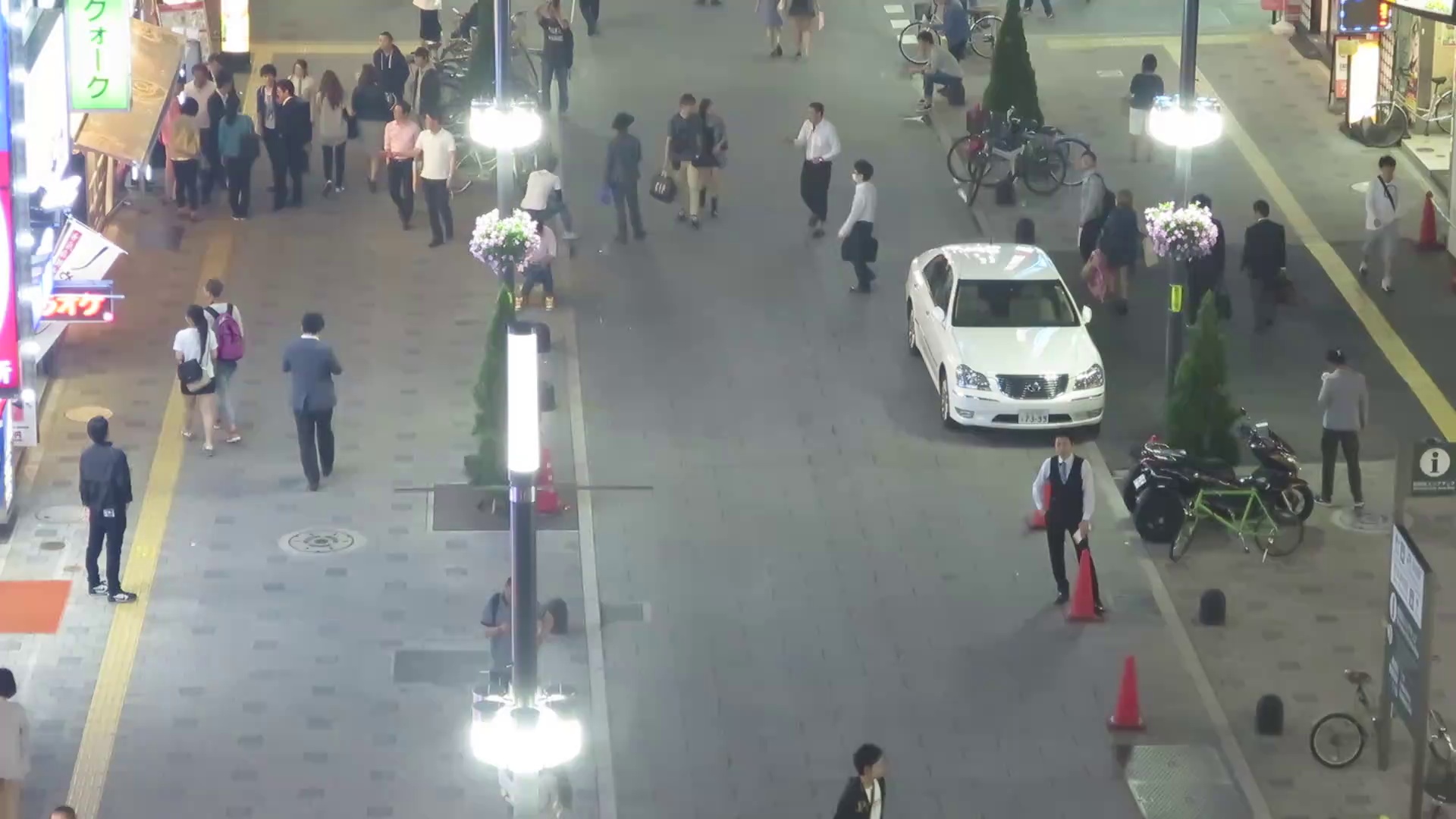}
    \includegraphics[width=0.15\textwidth]{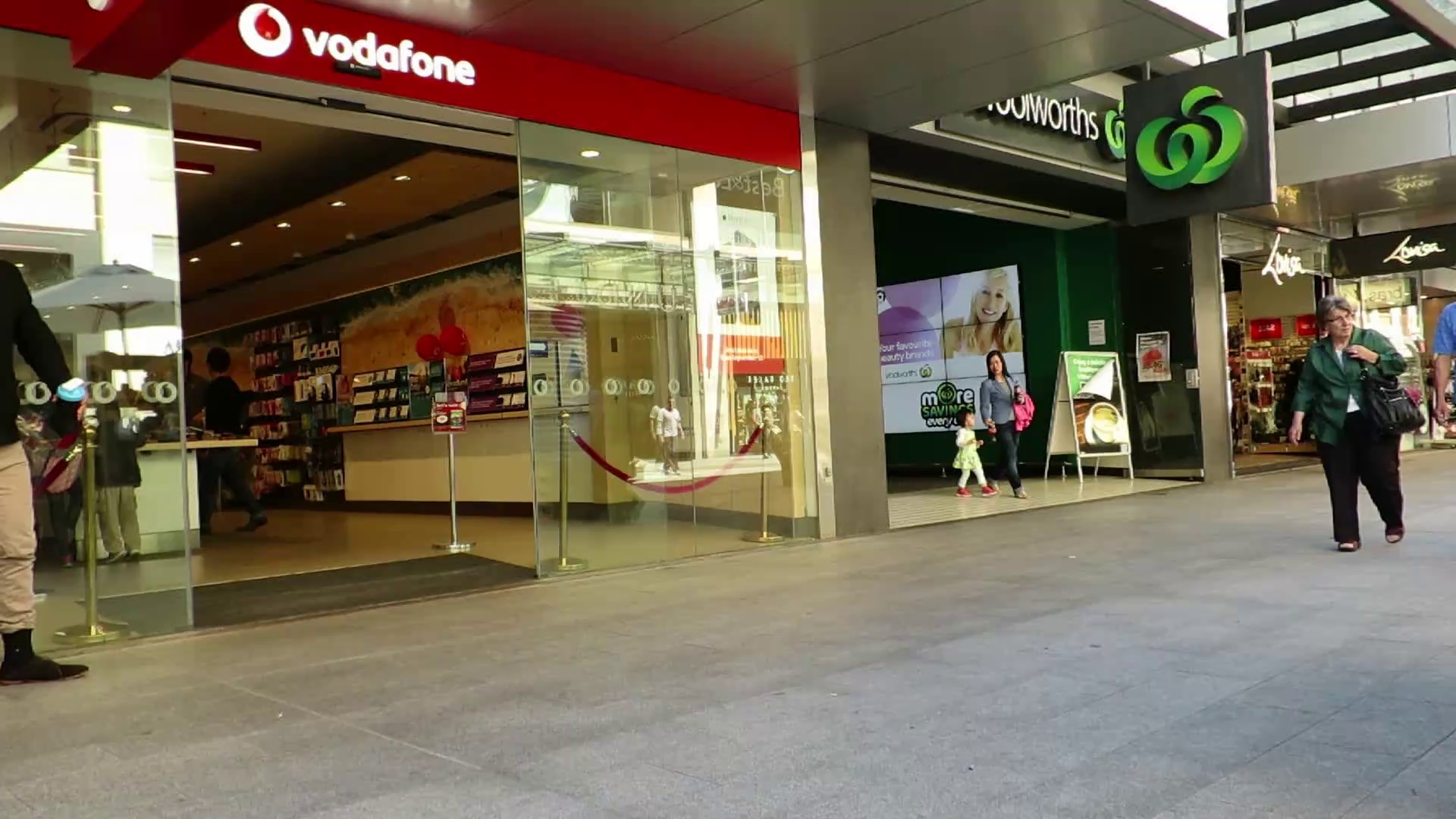}
    \includegraphics[width=0.15\textwidth]{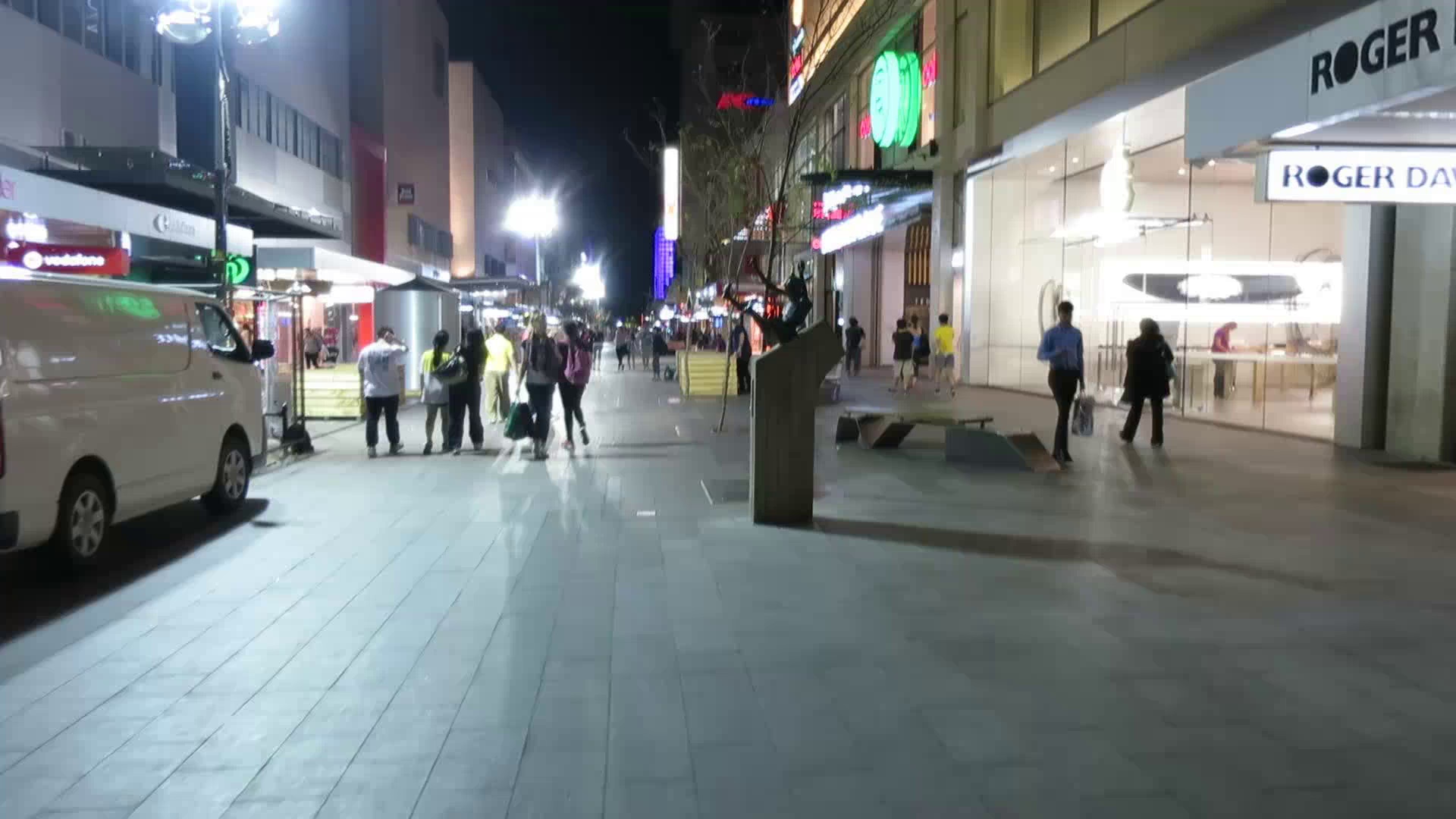}
    \includegraphics[width=0.15\textwidth]{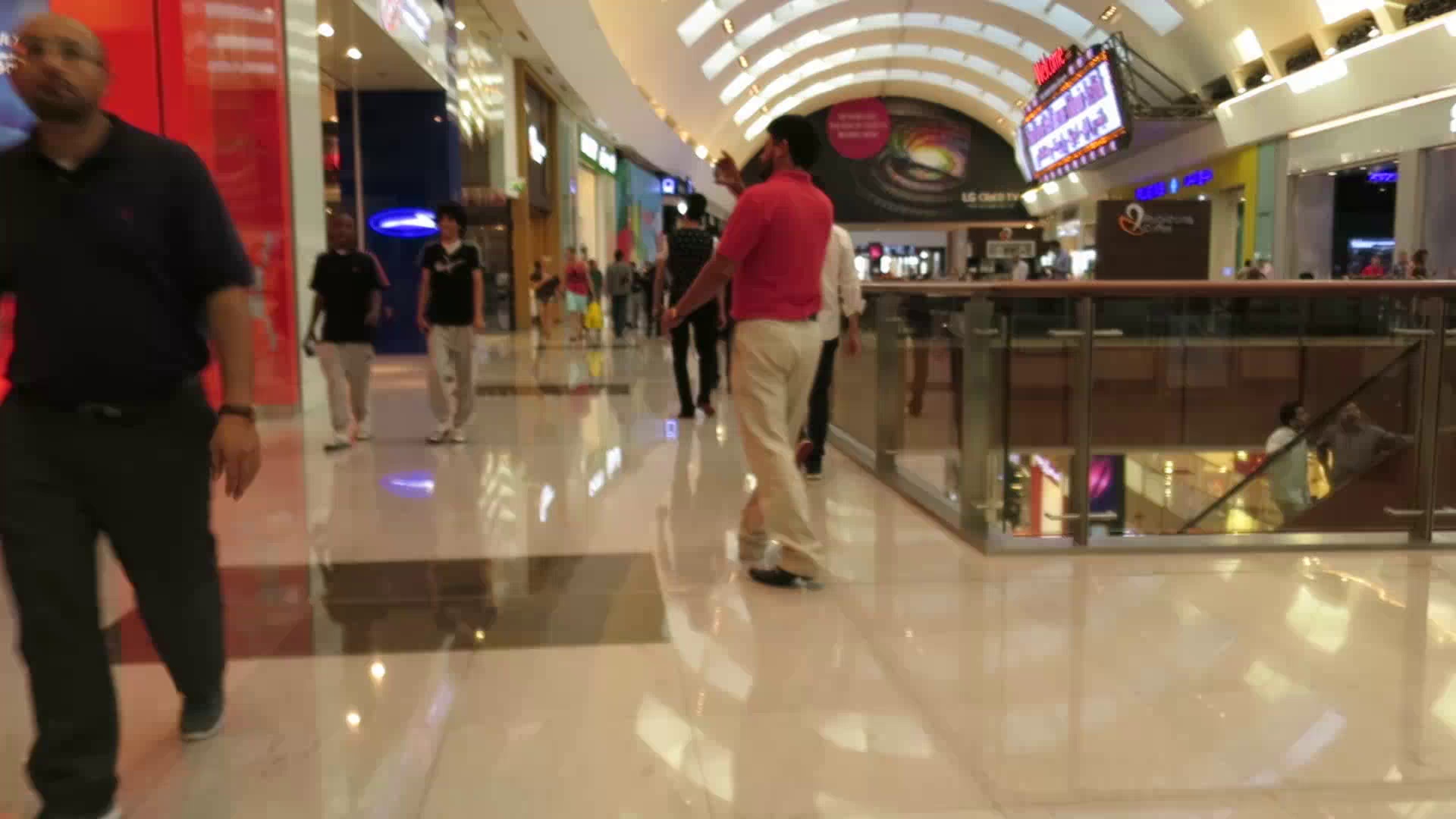}
    \includegraphics[width=0.15\textwidth]{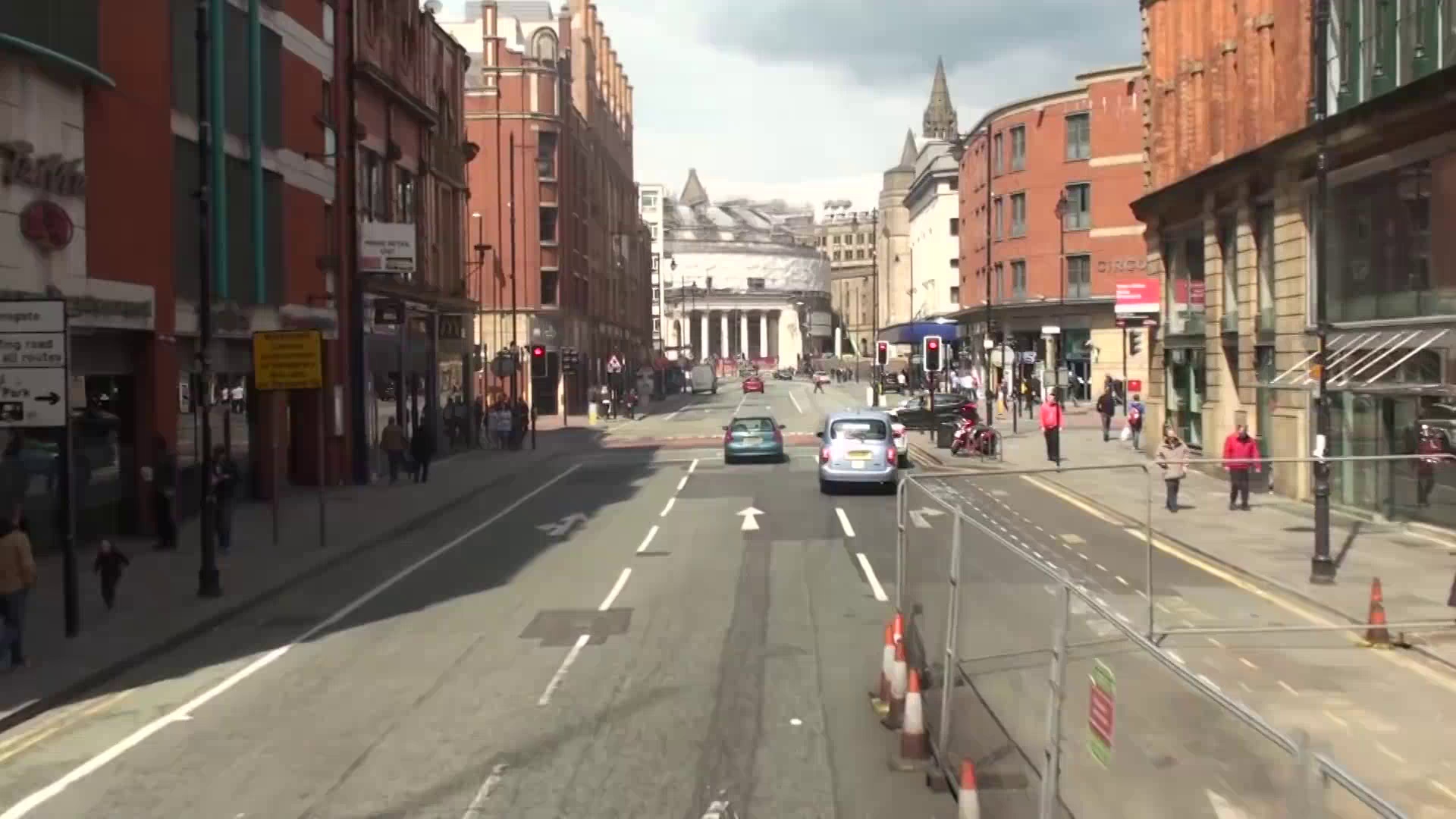}
    \\
    \raisebox{0.1\height}{\makebox[0.018\textwidth]{\rotatebox{90}{\makecell{\tiny \makecell{AccMPEG\\ (where2compress)}}}}}
    \includegraphics[width=0.15\textwidth]{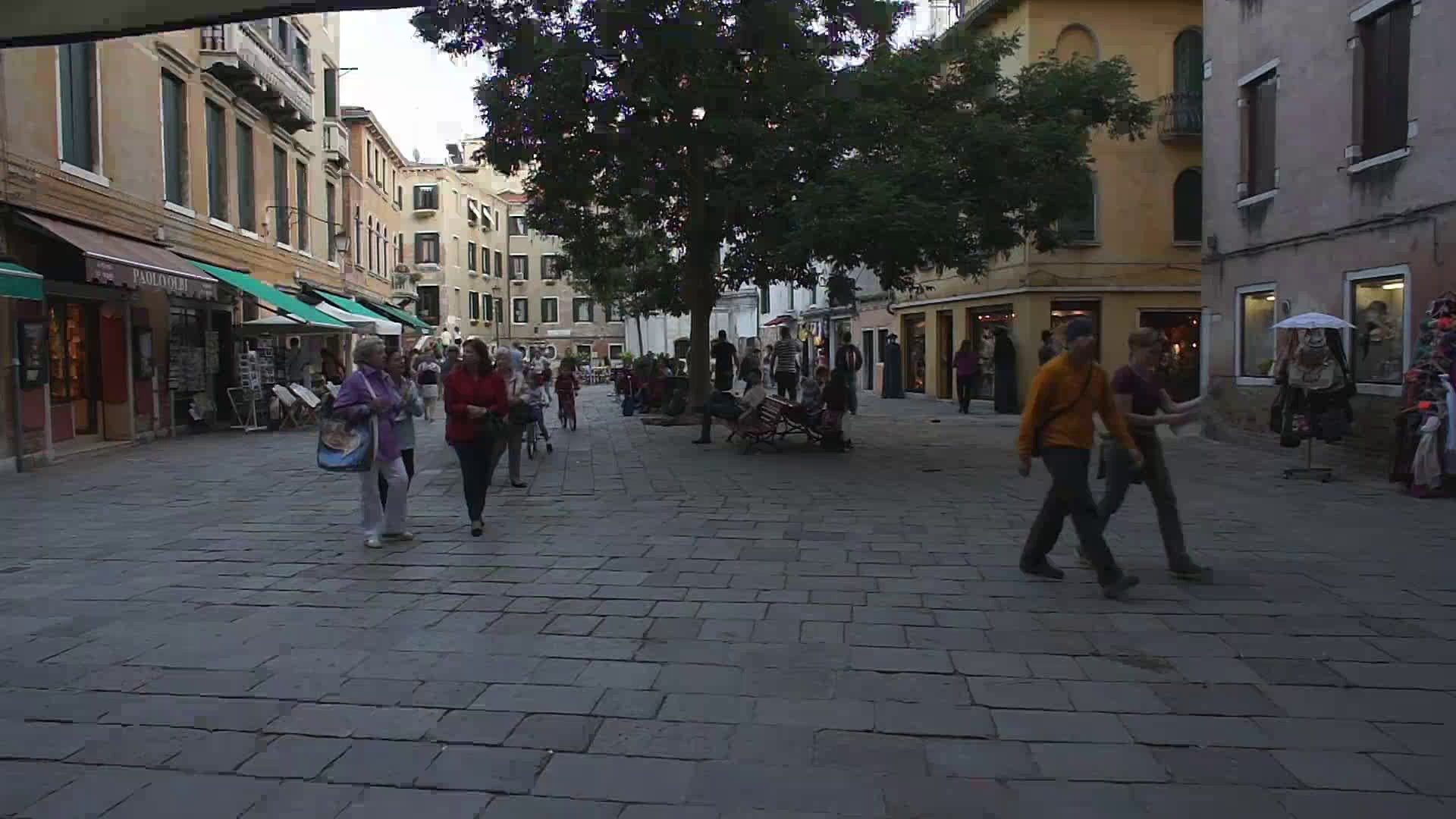}
    \includegraphics[width=0.15\textwidth]{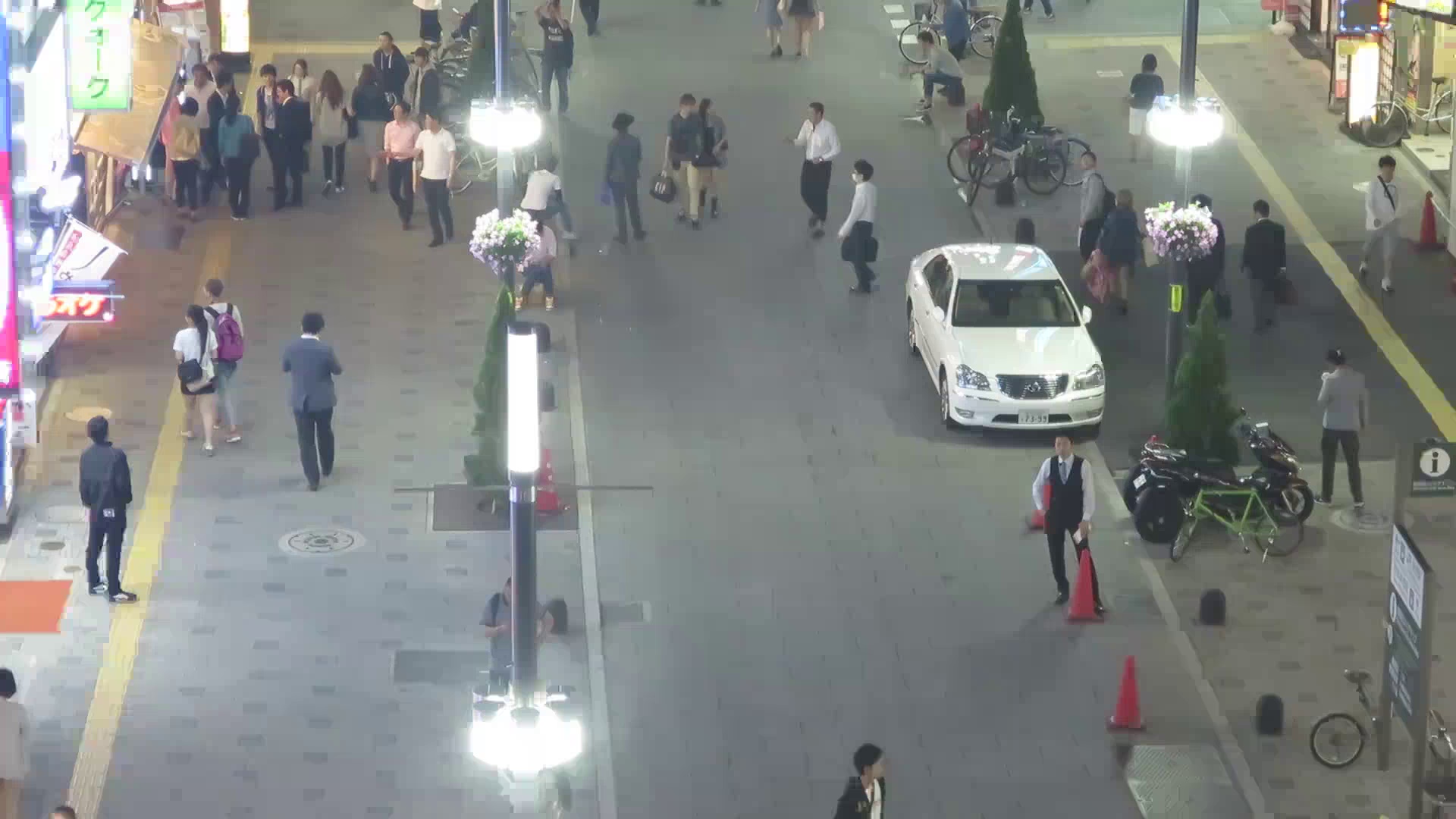}
    \includegraphics[width=0.15\textwidth]{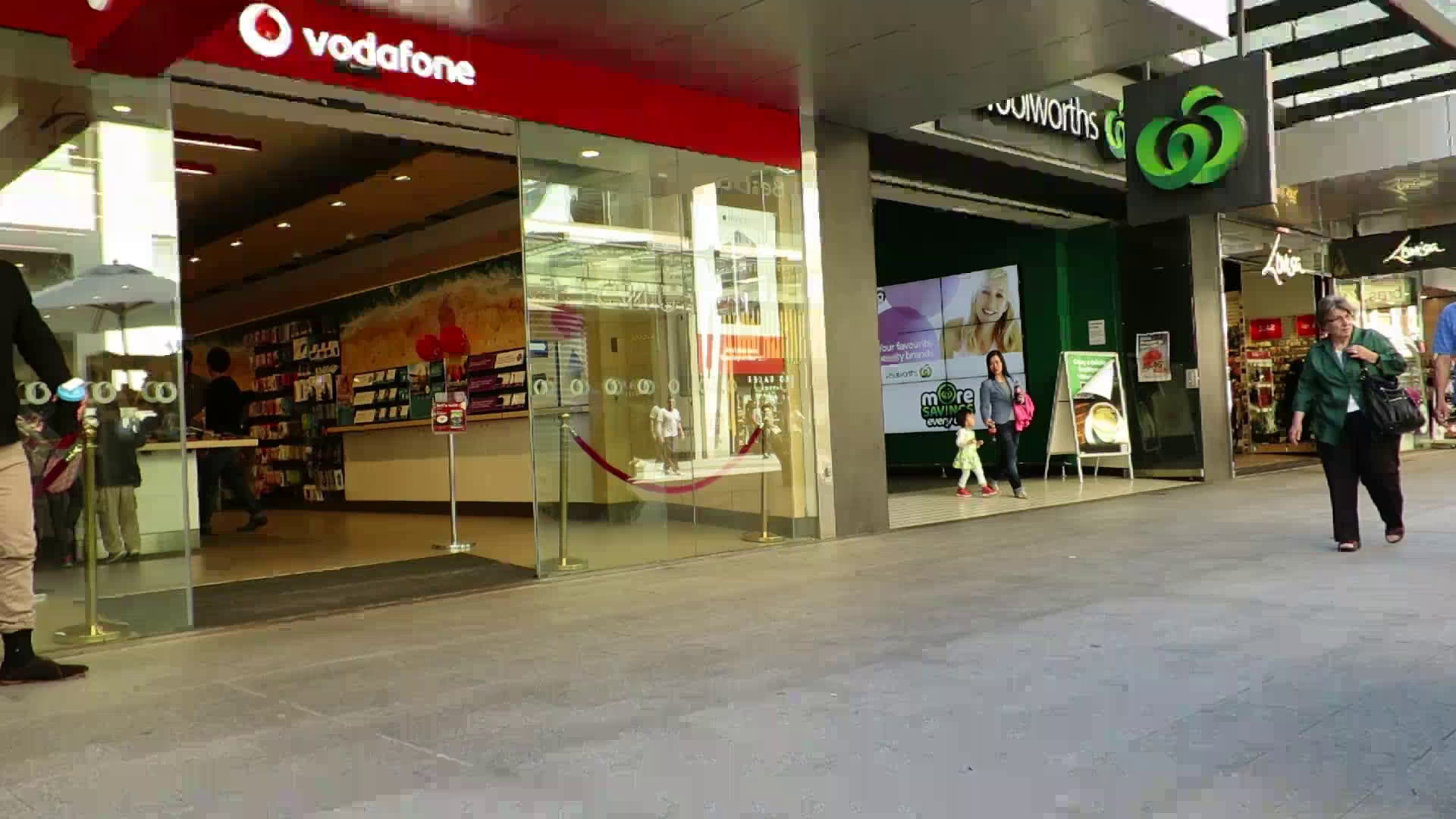}
    \includegraphics[width=0.15\textwidth]{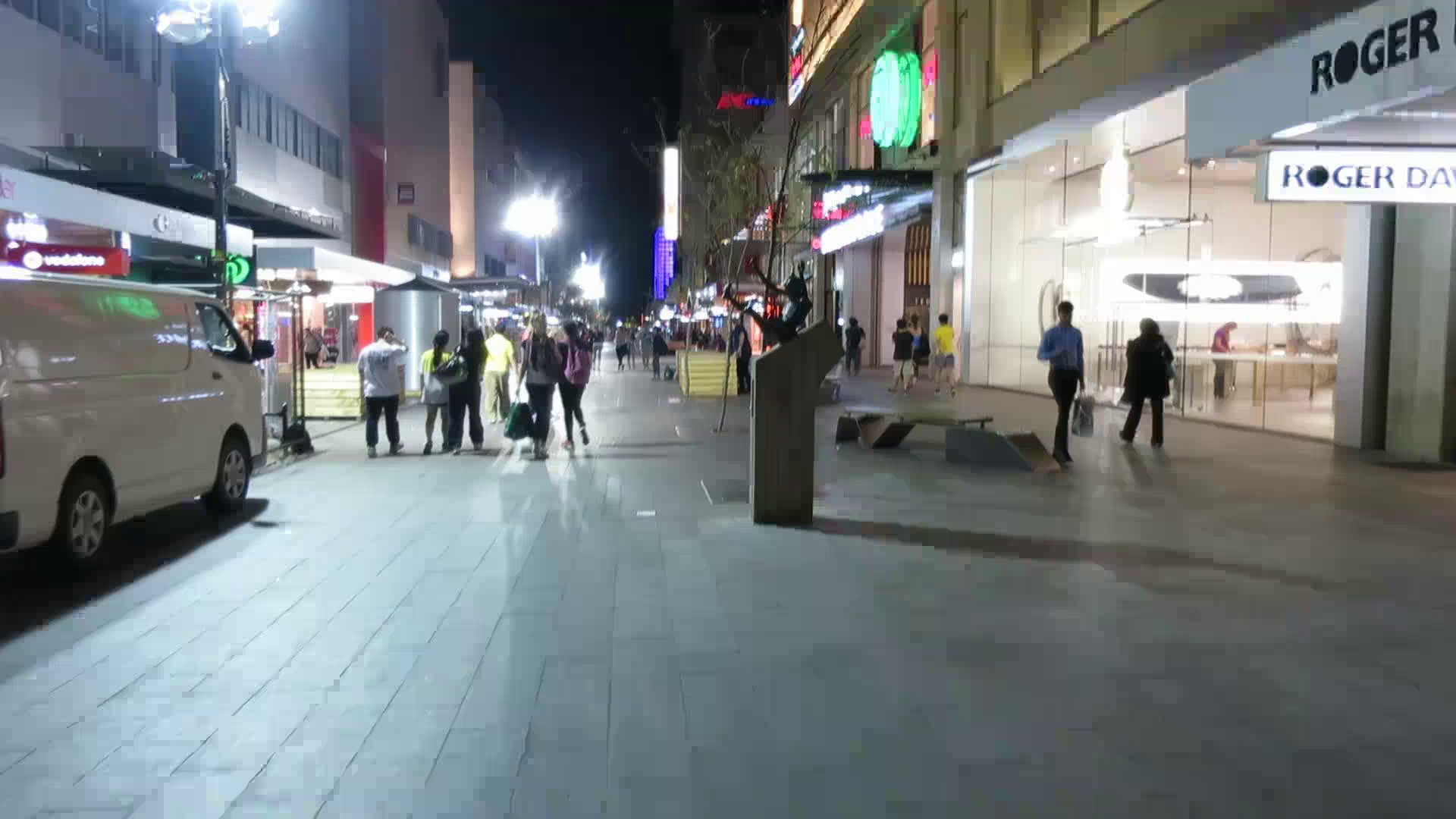}
    \includegraphics[width=0.15\textwidth]{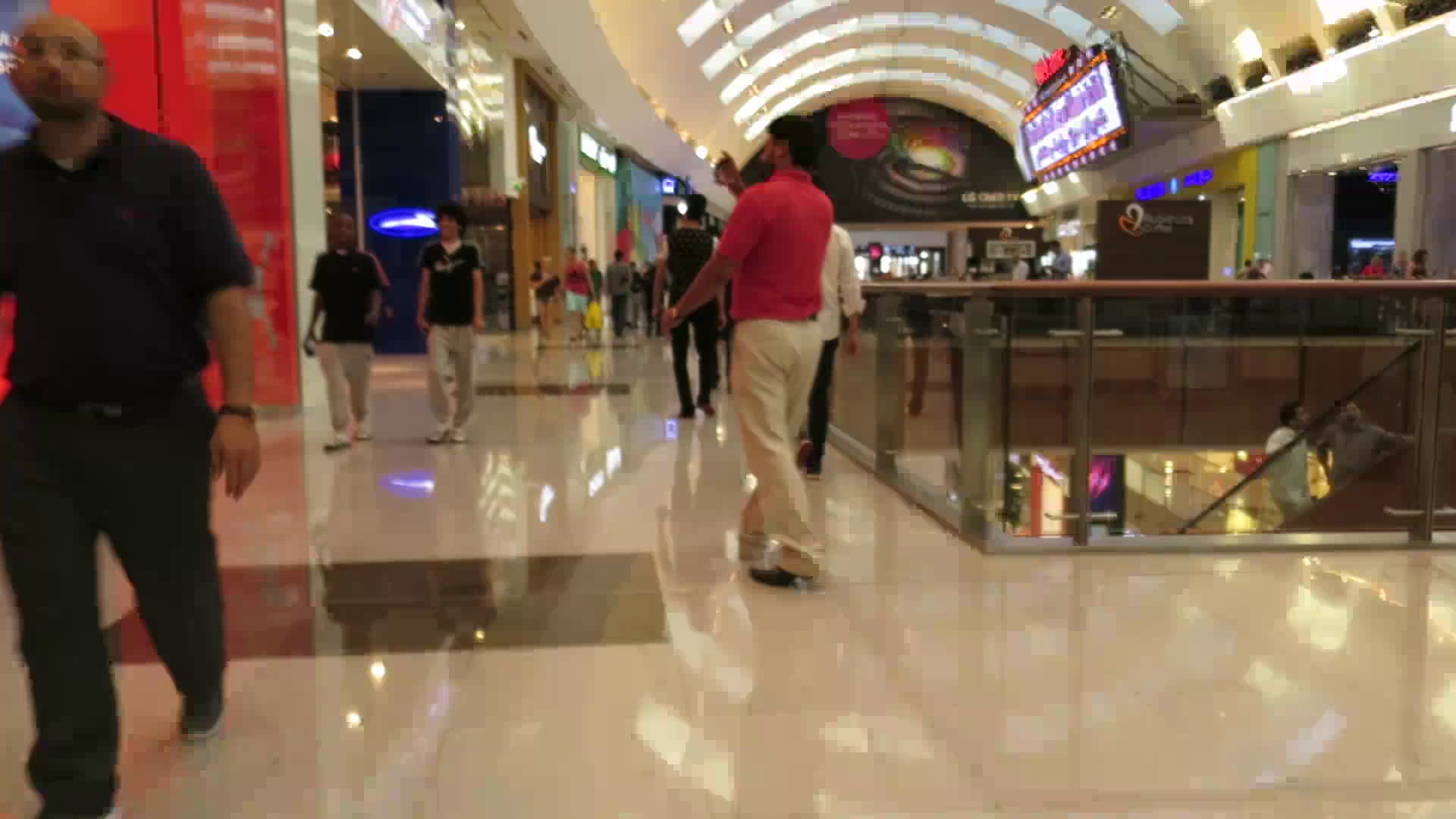}
    \includegraphics[width=0.15\textwidth]{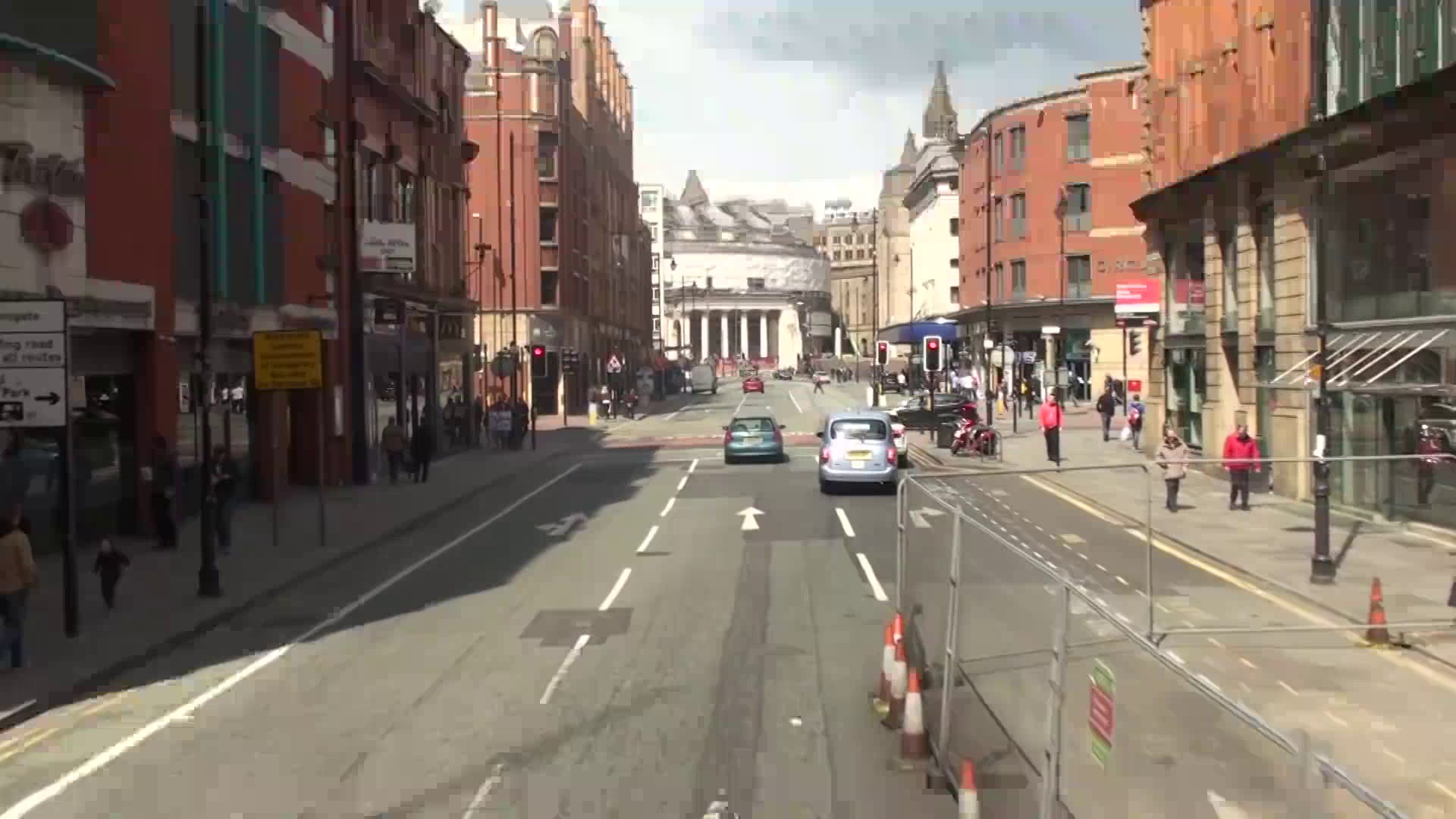}
    \\
    \raisebox{0.1\height}{\makebox[0.018\textwidth]{\rotatebox{90}{\makecell{\tiny \makecell{Adapt Quant \\ (Codec)}}}}}
    \includegraphics[width=0.15\textwidth]{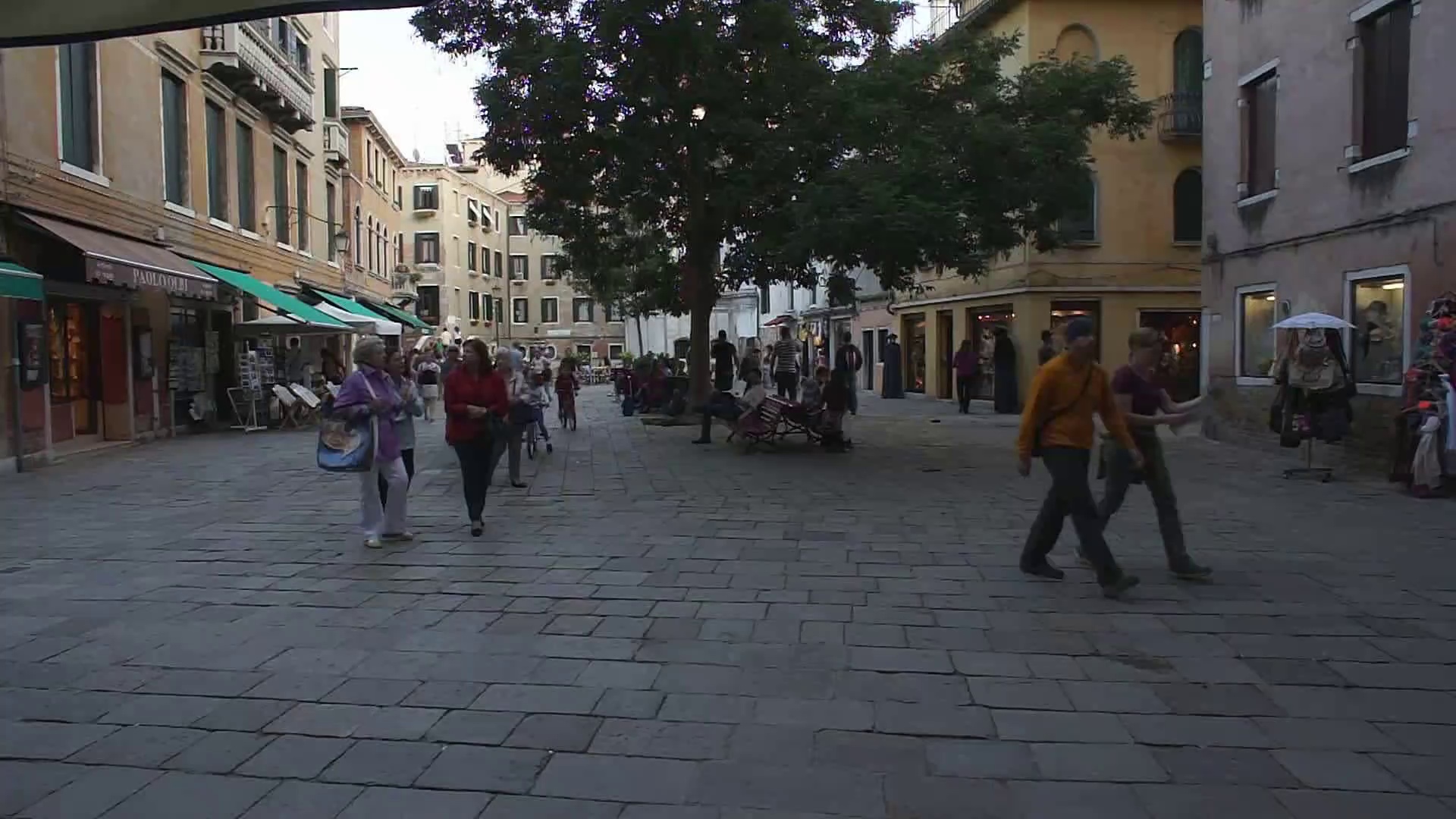}
    \includegraphics[width=0.15\textwidth]{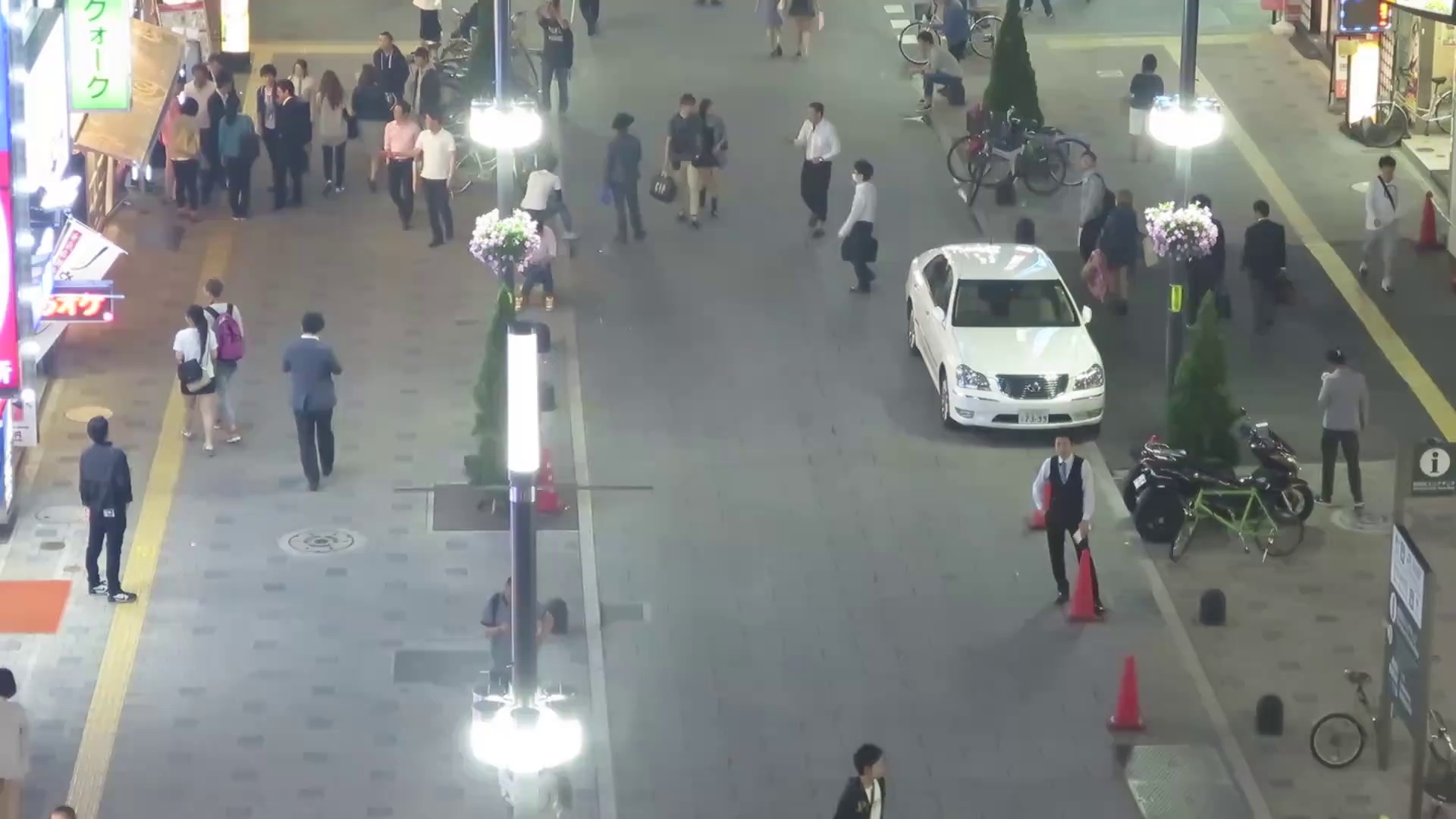}
    \includegraphics[width=0.15\textwidth]{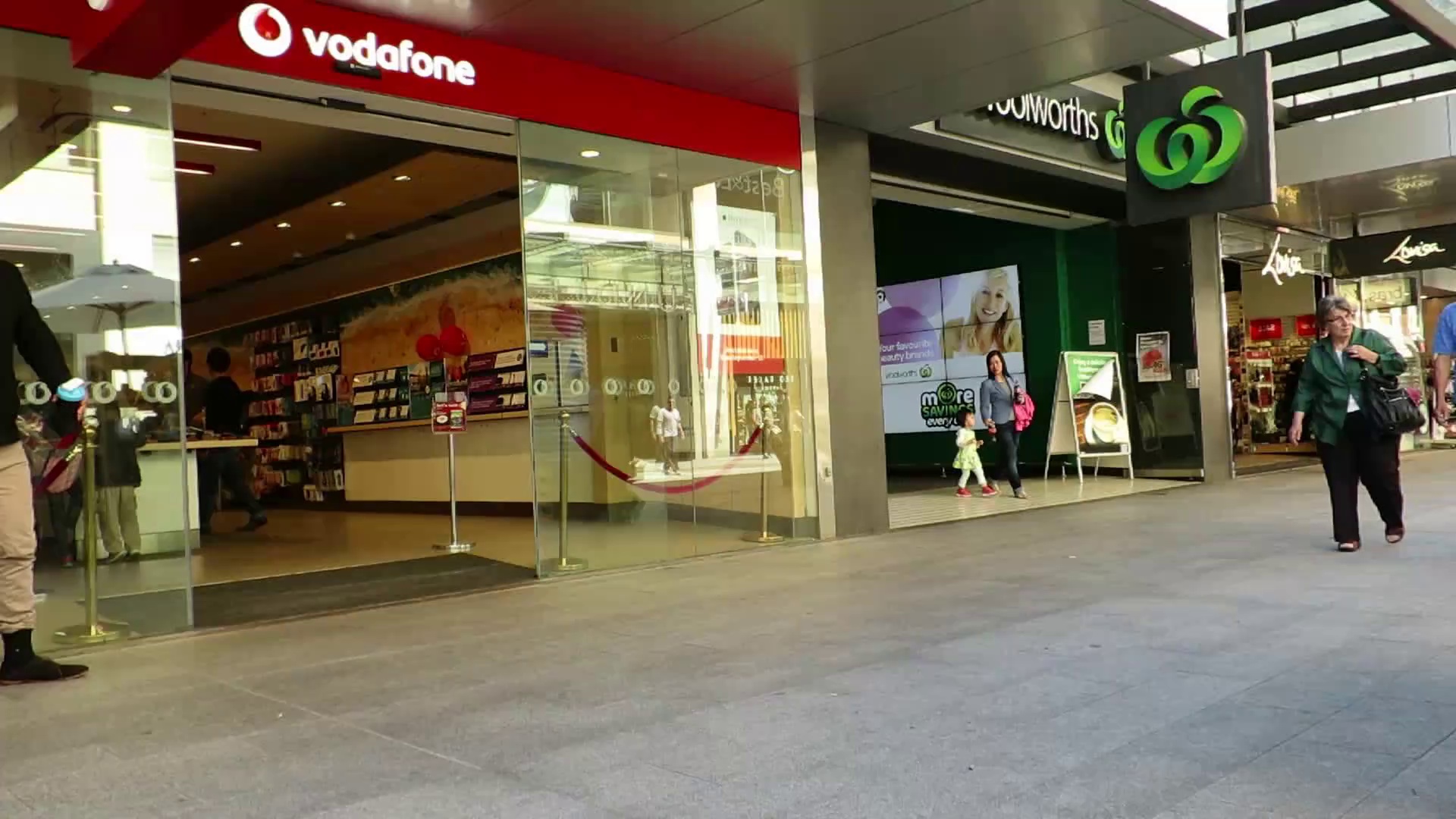}
    \includegraphics[width=0.15\textwidth]{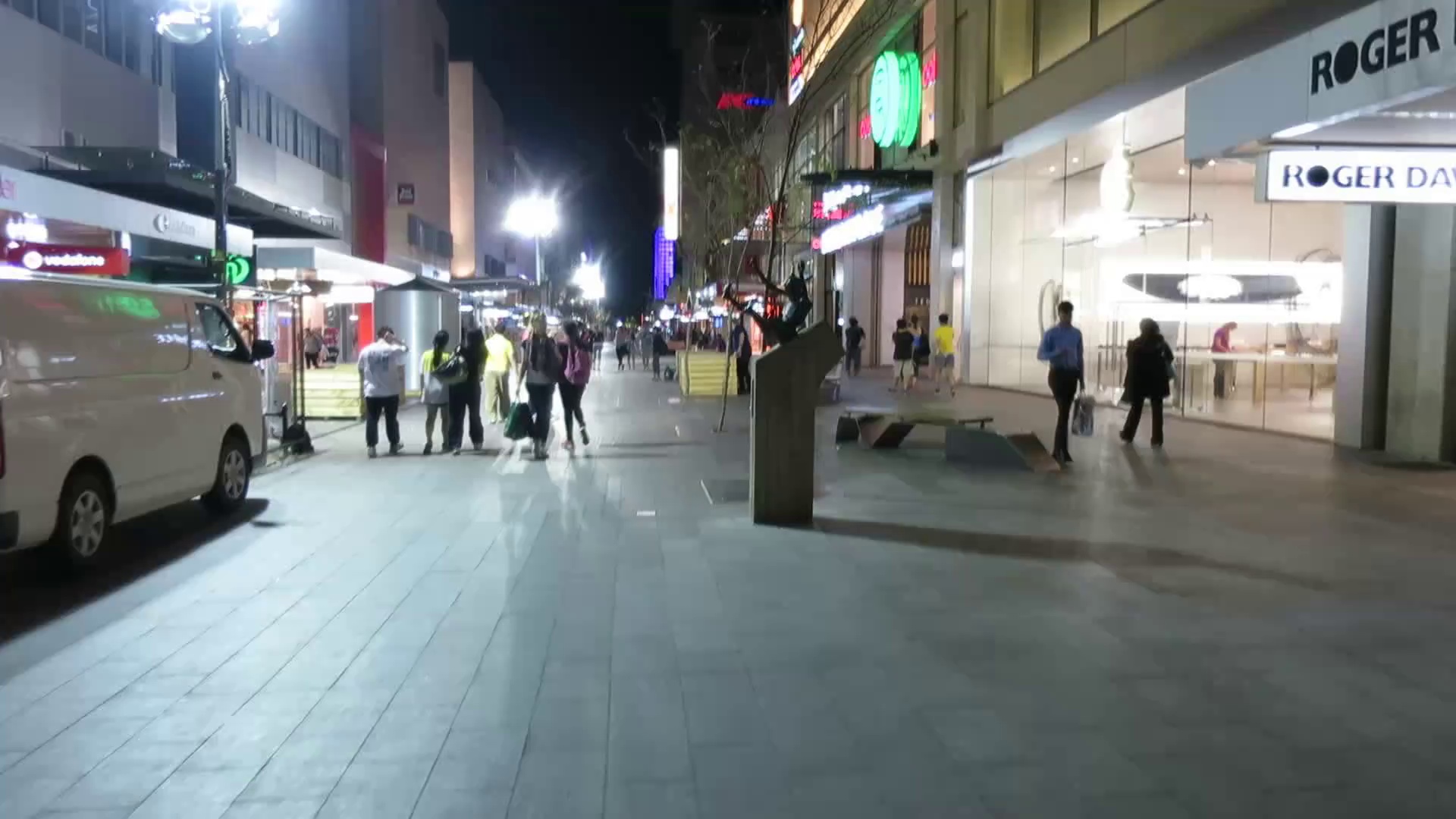}
    \includegraphics[width=0.15\textwidth]{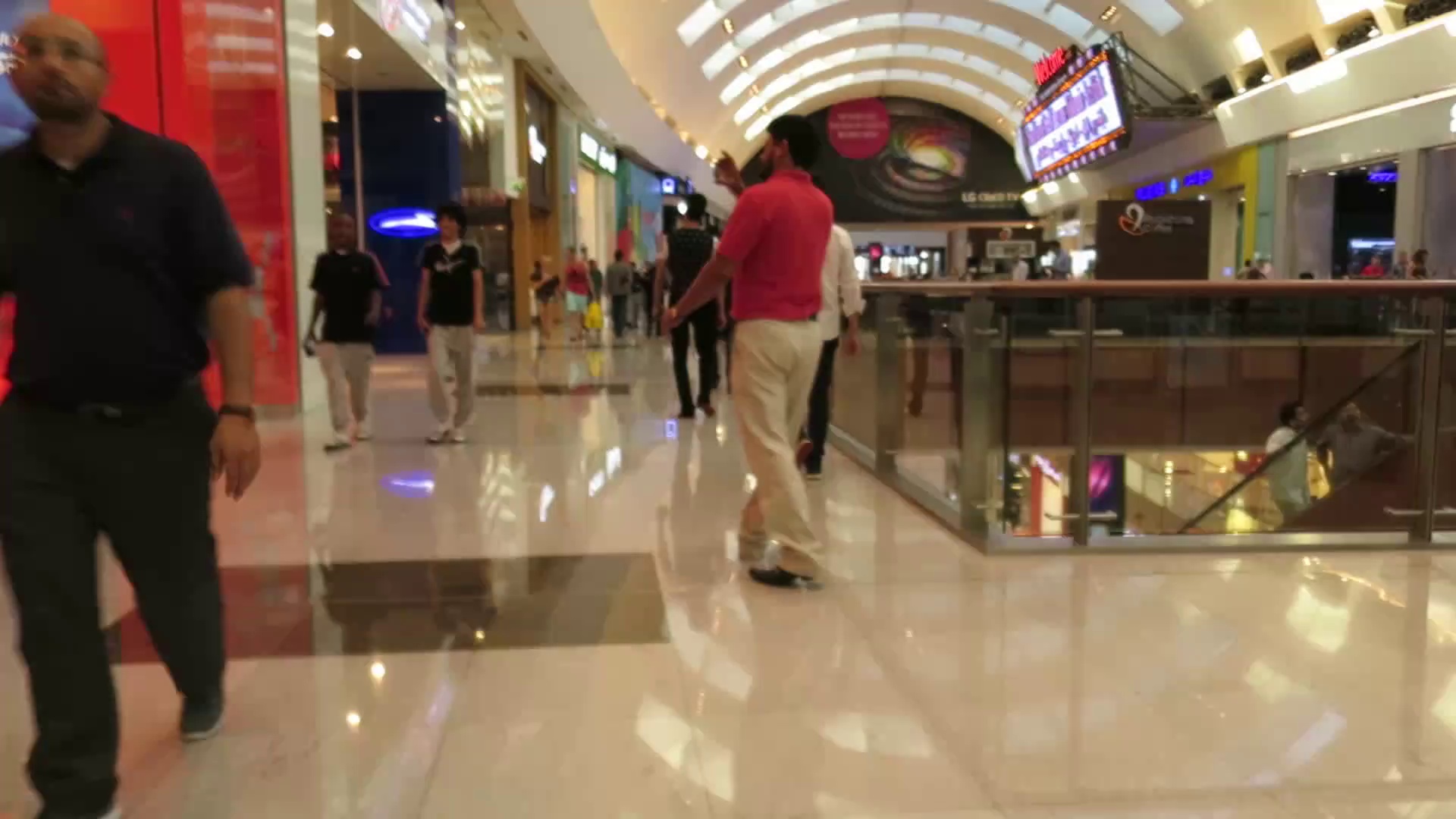}
    \includegraphics[width=0.15\textwidth]{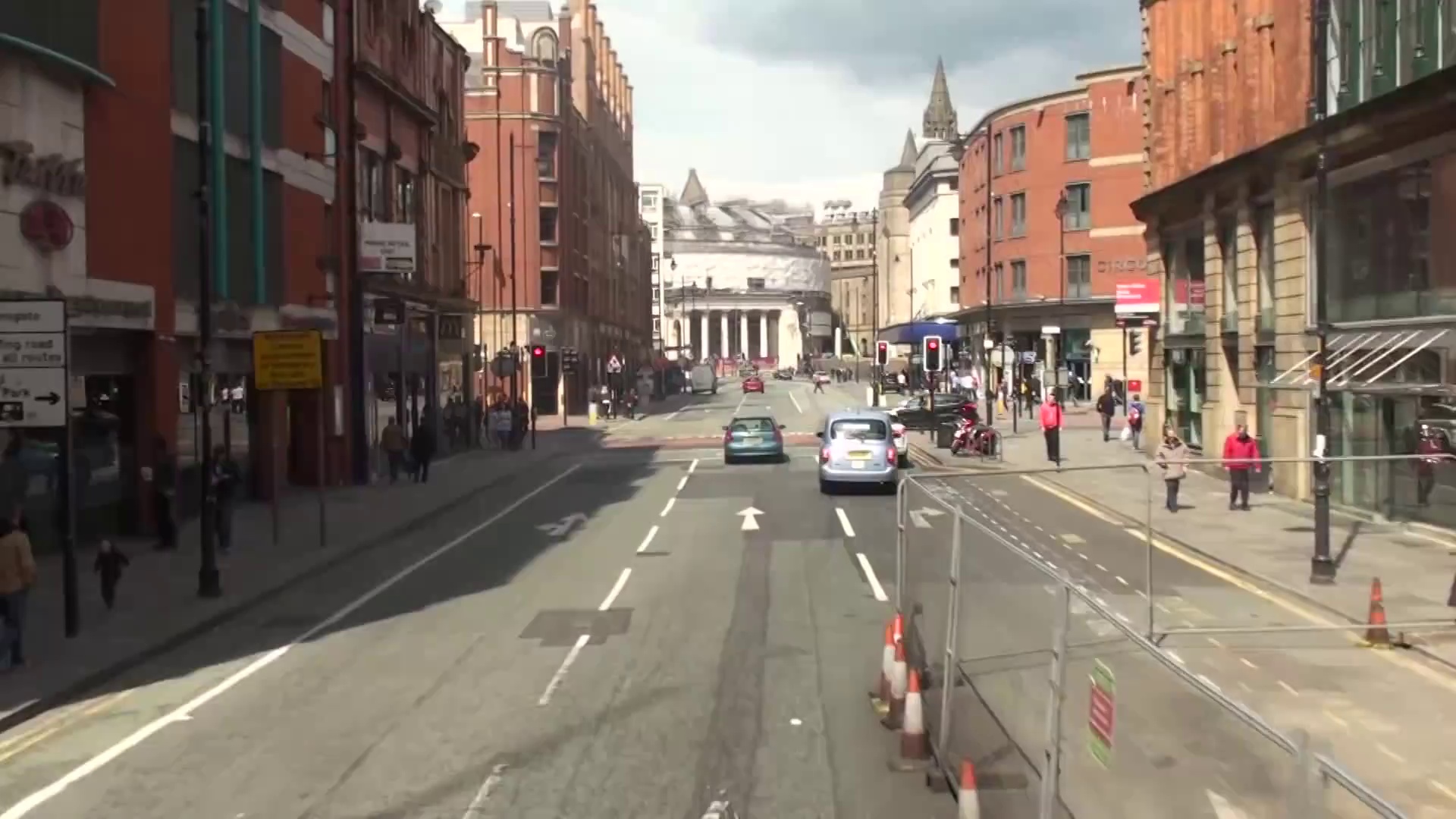}
    \\
    \raisebox{0.1\height}{\makebox[0.018\textwidth]{\rotatebox{90}{\makecell{\tiny How2Compress}}}}
    \includegraphics[width=0.15\textwidth]{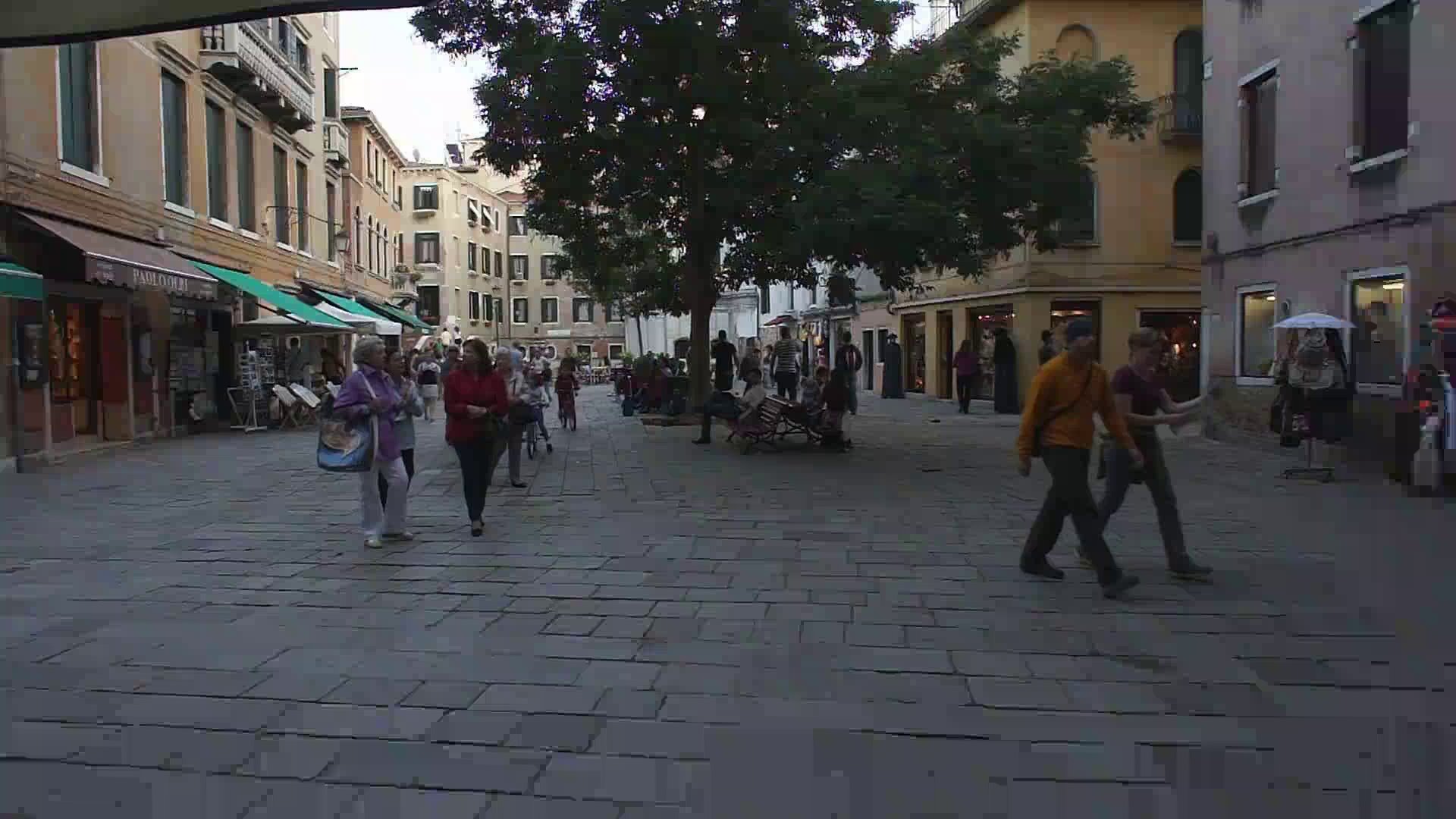}
    \includegraphics[width=0.15\textwidth]{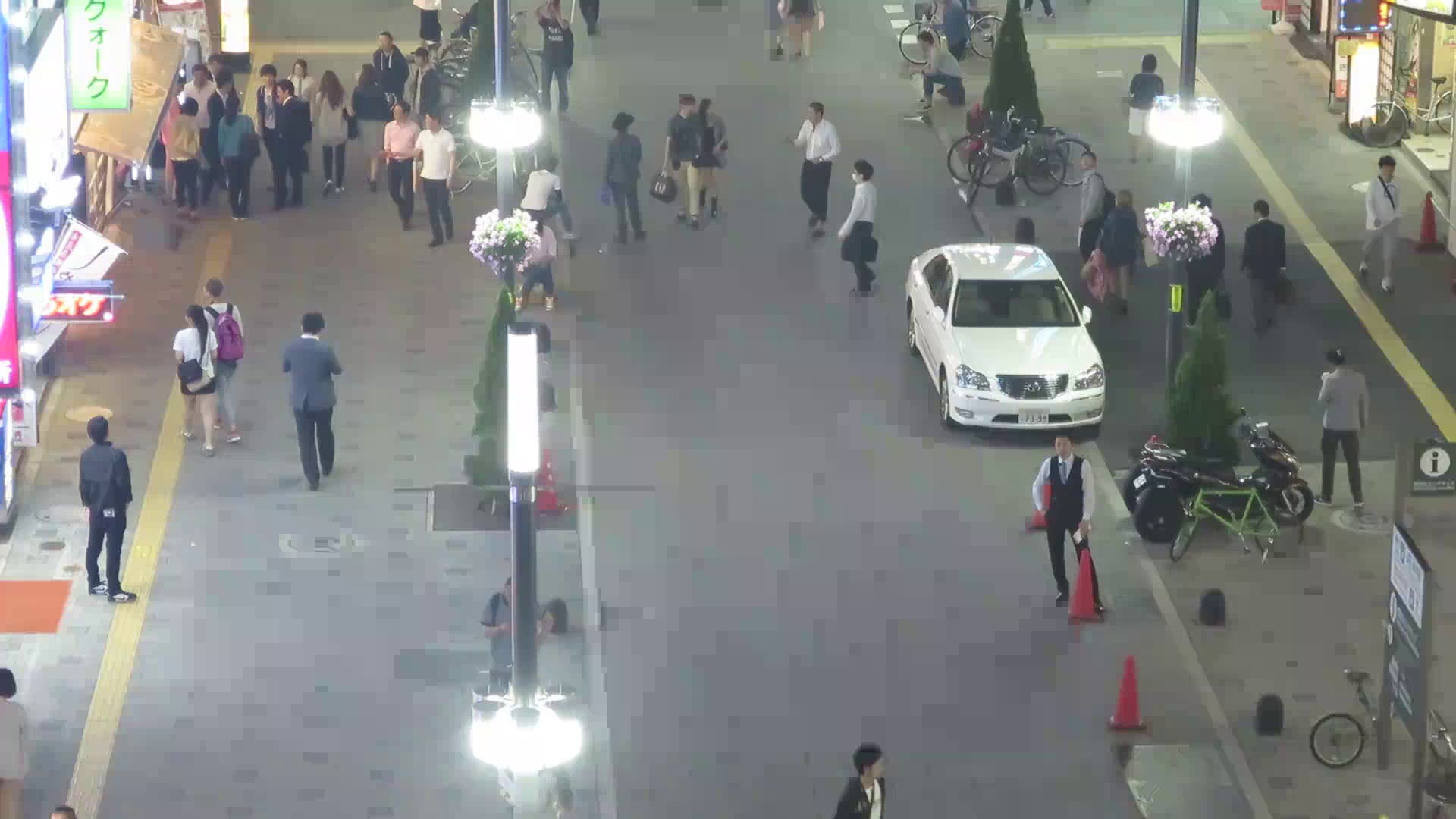}
    \includegraphics[width=0.15\textwidth]{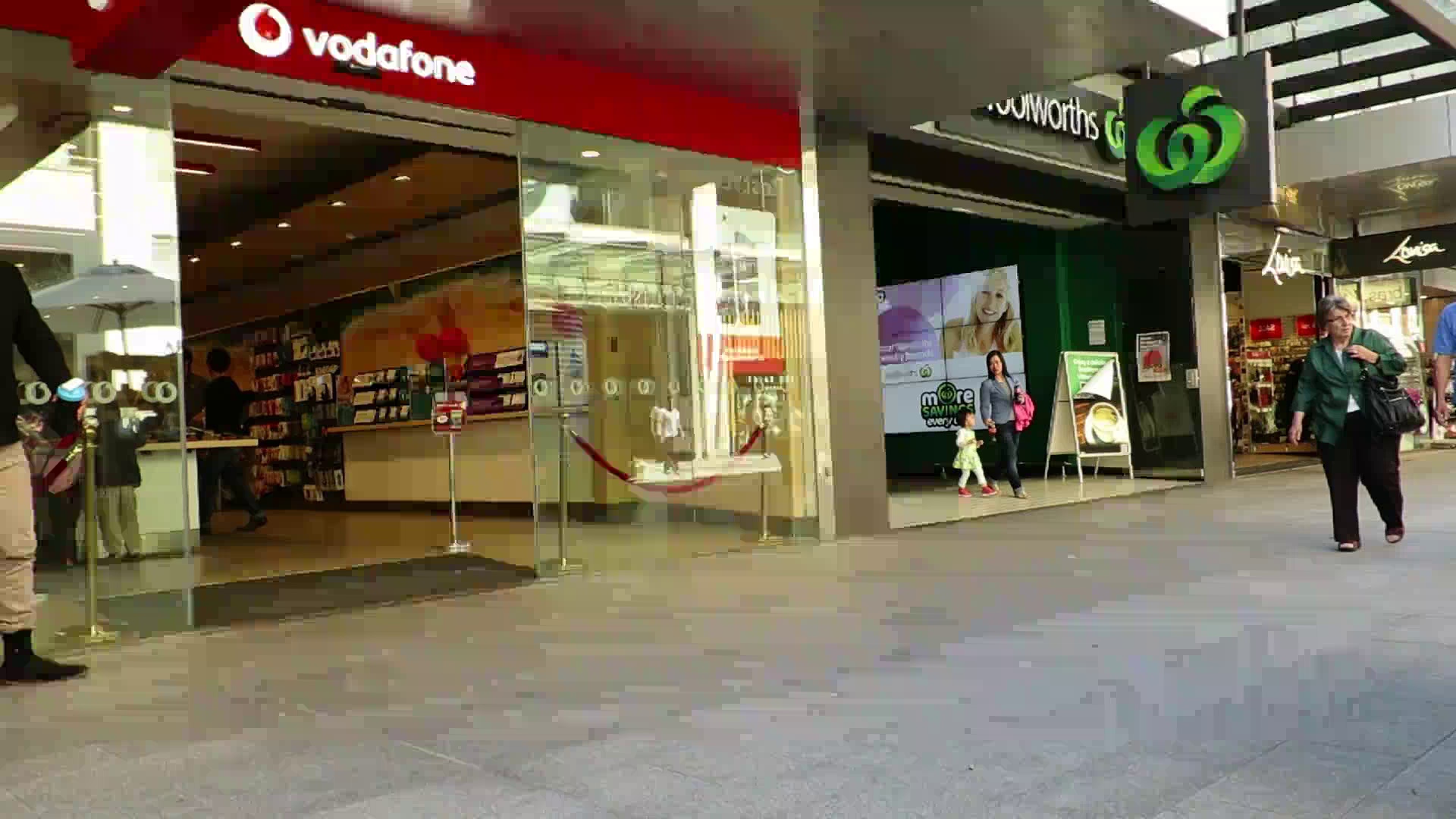}
    \includegraphics[width=0.15\textwidth]{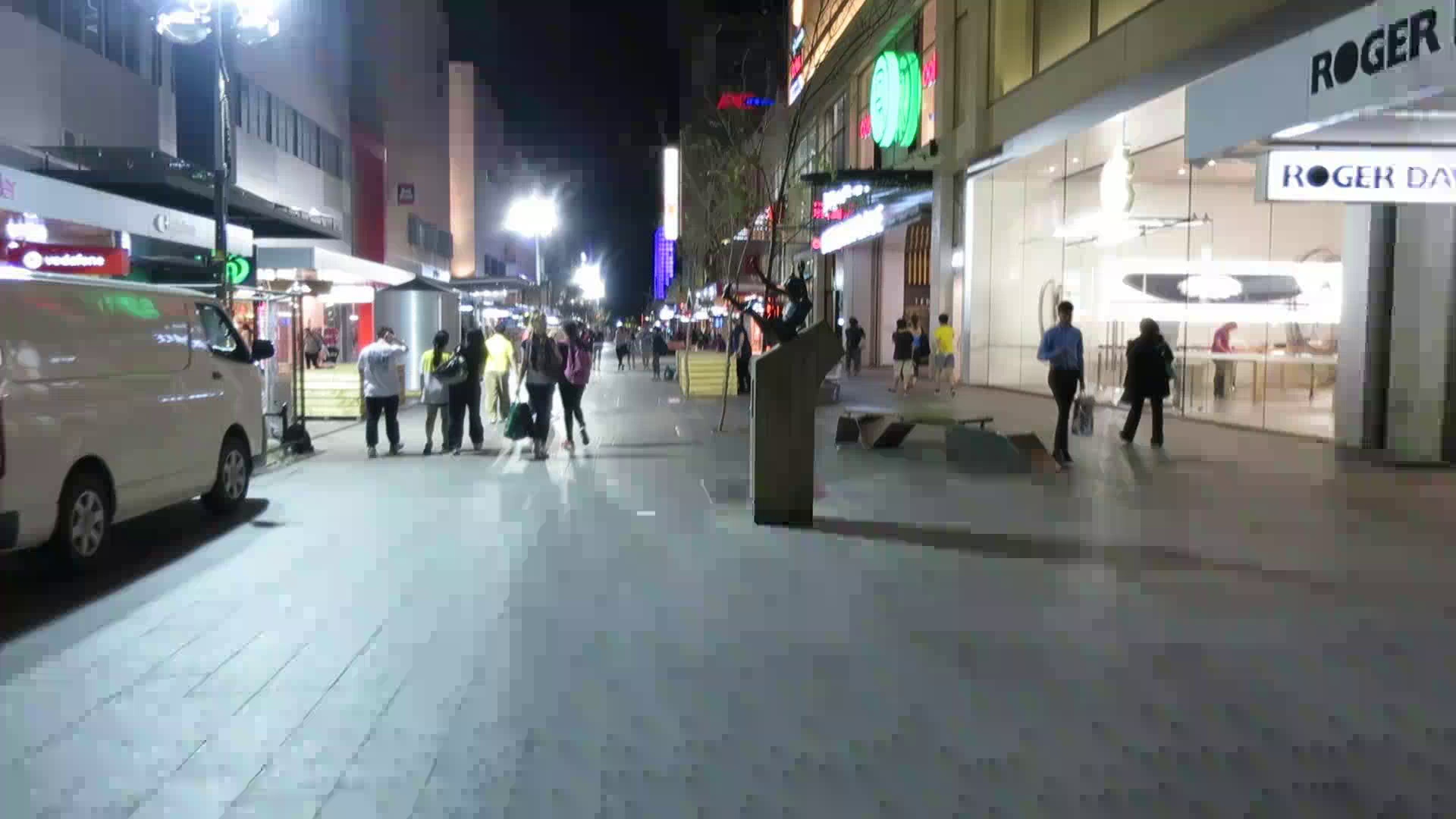}
    \includegraphics[width=0.15\textwidth]{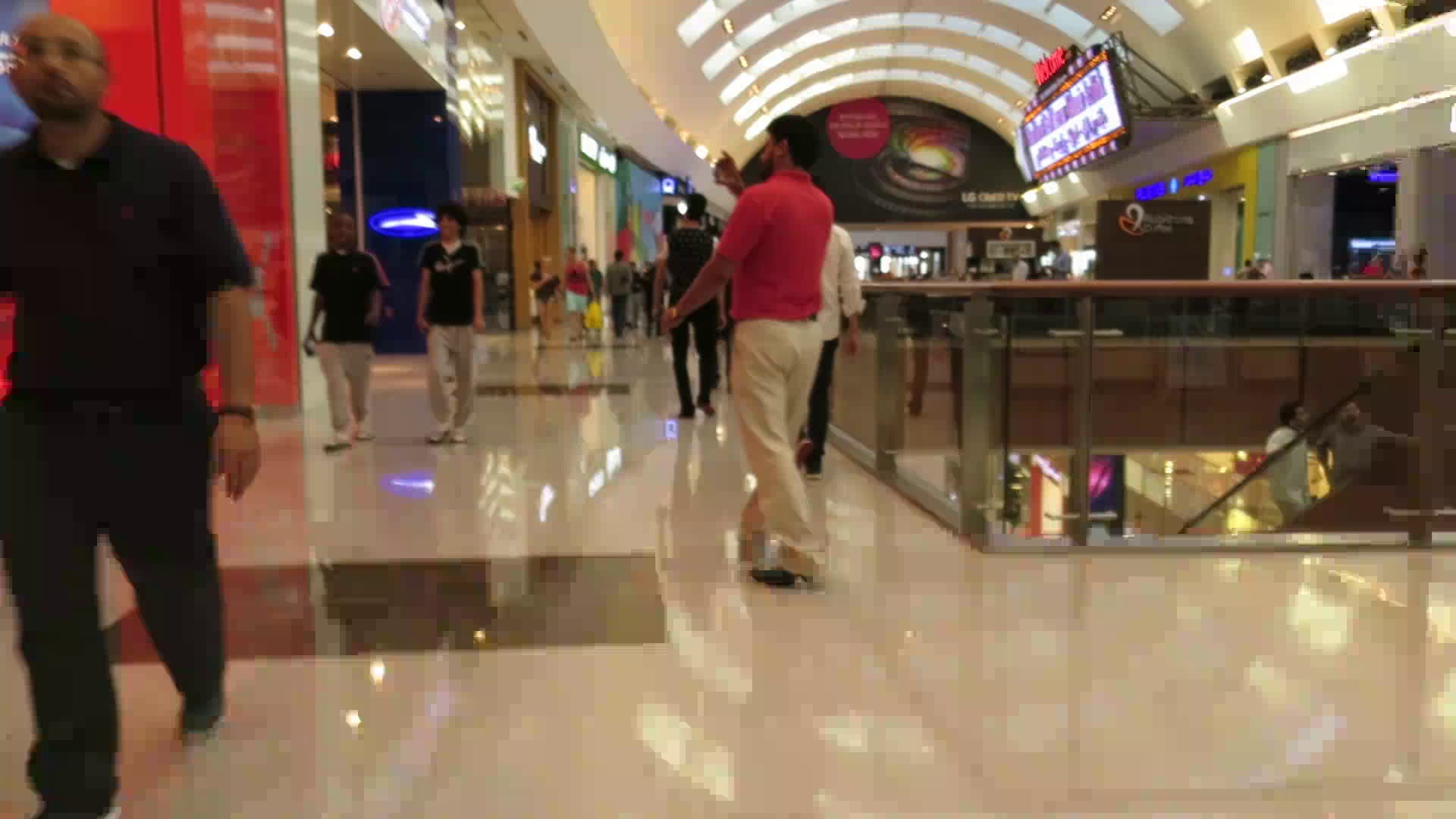}
    \includegraphics[width=0.15\textwidth]{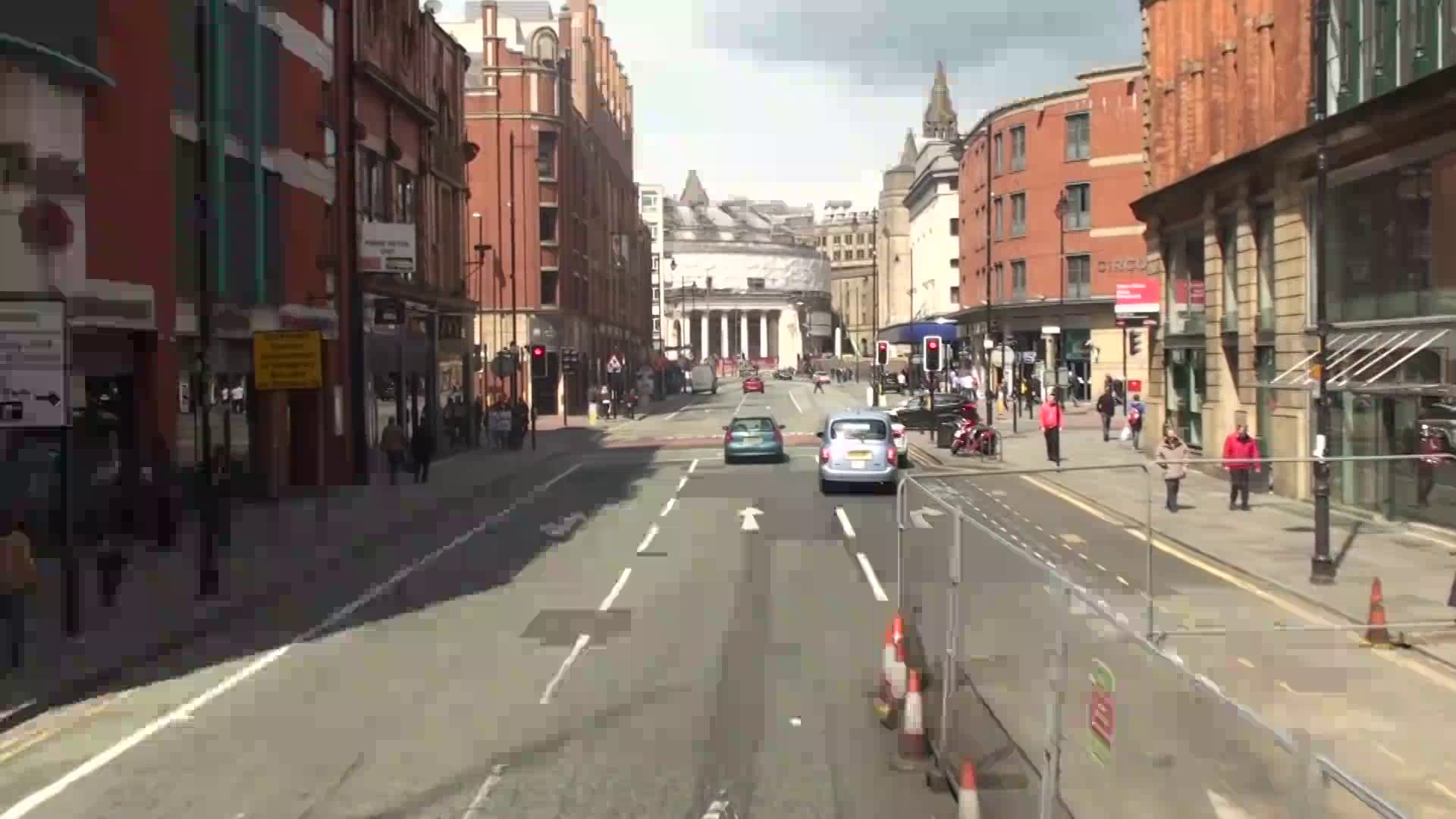}
    \\
    \caption{Qualitative visualization of post-compression frames across all methods. (zoom in for better visualization)} 
    \label{fig:visualize-all}
\end{figure*}

See Fig.~\ref{fig:visualize-all}.

\subsection{Lower SSIM means Better Compression?}
We elaborate the rationale in detail in Appendix D. In our framework, lower SSIM arises primarily from aggressive compression of task-irrelevant regions (\eg background), while preserving information in task-relevant areas (\eg object boundaries or foreground objects). This localized degradation leads to lower global SSIM, yet the information critical to the downstream task is retained. Because most vision models depend heavily on structural cues (\eg edges, contours), this selective compression often maintains (if not improves) task accuracy. Thus, in our context, lower SSIM is a sign of more efficient task-aware compression

\begin{figure*}[thb]
\centering

\begin{minipage}[t]{0.48\linewidth}
    \centering
    \makebox[0.99\textwidth]{
        \makecell{
            \includegraphics[width=0.44\textwidth,height=0.24\textwidth]{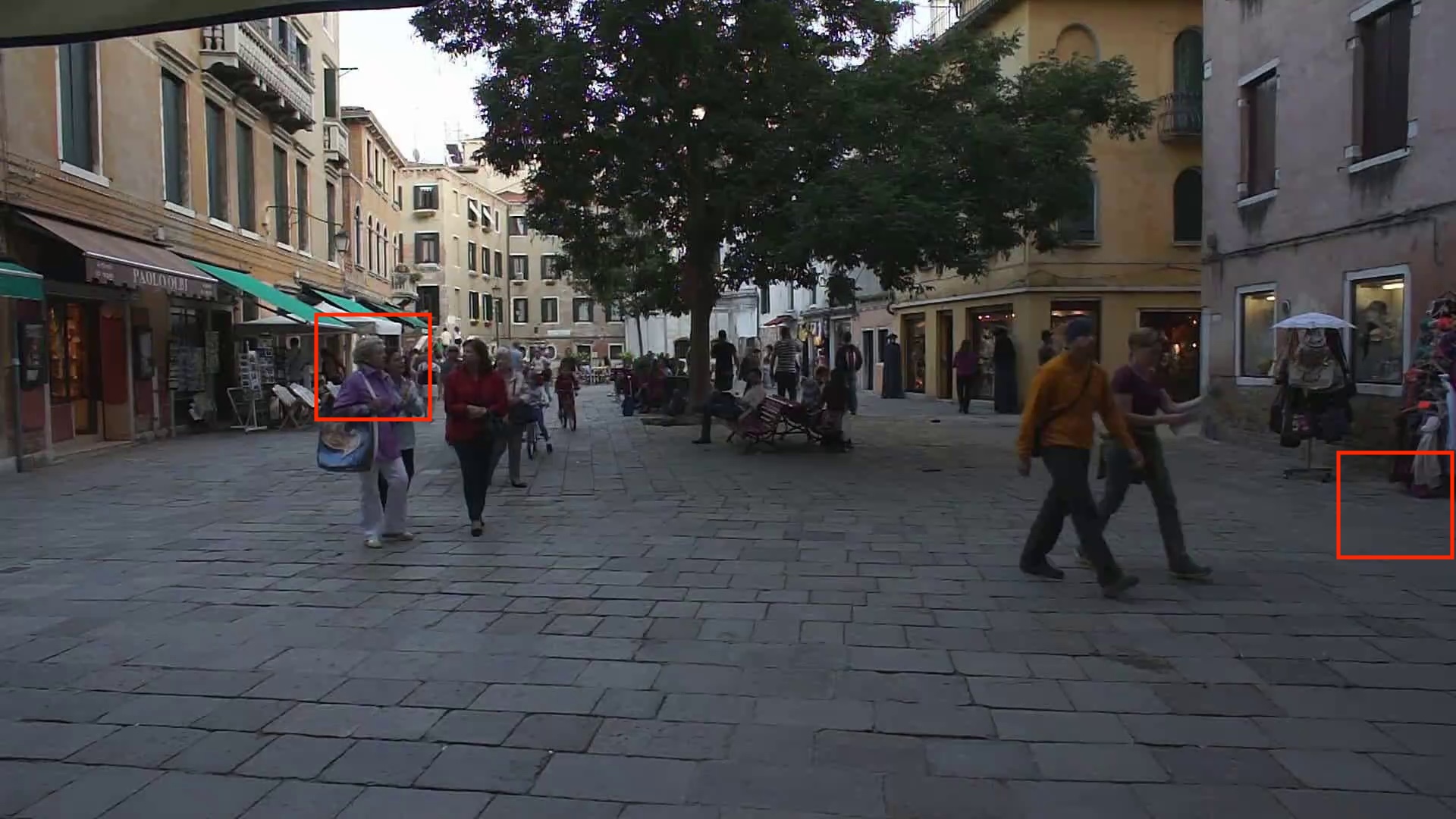}
        }
        \hspace{-1em}
        \makecell{
            \includegraphics[width=0.215\textwidth,height=0.12\textwidth]{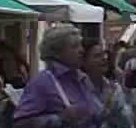}
            \includegraphics[width=0.215\textwidth,height=0.12\textwidth]{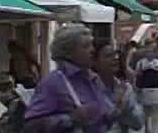}
            \\[-0.2em]
            \includegraphics[width=0.215\textwidth,height=0.12\textwidth]{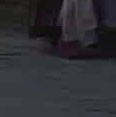}
            \includegraphics[width=0.215\textwidth,height=0.12\textwidth]{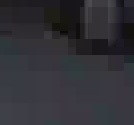}
        }
    }
    
    \makebox[0.5\textwidth]{\small Raw (before compression)}
    \makebox[0.2\textwidth]{\small AccMPEG}
    \makebox[0.28\textwidth]{\small \framework}
\end{minipage}
\hfill
\begin{minipage}[t]{0.48\linewidth}
    \centering
    \makebox[0.99\textwidth]{
        \makecell{
            \includegraphics[width=0.44\textwidth,height=0.24\textwidth]{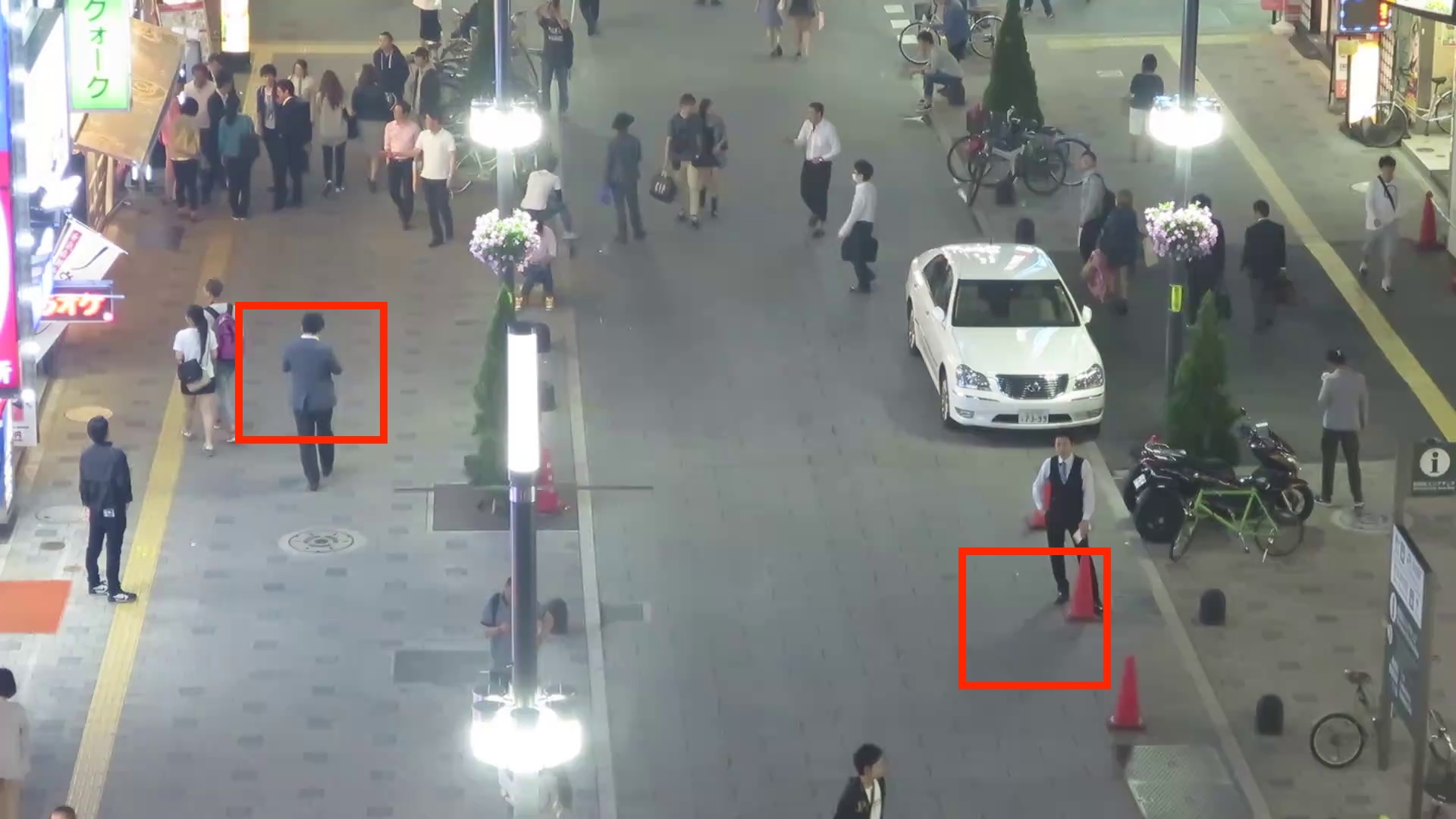}
        }
        \hspace{-1em}
        \makecell{
            \includegraphics[width=0.215\textwidth,height=0.12\textwidth]{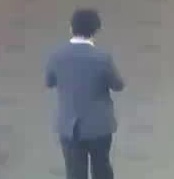}
            \includegraphics[width=0.215\textwidth,height=0.12\textwidth]{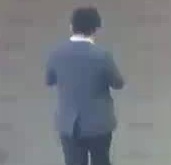}
            \\[-0.2em]
            \includegraphics[width=0.215\textwidth,height=0.12\textwidth]{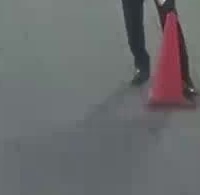}
            \includegraphics[width=0.215\textwidth,height=0.12\textwidth]{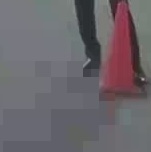}
        }
    }
    
    \makebox[0.5\textwidth]{\small Raw (before compression)}
    \makebox[0.2\textwidth]{\small AccMPEG}
    \makebox[0.28\textwidth]{\small \framework}
\end{minipage}

\end{figure*}

\begin{figure*}[thb]
\centering

\begin{minipage}[t]{0.48\linewidth}
    \centering
    \makebox[0.99\textwidth]{
        \makecell{
            \includegraphics[width=0.44\textwidth,height=0.24\textwidth]{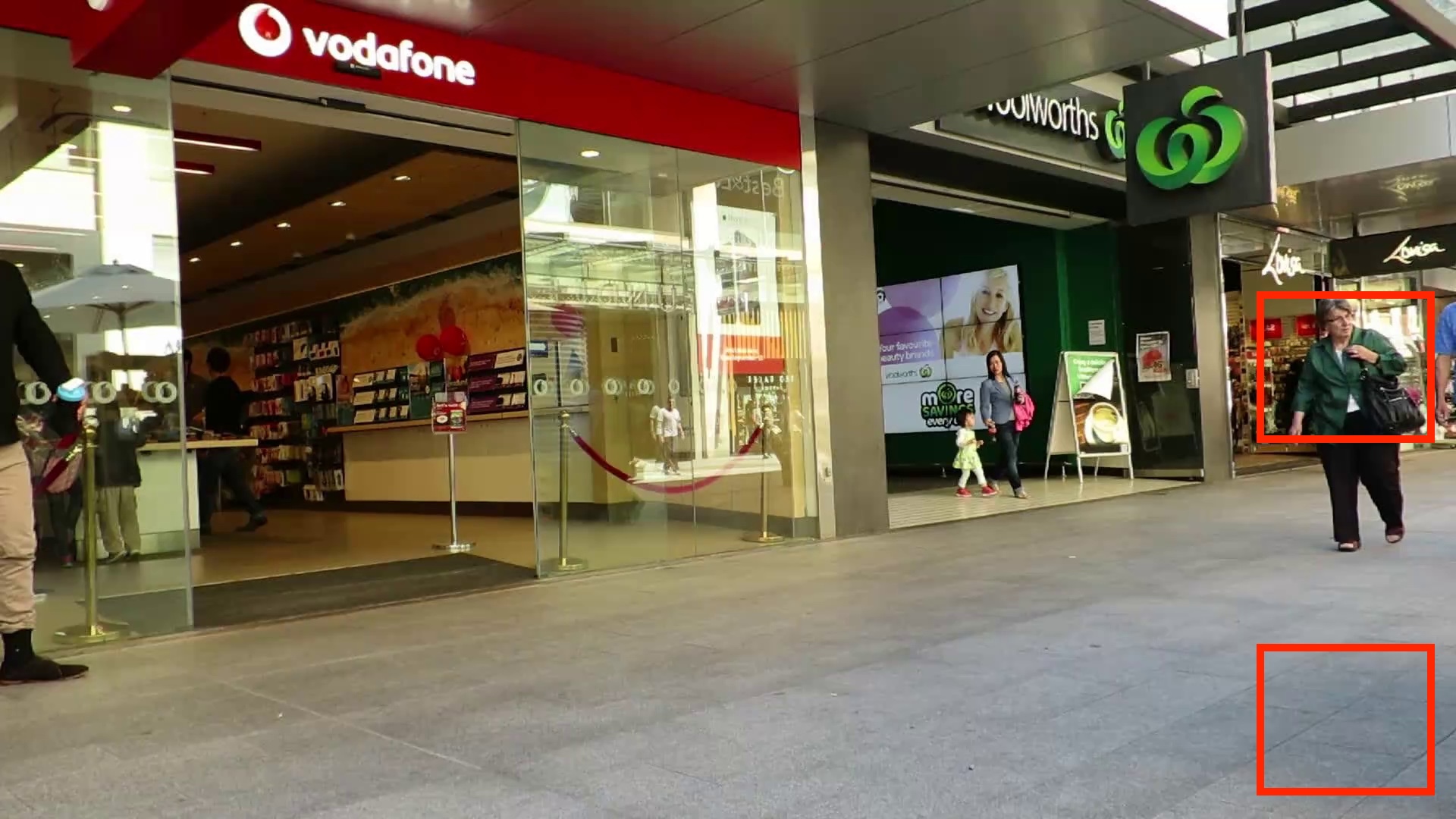}
        }
        \hspace{-1em}
        \makecell{
            \includegraphics[width=0.215\textwidth,height=0.12\textwidth]{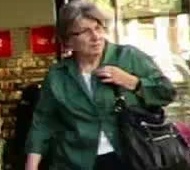}
            \includegraphics[width=0.215\textwidth,height=0.12\textwidth]{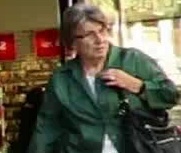}
            \\[-0.2em]
            \includegraphics[width=0.215\textwidth,height=0.12\textwidth]{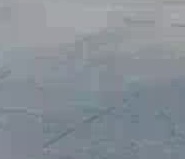}
            \includegraphics[width=0.215\textwidth,height=0.12\textwidth]{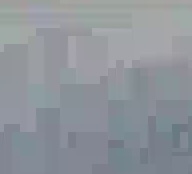}
        }
    }
    
    \makebox[0.5\textwidth]{\small Raw (before compression)}
    \makebox[0.2\textwidth]{\small AccMPEG}
    \makebox[0.28\textwidth]{\small \framework}

\end{minipage}
\hfill
\begin{minipage}[t]{0.48\linewidth}
    \centering
    \makebox[0.99\textwidth]{
        \makecell{
            \includegraphics[width=0.44\textwidth,height=0.24\textwidth]{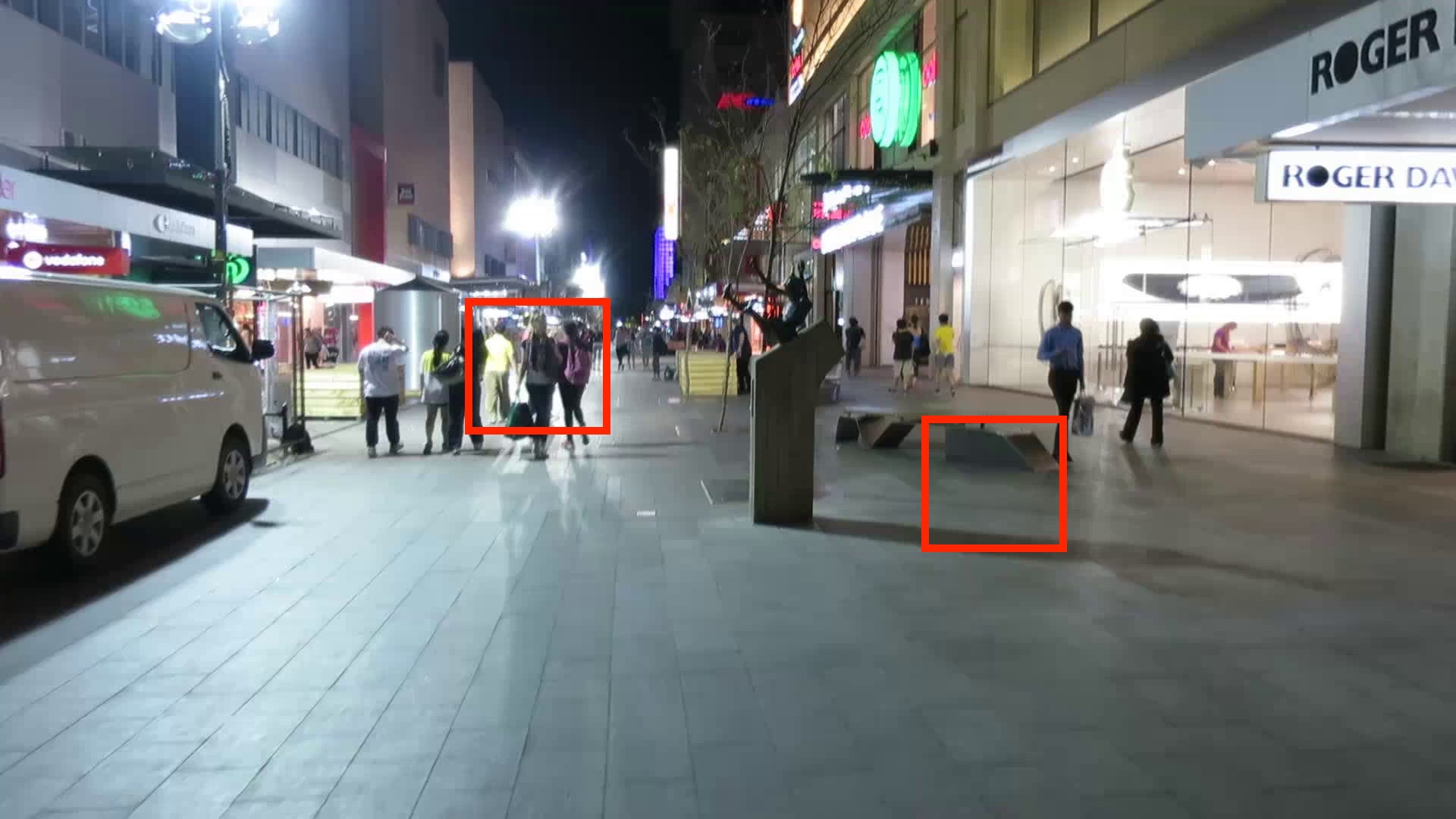}
        }
        \hspace{-1em}
        \makecell{
            \includegraphics[width=0.215\textwidth,height=0.12\textwidth]{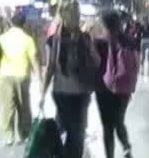}
            \includegraphics[width=0.215\textwidth,height=0.12\textwidth]{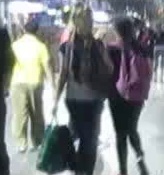}
            \\[-0.2em]
            \includegraphics[width=0.215\textwidth,height=0.12\textwidth]{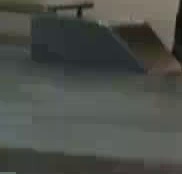}
            \includegraphics[width=0.215\textwidth,height=0.12\textwidth]{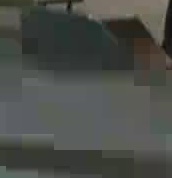}
        }
    }
    
    \makebox[0.5\textwidth]{\small Raw (before compression)}
    \makebox[0.2\textwidth]{\small AccMPEG}
    \makebox[0.28\textwidth]{\small \framework}
\end{minipage}

\end{figure*}

\begin{figure*}[thb]
\centering

\begin{minipage}[t]{0.48\linewidth}
    \centering
    \makebox[0.99\textwidth]{
        \makecell{
            \includegraphics[width=0.44\textwidth,height=0.24\textwidth]{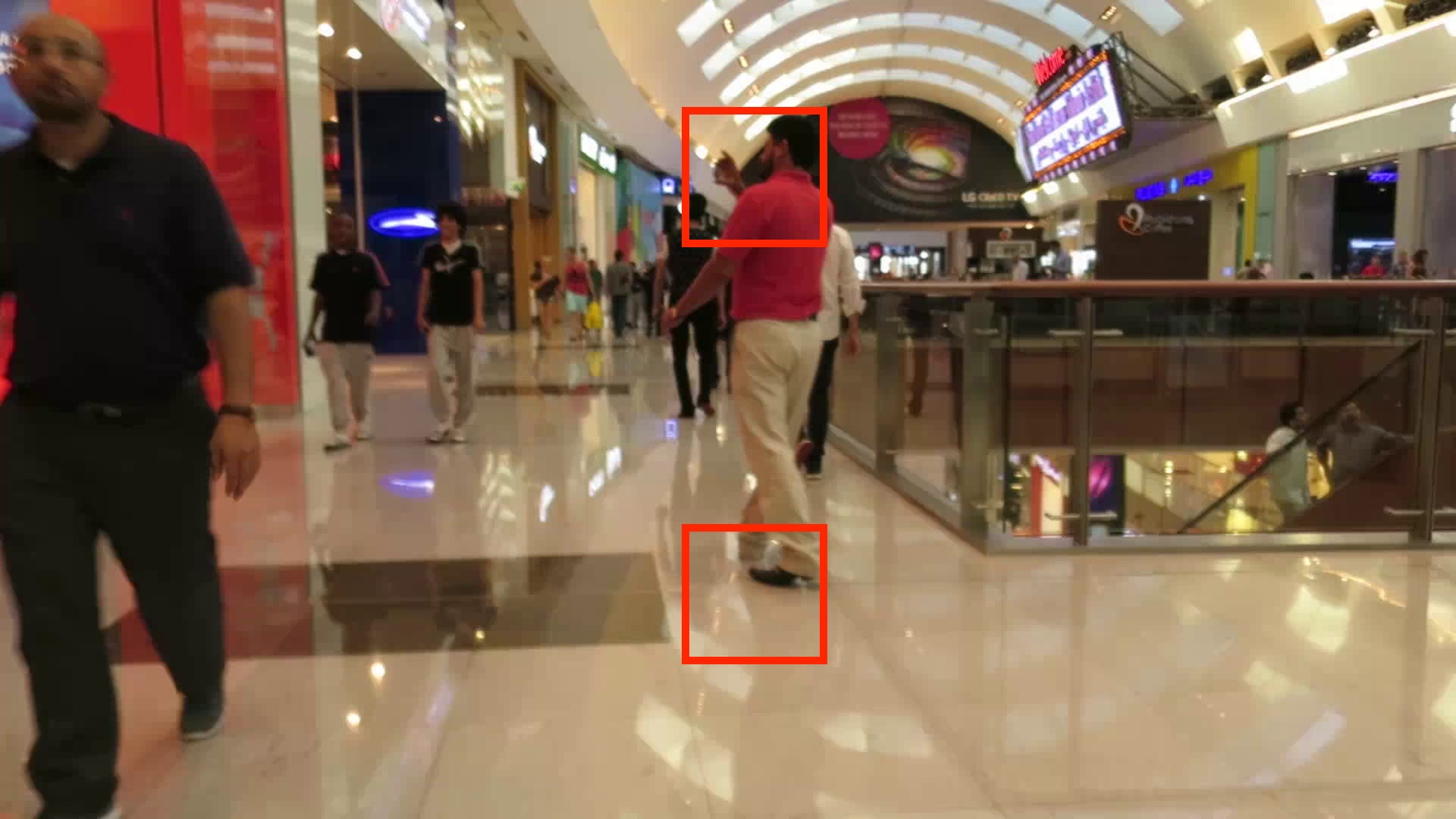}
        }
        \hspace{-1em}
        \makecell{
            \includegraphics[width=0.215\textwidth,height=0.12\textwidth]{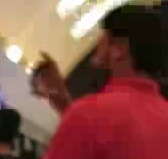}
            \includegraphics[width=0.215\textwidth,height=0.12\textwidth]{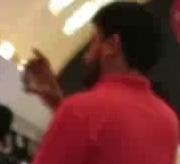}
            \\[-0.2em]
            \includegraphics[width=0.215\textwidth,height=0.12\textwidth]{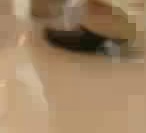}
            \includegraphics[width=0.215\textwidth,height=0.12\textwidth]{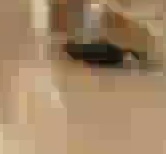}
        }
    }
    
    \makebox[0.5\textwidth]{\small Raw (before compression)}
    \makebox[0.2\textwidth]{\small AccMPEG}
    \makebox[0.28\textwidth]{\small \framework}
\end{minipage}
\hfill
\begin{minipage}[t]{0.48\linewidth}
    \centering
    \makebox[0.99\textwidth]{
        \makecell{
            \includegraphics[width=0.44\textwidth,height=0.24\textwidth]{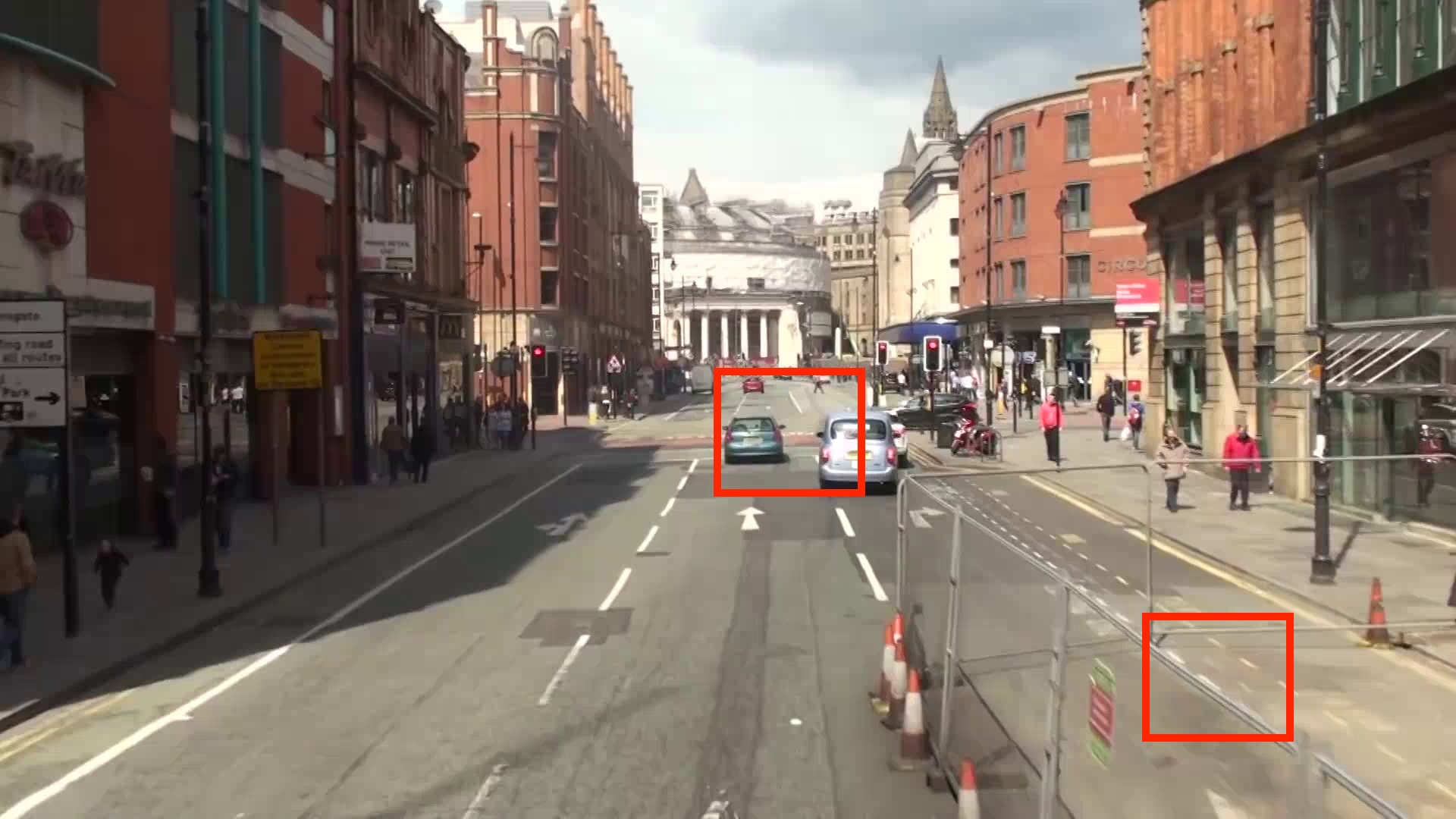}
        }
        \hspace{-1em}
        \makecell{
            \includegraphics[width=0.215\textwidth,height=0.12\textwidth]{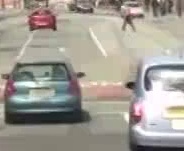}
            \includegraphics[width=0.215\textwidth,height=0.12\textwidth]{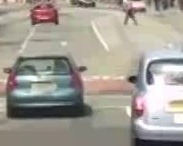}
            \\[-0.2em]
            \includegraphics[width=0.215\textwidth,height=0.12\textwidth]{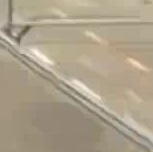}
            \includegraphics[width=0.215\textwidth,height=0.12\textwidth]{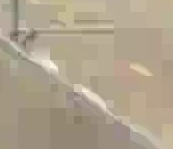}
        }
    }
    
    \makebox[0.5\textwidth]{\small Raw (before compression)}
    \makebox[0.2\textwidth]{\small AccMPEG}
    \makebox[0.28\textwidth]{\small \framework}
\end{minipage}
\caption{Qualitative visualization of compression artifacts. (Zoom in for better visualization)}
\label{fig:artifact-comparison}
\end{figure*}

Figure~\ref{fig:visualize-all} presents representative visualizations of frames from the MOT17 dataset after compression using different methods. These examples qualitatively illustrate how each method affects spatial quality and the allocation of compression across regions.

Our proposed method introduces more visually apparent artifacts and pushes a larger proportion of macroblocks toward lower quality levels. Despite this aggressive compression strategy, the object detection accuracy is well preserved, as demonstrated in Fig.~\ref{fig:artifact-comparison}. This result underscores the effectiveness of our framework in making precise, content-aware decisions regarding both the spatial distribution and magnitude of compression.


\section{Generalization}
\label{apx:generalization}

In the main paper, we implemented \framework using the H.264 codec, focusing on two of its most widely adopted implementations: libx264\cite{h264} and the NVIDIA Video SDK\cite{nvcodec}. Importantly, \framework is not inherently tied to H.264, as it operates independently of the core encoding logic and instead focuses on enhancing the codec’s quality assignment strategy. In this section, we first review the compression mechanisms of modern codecs, then illustrate how \framework can be seamlessly integrated into each of them with minimal modification.

\subsection{compression mechanism of different codecs}

Modern video codecs (\ie H.265 (HEVC)\cite{h265}, VP9\cite{vp9}, AV1~\cite{aom_av1}, and H.266 (VVC)) have evolved significantly from the rigid, fixed-block architectures of earlier standards. These codecs employ increasingly flexible and content-adaptive partitioning strategies that enable more efficient encoding of diverse visual scenes.

H.264/AVC typically uses a fixed $16 \times 16$ macroblock structure and follows a relatively rigid pipeline of intra/inter prediction, transformation, quantization, and entropy coding. While effective in its time, its limited spatial adaptivity reduces efficiency, especially for high-resolution or visually complex content.

H.265/HEVC introduces Coding Tree Units (CTUs) that support variable sizes up to $64 \times 64$. These CTUs are hierarchically partitioned into Coding Units (CUs), Transform Units (TUs), and Prediction Units (PUs), allowing finer spatial granularity. This design improves coding efficiency compared to H.264, particularly for scenes with heterogeneous texture or motion.

VP9 adopts a similar superblock-based design (up to $64 \times 64$) and further enhances adaptivity through variance-based adaptive quantization (AQ) and segmentation maps. Additional features such as tile-based parallel decoding and cyclic refresh support better temporal robustness and hardware parallelism.

AV1 extends these ideas by supporting $128 \times 128$ superblocks, multiple transform modes, and block-wise quantization control via segmentation and delta QIndex. VVC (H.266) further advances this trajectory with additional partitioning modes and generalized CTU structures, maintaining high efficiency across various content types.

Despite their advantages, the deployment of HEVC, VP9, AV1, and VVC in real-time edge environments is limited by their computational demands and the scarcity of mature, power-efficient hardware support. In contrast, H.264 continues to offer a favorable tradeoff between performance and efficiency. As shown in Fig.~\ref{fig:comparison-against-other-codecs}, our interleaved QP assignment further reduces H.264's encoding latency, enabling it to achieve faster encoding while maintaining competitive quality and bitrate.

\subsection{Integration with H.265 (HEVC)}

H.265~\cite{h265} supports spatially adaptive quantization via the \textit{delta QP (dQP)}. Each Coding Tree Unit (CTU) is assigned a base QP, and Coding Units (CUs) within the CTU can receive QP adjustments through signed dQP offsets. These offsets are determined based on visual features such as texture complexity or motion, allowing perceptually important regions to receive finer quantization.

\framework naturally aligns with this architecture by generating content-aware QP maps that can be translated into dQP values for each CU. These values are explicitly signaled in the bitstream, enabling precise control over spatial quality.

To implement this in practice, minor modifications to HEVC encoders are sufficient. For example, in the x265 implementation, \framework can override the QP assignment logic in the rate control module at \url{https://github.com/videolan/x265/blob/master/source/encoder/frameencoder.cpp#L604}. Similarly, in the reference HEVC Test Model (HM), CU-level QP control can be integrated by modifying \url{https://vcgit.hhi.fraunhofer.de/jvet/HM/-/blob/master/source/Lib/TLibEncoder/TEncCu.cpp#L1217}.

\subsection{Integration with H.266 (VVC)}

H.266~\cite{h266}, as an extension of HEVC, increases the maximum CTU size to $128 \times 128$ and retains native support for delta QP-based quantization control. Given \framework’s ability to predict per-block QP values, its integration into VVC is straightforward.

Several VVC implementations support external QP map injection. For example, the UVG266 encoder~\cite{uvg-vvc} allows custom QP assignments via the --roi flag, enabling direct input of region-based QP maps produced by \framework. This functionality is realized in the codebase through fine-grained QP control logic (uvg266.h at \url{https://github.com/ultravideo/uvg266/blob/master/src/uvg266.h#L418}).

Other implementations, such as VVENC~\cite{vvenc-aq}, do not provide direct support but can be adapted with minimal changes. In particular, the rate control logic at \url{https://github.com/fraunhoferhhi/vvenc/blob/master/source/Lib/EncoderLib/BitAllocation.cpp#L508} can be modified to accommodate externally specified QP values.

\subsection{Integration with VP9}

The VP9 codec~\cite{vp9}, as implemented in libvpx, supports adaptive quantization at the segment level through AQ modes. Internally, the encoder converts QP values into qindex parameters, which control quantization strength.

To integrate \framework, externally predicted QP values can be translated to qindex equivalents and injected into the encoder by modifying the rate control mechanism. Specifically, the quantization decision logic at \url{https://github.com/webmproject/libvpx/blob/main/vp9/encoder/vp9_quantize.c#L185} can be extended to reflect \framework’s per-segment quality assignments.

\subsection{Integration with AV1}

AV1, the successor to VP9, also supports block-wise quantization control using qindex values. \framework’s predicted QP values can be directly mapped to these indices to guide the encoder’s decisions.

The reference encoder libaom~\cite{aom_av1} provides mechanisms to adjust quantization at the block level, as defined in the AV1 specification at \url{https://aomediacodec.github.io/av1-spec/#quantizer-index-delta-syntax}. Practical integration can be realized by modifying the quantization logic in at \url{https://aomedia.googlesource.com/aom/+/refs/tags/v3.12.0/av1/encoder/av1_quantize.c#764}, allowing injection of \framework-generated quality signals.

\section{limitations}
\label{apx:limitations}

In this section, we discuss the limitations of both the existing frameworks and our proposed \framework. Additionally, we outline promising directions for addressing these limitations, which we leave as future work.

\noindent\textbf{Marginal Compression Gains in Certain Scenes.}
Although \framework consistently achieves notable bitrate savings without sacrificing accuracy in most scenarios, it demonstrates only marginal improvement for certain video content (\eg as shown in Table~\ref{tab:result} for the MOT17-04 sequence). In such cases, the built-in codec mechanisms may already achieve efficient compression due to their ability to organize QP values in a way that enhances inter-prediction and increases the use of skip mode macroblocks, thereby improving compression efficiency.
One of the key advantages of video compression over image compression lies in the exploitation of temporal redundancy through inter-frame prediction. In H.264, skip mode macroblocks (\ie inter-predicted macroblocks that omit motion vectors and residuals) can be used when the content exhibits minimal motion and negligible prediction error. However, externally controlling QP at a fine granularity may disrupt the codec’s internal optimization pipeline, such as motion estimation and residual prediction, ultimately reducing the proportion of skip mode macroblocks.
We believe this issue can be mitigated by introducing temporally-aware QP adjustment mechanisms, which jointly consider visual saliency and motion consistency to align better with the codec’s prediction model, thereby improving the ratio of skipped macroblocks without compromising accuracy.

\noindent\textbf{Unstable Bitrate Behavior in Streaming.}
In real-world video streaming scenarios, especially under bandwidth-constrained or latency-sensitive conditions, maintaining a stable and predictable bitrate is crucial for smooth transmission and adaptive bitrate (ABR) control. However, external fine-grained control of QP, while effective for accuracy-aware compression, which may lead to large fluctuations in bitrate across frames, depending on the spatial and temporal complexity of the scene. This variability may result in buffer underflows, unstable transmission, or degraded Quality of Experience (QoE) in live offloading scenarios. Incorporating bitrate regularization techniques or post-hoc rate control can help address this issue.

\end{document}